\documentclass[fleqn,usenatbib]{mnras}

\usepackage{newtxtext,newtxmath}

\usepackage[T1]{fontenc}

\DeclareRobustCommand{\VAN}[3]{#2}
\let\VANthebibliography\thebibliography
\def\thebibliography{\DeclareRobustCommand{\VAN}[3]{##3}\VANthebibliography}

\usepackage{graphicx}	
\usepackage{amsmath}	
\usepackage[normalem]{ulem}
\usepackage{multirow}
\usepackage[version=4]{mhchem}
\usepackage{array}
\usepackage{xcolor}

\def\e#1{\times 10^{#1}}

\def\msol{\mathrm{M}_\odot}
\def\lsol{\mathrm{L}_\odot}

\def\h2{$\mathrm{H}_2$}
\def\so2{$\mathrm{SO}_2$}
\def\spy{\;\msol~\mathrm{ yr}^{-1}}

\def\citeapos#1{\citeauthor{#1}'s (\citeyear{#1})}  
\def\hc3n{HC$_3$N}
\def\deg{^\circ}

\newcommand*\ohth{{OH~30.1$\,-$0.7}}
\newcommand*\ohtw{{OH~26.5$\,+$0.6}}
\newcommand\kms{km\,s$^{-1}$}

\newcolumntype{x}{>{\setbox0=\hbox\bgroup}c<{\egroup}@{}}

\title[The salty emission of OH~30.1$-$0.7]{The salty emission of the intermediate-mass AGB star OH~30.1$-$0.7}

\author[Danilovich et al]{T. Danilovich$^{1,2,3}$\thanks{E-mail: taissa.danilovich@monash.edu},
A. M. S. Richards$^{4}$,
M. Van de Sande$^{5}$,
C. A. Gottlieb$^{6}$,
T. J. Millar$^{7}$,
A. I. Karakas$^{1,2}$,
\newauthor
H. S. P. M\"uller$^{8}$, 
K. Justtanont$^{9}$,
J. M. C. Plane$^{10}$,
S. Etoka$^{4}$,
S. H. J. Wallstr\"om$^{3}$,
L. Decin$^{3,10}$,
D. Engels$^{11}$,
\newauthor
M. A. T. Groenewegen$^{12}$,
F. Kerschbaum$^{13}$,
T. Khouri$^{9}$,
A. de Koter$^{14,3}$,
H. Olofsson$^{9}$,
C. Paladini$^{15}$,
\newauthor
R. J. Stancliffe$^{16}$
\\
$^{1}$School of Physics \& Astronomy, Monash University, Wellington Road, Clayton 3800, Victoria, Australia\\
$^{2}$ARC Centre of Excellence for All Sky Astrophysics in 3 Dimensions (ASTRO 3D), Clayton 3800, Australia\\
$^{3}$Institute of Astronomy, KU Leuven, Celestijnenlaan 200D,  3001 Leuven, Belgium\\
$^{4}$JBCA, Department Physics and Astronomy, University of Manchester, Manchester M13 9PL, UK\\
$^{5}$Leiden Observatory, Leiden University, P.O. Box 9513, 2300 RA Leiden, The Netherlands\\
$^{6}$Harvard-Smithsonian Center for Astrophysics, 60 Garden Street, Cambridge, MA 02138, USA\\
$^{7}$Astrophysics Research Centre, School of Mathematics and Physics, Queen's University Belfast, University Road, Belfast BT7 1NN, UK\\
$^{8}$Astrophysik/I. Physikalisches Institut, Universit\"at zu K\"oln, 50937 K\"oln, Germany\\
$^{9}$Department of Space, Earth and Environment, Chalmers University of Technology, Onsala Space Observatory, 43992 Onsala, Sweden\\
$^{10}$School of Chemistry, University of Leeds, Leeds LS2 9JT, UK\\
$^{11}$Hamburger Sternwarte, Universit\"at Hamburg, Gojenbergsweg 112, 21029 Hamburg, Germany\\
$^{12}$Koninklijke Sterrenwacht van Belgi\"e, Ringlaan 3, 1180 Brussels, Belgium\\
$^{13}$Department of Astrophysics, University of Vienna, T\"urkenschanzstrasse 17, A-1180 Vienna, Austria\\
$^{14}$Anton Pannekoek Institute for Astronomy, University of Amsterdam, 1090 GE Amsterdam, The Netherlands\\
$^{15}$European Southern Observatory, Alonso de Cordova 3107, Vitacura, Santiago, Chile\\
$^{16}$H.H. Wills Physics Laboratory, University of Bristol, Tyndall Avenue, Bristol BS8 1TL, UK
}

\date{Accepted 2024 November 12. Received 2024 November 11; in original form 2024 August 30}

\pubyear{2023}

\begin{document}
\label{firstpage}
\pagerange{\pageref{firstpage}--\pageref{lastpage}}
\maketitle

\begin{abstract}
We analyse continuum and molecular emission, observed with ALMA, from the dust-enshrouded intermediate-mass AGB star \ohth. We find a secondary peak in the continuum maps, ``feature B'', separated by 4.6\arcsec\ from the AGB star, which corresponds to a projected separation of $1.8\e{4}$~au, placing a lower limit on the physical separation. This feature is most likely composed of cold dust and is likely to be ejecta associated with the AGB star, though we cannot rule out that it is a background object. The molecular emission we detect includes lines of CO, SiS, CS, \so2, NS, NaCl, and KCl. We find that the NS emission is off centre and arranged along an axis perpendicular to the direction of feature B, indicative of a UV-emitting binary companion {(e.g. a G-type main sequence star or hotter)}, perhaps on an eccentric orbit, contributing to its formation. However, the NaCl and KCl emission constrain the nature of that companion to not be hotter than a late B-type {main sequence} star. We find relatively warm emission arising from the inner wind and detect several vibrationally excited lines of SiS ($\varv=1$), NaCl (up to $\varv=4$) and KCl (up to $\varv=2$), {and emission from {low energy levels} in the mid to outer envelope, as traced by \so2}. The CO emission is abruptly truncated around $3.5\arcsec$ or 14,000~au from the continuum peak, suggesting that mass loss at a high rate may have commenced as little as 2800 years ago.


\end{abstract}




\section{Introduction}

The asymptotic giant branch (AGB) phase is a late evolutionary stage that low- and intermediate-mass stars (with initial masses $0.8~\msol\lesssim M_i \lesssim 8~\msol$) pass through towards the end of their lives. During this phase, the stars eject a significant amount of material through a stellar wind, with mass-loss rates in the range $\sim10^{-8}$ to $10^{-4}\spy$.
Some lower-mass stars, with initial masses $\sim 1.5$ to $4~\msol$, are expected to end their lives as carbon stars, after dredging up newly-synthesised carbon from their interiors to their surfaces \citep{Herwig2005,Karakas2014}.
Intermediate-mass stars, with initial masses in the range $\sim4$ to $8~\msol$, are not expected to end their lives as carbon stars \citep{Herwig2005,Karakas2014,Ventura2018,Marigo2020} because they are massive enough to initiate hot bottom burning (HBB), a phenomenon where the star's convective envelope penetrates the hydrogen-burning shell and enables additional (proton-capture) nucleosynthesis processes \citep{Karakas2014}. In particular, HBB destroys \ce{^12C} and \ce{^18O}. The destruction of \ce{^12C} prevents the formation of carbon stars and results in low \ce{^12C}/\ce{^13C} ratios, and the destruction of \ce{^18O} results in very large \ce{^17O}/\ce{^18O} ratios, allowing us to observationally identify stars in which HBB has taken place \citep{Justtanont2015}. We also note that the precise mass ranges of stars that become carbon stars or which undergo HBB depends on their initial metallicity and on some numerical details of the models \citep[such as the treatment of convection and mass-loss rates,][]{Karakas2016}. The numbers given here are for solar metallicity.

A small sample of stars have been shown to have intermediate initial masses based on the prevalence of \h2$^{17}$O emission and the absence of \h2$^{18}$O emission in \textsl{Herschel} observations \citep{Justtanont2015}. Some of these stars have also been found to have low \ce{^12C}/\ce{^13C} ratios \citep[e.g.][]{Delfosse1997}, providing further evidence for HBB. These intermediate-mass stars have all previously been found to have high mass-loss rates, though the precise values of the mass-loss rates have been the subject of much debate since different values are found using different methods \citep[see for example][]{Heske1990,Justtanont2006}.
Several explanations have been put forward to explain these discrepant results, including the concept of a ``superwind'': a recently-initiated period of enhanced mass-loss \citep{Heske1990,Delfosse1997,Justtanont2013}. This scenario could explain the relatively high mass-loss rates obtained from spectral energy distributions (SEDs) and dust modelling, thought to represent the more recent mass loss, while the lower mass-loss rates from CO are thought to be the remnant of an earlier period of lower mass loss.

Thus far, no studies of the circumstellar chemistry of confirmed intermediate-mass AGB stars have been presented.
In this work we focus on the intermediate-mass AGB star \ohth, thought to have a very high mass-loss rate \citep[$>10^{-4}\spy$ based on SED modelling,][]{Justtanont2013,Justtanont2015}. This star has featured in several other studies of extremely dusty stars based on various (spatially unresolved) infrared and radio observations  \citep{Heske1990,Delfosse1997,Justtanont2006}. Although several previous studies have considered \ohth\ to have a superwind, \cite{Decin2019} argued for a binary scenario rather than a superwind.

Atacama Large Millimetre/sub-millimetre Array (ALMA) observations of CO around two intermediate-mass stars (\ohth\ and \ohtw) were presented by \cite{Decin2019} who attributed structures in their circumstellar envelopes (CSEs) to interactions with binary companions. \cite{Decin2019} further argue that the apparently high mass-loss rates could be explained by an equatorial density enhancement of gas and dust arising from gravitational shaping by a companion. This would then change the apparent mass-loss rate depending on the orientation of the system and the line of sight through the CSE.
Certainly some of the structures observed in the winds of both of these stars are similar to structures seen around other AGB stars that are also attributed to binary interactions \citep[for example][]{Cernicharo2015a,Kim2017,Decin2020}.

For AGB stars more generally, it has recently been shown that binary companions, especially solar-like stars emitting notable UV flux, can have an impact on circumstellar chemistry \citep{Van-de-Sande2022}. Observational studies have found some molecular emission distributions that are best explained through the presence of a solar-type companion \citep{Siebert2022,Coenegrachts2023,Danilovich2024}. In particular, \cite{Danilovich2024} showed that asymmetric distributions of SiN, SiC and NS could be attributed to a main sequence companion on a highly eccentric orbit passing through the inner wind of W~Aql every $\sim1100$ years. \cite{Coenegrachts2023} found asymmetric clumps of NaCl emission forming an approximate spiral in the inner wind of the oxygen-rich AGB star IK~Tau, which they attributed to an unseen companion close to the AGB star.

We present here a complete molecular inventory of the existing ALMA data of \ohth\ (also known as V1362~Aql, IRAS 18460 --0254, AFGL 5535), where previously only CO observations were published. In Sect.~\ref{sec:obs} we describe our re-reduction of the ALMA data, present the continuum, and describe our analysis of the molecular lines. In Sect.~\ref{sec:results} we present the detected molecular lines and discuss the implications of these results in Sect.~\ref{sec:discussion}, in particular how these observations support the presence of a companion star. We summarise our conclusions in Sect.~\ref{sec:conclusions}.

\section{Observations}\label{sec:obs}


\ohth\ was observed with two ALMA receivers, Bands 3 and 6, in 2016 and 2017, as part of project 2016.1.00005.S. The frequency ranges, restoring beams, observation dates and ALMA configurations are given for each tuning in Table \ref{tab:oh30obs}. The Band 3 data were observed with a single ALMA configuration (angular resolution $\sim1.3\arcsec$), whereas the Band 6 data were observed with two antenna configurations and, for convenience, we refer to these as the extended (angular resolution $\sim0.3\arcsec$) and compact (angular resolution $\sim1.5\arcsec$) configurations. The Band 3 data were observed with velocity resolutions of 2.6 to 2.9~\kms\ (1~MHz) and the Band 6 data were observed with a velocity resolution around 2.5~\kms\ (2~MHz).
More precise velocity and frequency resolutions are noted in Table \ref{tab:oh30obs} for each tuning. At the time of the ALMA observations, the star was close to minimum light (phase $\phi \sim 0.45$) in its variability cycle.

\subsection{Data reduction}\label{sec:datared}

\subsubsection{Imaging}

Rather than using the data products from the standard ALMA pipeline, we performed an additional self-calibration for \ohth.
First we downloaded the standard ALMA data products for project 2016.1.00005.S from the ALMA Science Archive \citep{Stoehr2017} and reconstituted the calibrated Measurement Sets (MS) for each array configuration and spectral tuning.
We mainly followed the data processing steps as listed for ATOMIUM in Sect.~3.2.1 of \cite{Gottlieb2022}. In summary,
we split out each spectral window of the calibrated target data, adjusting the frequencies to constant velocity with respect to the Local Standard of Rest ($\upsilon_\mathrm{LSR}$) in the target direction.
We identified the line-free channels by inspection of the visibility spectra and (where available) the archive example image cubes. We made continuum-only images for each configuration and used these as a starting model for self-calibration. 
Continuum subtracted image cubes were made for each spectral window as listed in Table \ref{tab:oh30obs}.
The Band 6 data for \ohth\ were observed in two configurations with equivalent spectral tunings, and these were combined as described in Sect.~3.2.3 of \cite{Gottlieb2022} to make combined continuum images and combined spectral cubes. 

The details of the various continuum images we made are given in Table~\ref{tab:continuum}. These include continua for each individual configuration, the combined images created using different weightings, and an additional Band 3 image created with the same beam parameters as the compact Band 6 continuum image, to allow for a more direct comparison, as discussed below.
In all imaging, Briggs weighting \citep{Briggs1999} was used with robust 0.5 to interpolate across unsampled regions of the visibility plane, giving more weight to cells closer to measured samples.  In combining data from extended and compact configurations, we gave both data sets equal weight which gives a full resolution close to that of the extended configuration but with better sensitivity to large angular scales. However, the sampling of the visibility plane is sparsest for the longest baselines, which can lead to slight artefacts.  
To mitigate this we applied a Gaussian taper to the visibility weights as a function of projected baseline length to reduce the weight of data from the longest baselines. The FWHM (full width half maximum) of the taper is selected to correspond to a slightly smaller angular resolution than the desired synthesised beam. This gives a coarser resolution but reduces errors due to unsampled scales. The resulting beam sizes and tapers for the combined continuum data are given in Table~\ref{tab:continuum}. No taper was used for the line data and beam sizes for those cubes are given in Table \ref{tab:oh30obs}.

\subsubsection{Accuracy}\label{sec:accuracy}

The factors contributing to absolute flux scale and astrometric accuracy are described in the ALMA Technical Handbook \citep{Cortes2023}, as applied to similar data in Sect. 3.2.2 of \cite{Gottlieb2022}.
We matched the continuum flux densities on baselines of similar length for the two band 6 configurations to remove relative flux errors to the level of the noise (see Table \ref{tab:continuum}).  The ALMA flux scale in Band 6, for a single observation, is accurate to 5\%, or up to $\sim10\%$ for combined images, and somewhat better in Band 3.

After fitting a 2D Gaussian component to a compact source, the
relative (stochastic) position error is proportional to the beam size divided by the signal to noise (S/N). The constant of proportionality is 0.5 for a well-filled array \citep{Condon1997},  rising for sparser $uv$ coverage such as ALMA at higher frequencies and a more extended configuration. This gives stochastic errors for the position of the stellar peaks of $\sim12$, $\sim7$ and $\sim1$ mas, for Band 3 and the Band 6 lower and higher resolutions, respectively (using the values in Table \ref{tab:continuum} and constants of 0.5, 0.75 and 1). These are relevant for relative separations e.g. from a putative companion.
The phase-reference source was J1851+0035, 3\farcs5 from the target. The main contribution to astrometric uncertainty is errors in transferring the phase reference solutions to the target (approximately [phase rms]/360$\times$ beam). 
The phase rms on a point source was about $25\deg$, $15\deg$, and $10\deg$ at Band 3 and the Band 6 compact and extended configurations, respectively. This leads to astrometric errors of approximately [beam size]/15, /25, and /35, respectively. Using the mean beam sizes (Table \ref{tab:continuum}) this corresponds to astrometric errors of approximately 90 mas, 60 mas, and 10 mas for Band 3 and the Band 6 lower and higher resolutions, respectively.
This is in addition to S/N (stochastic) errors; other contributions such as antenna position errors are negligible except at the highest resolution, giving  total astrometric errors for the continuum peaks of about 91, 63 and 10 mas, respectively. The position of \ohth\ based on our observations, using a weighted mean and taking these astrometric uncertainties into account, is given in Table~\ref{tab:metry}. We find that the measured positions are offset from the observing positions (also given in Table \ref{tab:metry}) by $\sim0\farcs6$ in each coordinate.


\begin{table}
	\centering
	\caption{Sky positions and key parameters of \ohth.}
	\label{tab:metry}
	\begin{tabular}{lll} 
		\hline
		& \multicolumn{1}{c}{Right Ascension} & \multicolumn{1}{c}{Declination} \\
		\hline
Observed position B6	&	18:48:41.91	&	$-$2:50:28.3	\\
Observed position B3	&	18:48:41.91	&	$-$2:50:28.3	\\
\hline					
Band 6 extended$^{a}$	&	18:48:41.94961	&	$-$2:50:28.9241	\\
Band 6 compact$^{a}$	&	18:48:41.94932	&	$-$2:50:28.8653	\\
\smallskip
Band 3$^{a}$	&	18:48:41.95096	&	$-$2:50:28.8333	\\
Mean position$^{b}$	&	18:48:41.9496	&	$-$2:50:28.921	\\
	&	\multicolumn{1}{r}{$\pm 1.3$~mas}	&	\multicolumn{1}{r}{$\pm 7.8$~mas}	\\
\hline
Distance$^c$ & \multicolumn{2}{c}{$3900\pm800$ pc}\\
$\upsilon_\mathrm{LSR}$$^d$ &  \multicolumn{2}{c}{$100.3\pm1.2$~\kms}\\
		\hline
	\end{tabular}
\begin{flushleft}
\textbf{Notes.} ($^{a}$) Astrometric positions in ICRS J2000 obtained from fits to the continuum peaks in this work. Uncertainties as described in Sect.~\ref{sec:accuracy}. ($^{b}$) The weighted mean position, assuming negligible proper motion between epochs at the distance of \ohth, with the associated uncertainty calculated from the standard deviation. ($^{c}$) Distance calculated using the maser phase-lag method by \cite{Engels2015}. ($^d$) LSR velocity calculated in Sect. \ref{sec:lines}.
\end{flushleft}
\end{table}

\subsection{Continua} \label{sec:cont}

\begin{figure*}
	\includegraphics[width=0.45\textwidth]{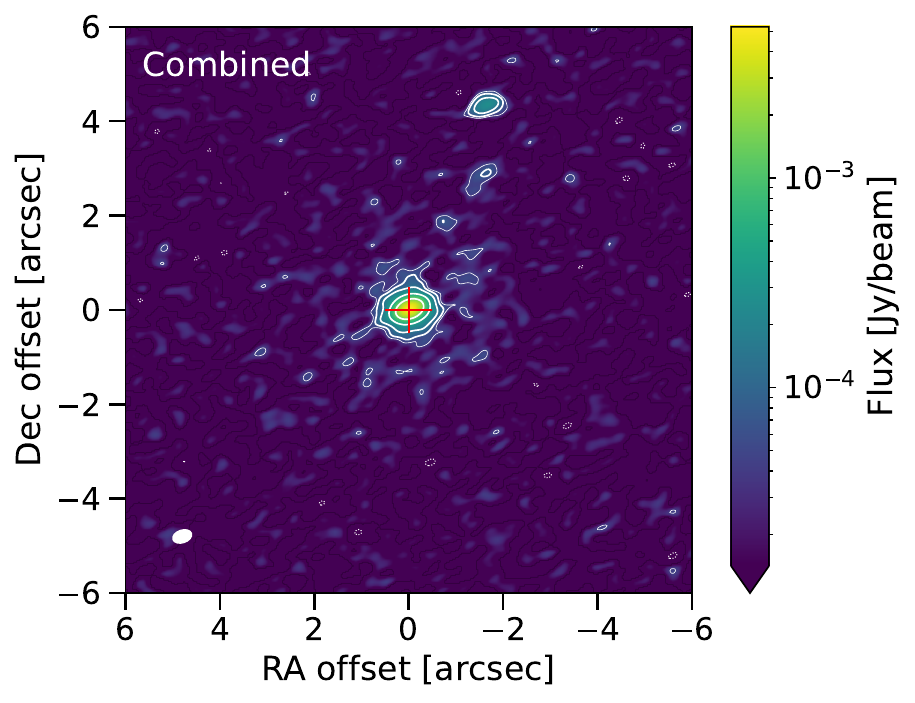}
	\includegraphics[width=0.45\textwidth]{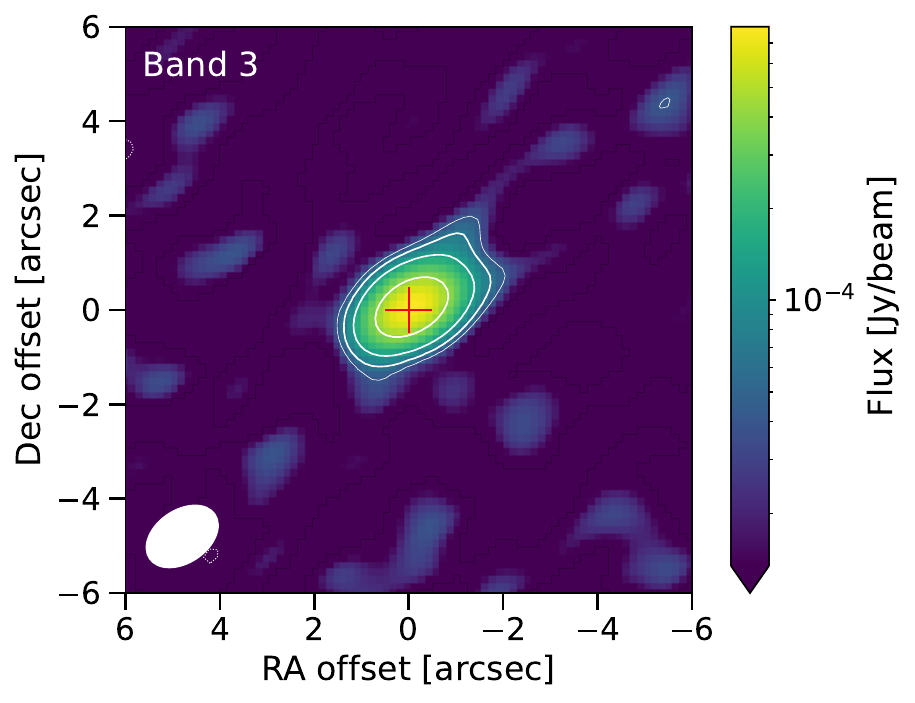}
	\includegraphics[width=0.45\textwidth]{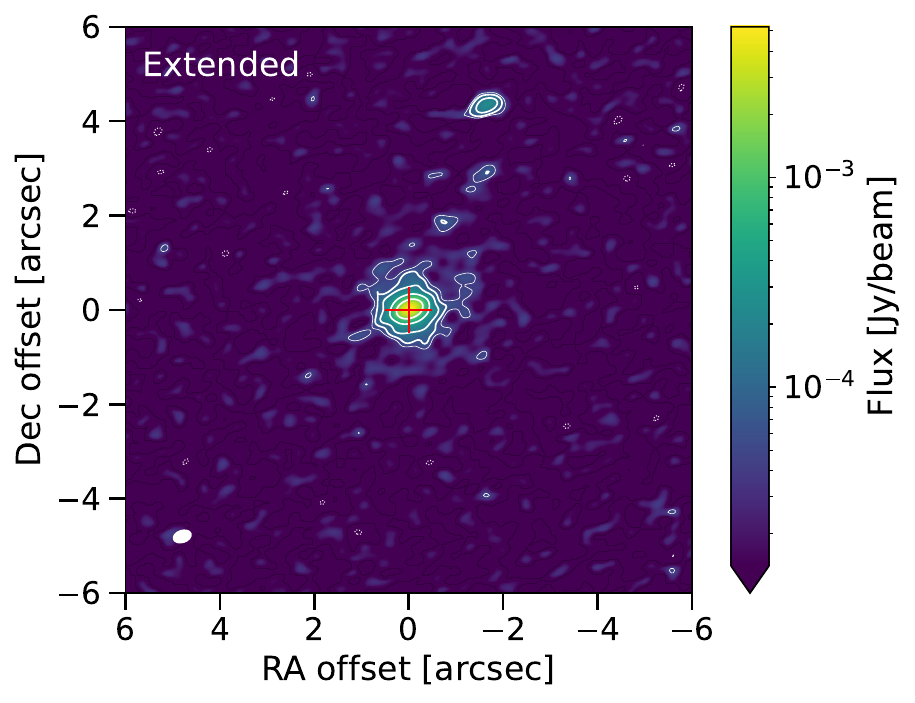}
	\includegraphics[width=0.45\textwidth]{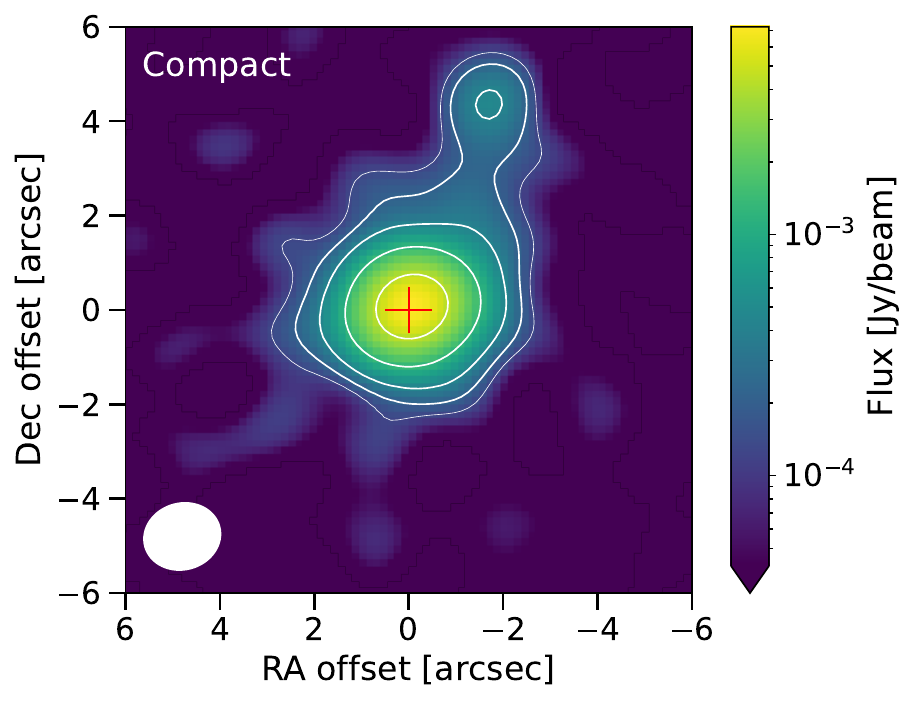}
    \caption{Various continuum images for \ohth, plotted using logarithmic flux scales. The configurations are given in the top left corner of each image and refer to the Band 6 observations unless Band 3 is indicated. The contours indicate levels of 3$\sigma$ (thinnest white line), and 5, 10, 30 and $100\sigma$ (thicker white lines). The red crosses indicate the position of the continuum peak; see Table~\ref{tab:metry} for the astrometric precision of the position. The beams are shown as white ellipses in the bottom left corners. North is up and east is left. Additional details are given in Table~\ref{tab:continuum}.}\label{fig:continua}
\end{figure*}



In Fig.~\ref{fig:continua} we plot the continuum images for each configuration, and the combined Band 6 continuum image. The central continuum emission is resolved in the Band 6 images, and we detect some extended emission. 
The red crosses in Fig.~\ref{fig:continua} indicate the peaks of the continuum emission, which we found through fitting a 2D Gaussian to the extended configuration continua using the CASA routine \texttt{imfit}. In the continuum plots (and all subsequent plots of molecular emission) the data is normalised such that the continuum peaks lie at (0,0). The precise location of this peak, in J2000 coordinates, is given in Table~\ref{tab:metry} as are the positions measured from the other ALMA configurations. We find that the slight differences between positions are close to the predicted astrometric errors.

Considering the combined Band 6 continuum emission, the $5\sigma$ contour shows some ellipticity with a major axis running from southeast to northwest and varying in extent from 0.57\arcsec\ to 0.73\arcsec. The $3\sigma$ contour has a less regular shape, encompassing more faint emission with a greater extent of 1.06\arcsec\ to the southeast and a minimum extent of 0.64\arcsec\ to the northwest. These measurements consider only regions of flux enclosing the continuum peak and neglect isolated islands of emission. We use a similar method to measure extents (including the extents of molecular emission) as that described in \cite{Danilovich2021} for angular sizes. Based on the distance of 3900 pc \citep{Engels2015}, these continuum extents correspond to dust at projected separations of 2200 -- 4100~au from the continuum peak of \ohth. Although we mainly analysed the combined (equal weighting) continuum, we find essentially identical results when considering the continuum from the extended array only. 

\subsubsection{Secondary peak: feature B}

We detect a secondary peak to the north of the primary peak, with a certainty $>10\sigma$. This lies well beyond the extent of the emission centred on the primary peak. As can be seen in Fig.~\ref{fig:continua}, there are some less intense regions of continuum emission (detected at $>3\sigma$) lying between the primary and secondary peaks, hinting at a connection between them. To check whether the secondary peak is truly associated with the AGB star, rather than being a coincidental background or foreground source, we checked the individual configuration data observed with Band 6 and the Band 3 data. The secondary peak is clearly detected in data from both the extended and compact Band 6 configurations. However, the Band 3 continuum for \ohth\ has a lower signal to noise ratio and we do not detect the secondary peak. In Fig.~\ref{fig:oh30-b3-cont} we plot the Band 3 continuum next to the compact configuration Band 6 continuum, with the former processed to have the same restoring beam as the Band 6 data. In the compact Band 6 data, there appears to be a bridge connecting the primary and secondary peaks, traces of which are seen in the combined and extended continuum data in Fig.~\ref{fig:continua}. This suggests an association between the two peaks. For simplicity we will henceforth refer to the secondary continuum peak as ``feature B'' and will assume that the AGB star coincides with the primary continuum peak.
We find the separation between the two peaks to be 4.6\arcsec. At a distance of 3.9~kpc, this gives a projected separation of $1.8\e{4}$~au ($2.7\e{17}$~cm), corresponding to a lower limit on the physical separation. 
Feature B lies beyond the extent of the CO $J=2-1$ and $J=1-0$ emission, as discussed in Sect.~\ref{sec:oh30-sec-peak}. 

\subsubsection{Spectral indices}\label{sec:specind}

Since we have observations from two different ALMA Bands for \ohth, we next calculate the spectral indices of the AGB star and of feature B. This is done using the continuum images from the compact Band 6 configuration and the Band 3 data processed to have the same beam as the compact Band 6 data (see Fig.~\ref{fig:oh30-b3-cont} and Table \ref{tab:continuum}).
The nominal centre frequencies for the Band 3 and Band 6 continua are 107.86 and 236.02 GHz, and the rms are 0.013 and 0.042 mJy respectively. The continuum peaks are 0.788 and 7.26 mJy~beam$^{-1}$. Allowing for the rms and a 7\% flux scale uncertainty as applied to the target, per band, we find a spectral index of $2.8 \pm 0.3$ for the AGB star. 

For feature B, the Band 6 flux is 0.46 mJy~beam$^{-1}$ and for Band 3 the flux in the same region is 0.028 mJy~beam$^{-1}$ (i.e. $<3\sigma$). Using the Band 3 flux as an upper limit, this gives a lower limit on the spectral index of $>3.6$, even steeper than for the AGB star itself. These results are discussed in more detail in Sect.~\ref{sec:oh30-sec-peak}.


\subsection{Line identification}\label{sec:lineid}

\begin{figure*}
\includegraphics[width=\textwidth]{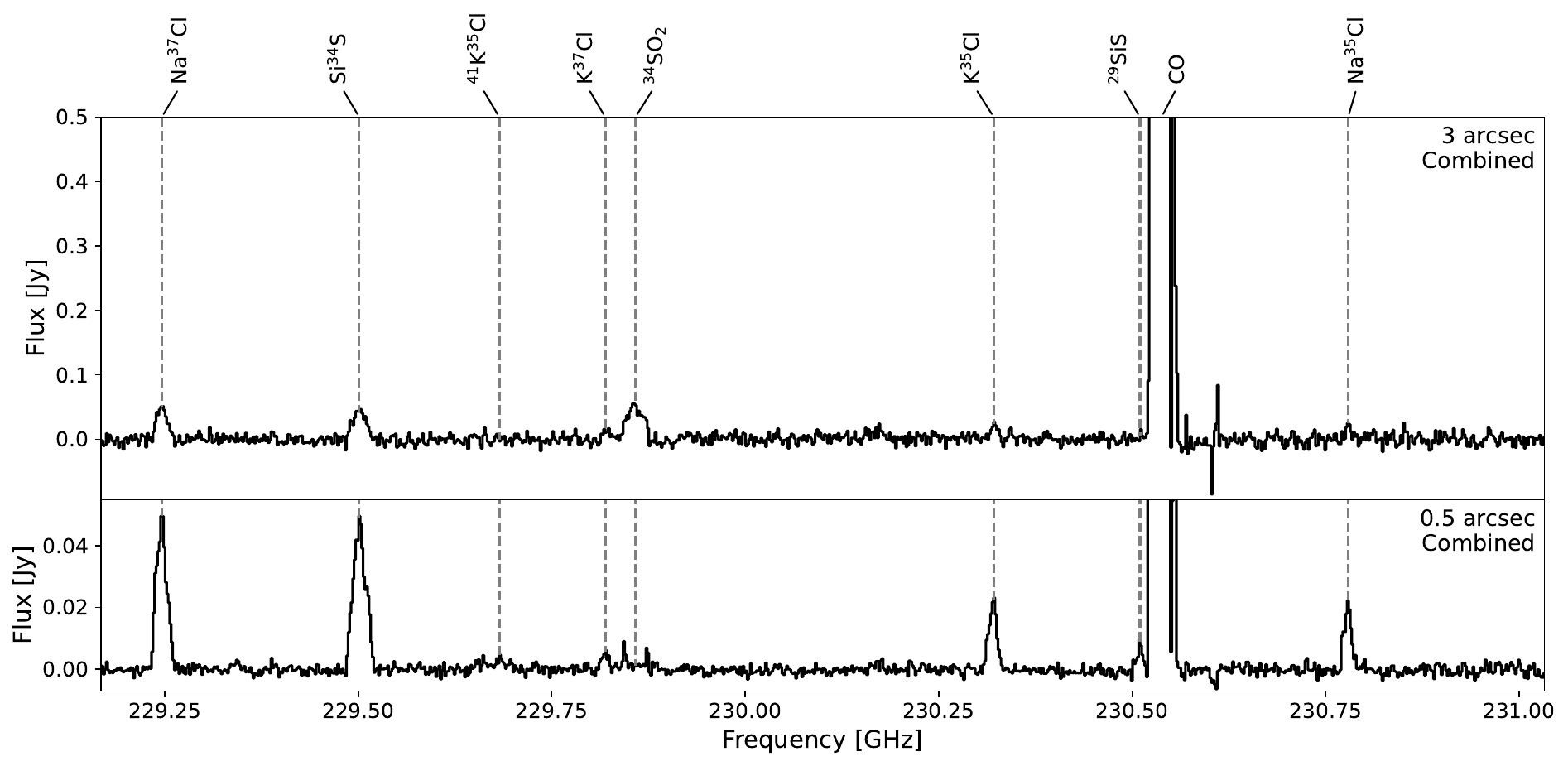}
    \caption{ALMA spectra of one tuning towards \ohth, extracted for two apertures, as indicated in the top right of each subplot. The molecular carriers of various spectral lines are indicated at the top of each plot; see Table \ref{tab:oh30ids} for further details. The narrow feature at 230.6 GHz is interstellar CO contamination (see Sect.~\ref{sec:contam}).}\label{fig:lineidexample}
\end{figure*}

The line identification was initially carried out by extracting spectra in circular apertures centred on the stellar continuum peak. Spectra were extracted for apertures of different radii to ensure that lines excited in different parts of the circumstellar envelope were not overlooked. For example, compact lines, excited close to the star, are most easily detected in spectra extracted for smaller apertures, whereas more extended lines, especially those with shell-like emitting regions, are more easily detected in larger extraction apertures, where the aperture encompasses large fractions of their fluxes. In cases where the line detection appeared marginal in the spectra, we constructed zeroth moment maps (integrated intensity maps) to check for emission. The lines identified towards \ohth\ are listed in Table \ref{tab:oh30ids}. Lines are included in our list of detections if they are seen at levels of $\geq3\sigma$ above the noise in at least one of the spectra and/or the zeroth moment map. Lines marked as tentative either fulfil only one of these criteria or are blended with another line. Zeroth moment maps were constructed to include all channels with emission $>3\sigma$ in the spectra plus an additional 1--2 channels on either side of the line. An exception are the partially blended lines Na$^{37}$Cl and $^{30}$SiS at 227.560 and 227.590 GHz, for which we carefully chose a minimal channel between the two peaks that was not then included in either zeroth moment map, to avoid cross-contamination.

We used data from the Cologne Database of Molecular Spectroscopy \citep[CDMS,][]{Muller2001,Muller2005} to assign carriers to detected lines.
Once we had constructed our initial list of identified lines, we checked which lines from each species were covered by our observations, including lines in vibrationally excited states. The detailed results of these checks are given in Sect.~\ref{sec:results}, but we note here that in some cases we identified additional lines in this manner that are blended with other, often brighter, lines. In those cases, we note the potentially blended lines in the notes columns of Table \ref{tab:oh30ids}.

An example spectrum from the Band 6 data, plotted for two extraction apertures, is given in Fig.~\ref{fig:lineidexample} with the identified lines labelled. The remaining Band 6 spectra are shown in Fig.~\ref{fig:oh30specb6} and the Band 3 spectra are shown in Fig.~\ref{fig:oh30specb3}.

\subsection{Interstellar contamination and resolved-out flux}\label{sec:contam}

Previous observations of CO towards \ohth\ have encountered problems with contamination by interstellar features, see for example \cite{De-Beck2010}. In short, because this star is located in the direction of a densely populated region of the galactic disc, a larger telescope beam --- such as one from a single antenna --- is prone to capture emission, especially of low-energy CO lines, from background or foreground sources along the line of sight, in addition to the emission originating from the target AGB star. The contaminating sources are commonly molecular clouds with relatively narrow lines, compared with the wide lines of AGB circumstellar envelopes, making them easy to distinguish, though not necessarily easy to disentangle from the AGB emission. In the particular case of \ohth, we have identified the massive star-forming region W43 as the likely source of background contamination, which has been found to have gas velocities comparable to the velocities at which we see contamination \citep{Nguyen-Luong2011}. \citep[See also][where \ohth\ can be seen as a bright orange point source on the edge of W43 in their Fig.~2.]{Saral2017}

ALMA has an advantage because, although the ALMA field of view can be similar to the half-power beam width (HPBW) of a single antenna, the interferometer acts as a spatial filter, resolving out smooth large-scale emission, including that from large molecular clouds. {Unfortunately, some contamination persists, as can be seen in our CO channel maps in Figs.~\ref{fig:cochans} and \ref{oh30-co-b3-chan}. A more thorough discussion of the contamination is given in Appendix~\ref{app:contam}.}

{A common problem with observing using an interferometer such as ALMA is the possibility of resolved-out flux -- that is, smooth large-scale flux is filtered out of the observations. However, the maximum recoverable scale of our lowest resolution Band 6 observations is 12.8\arcsec, larger than the CO emitting region (see Fig.~\ref{fig:cochans}), and the spectra of the low and high resolution Band 6 observations are in good agreement, indicating that no flux has been resolved out.}

\subsection{Integrated line fluxes and central velocities}\label{sec:lines}

Having identified the carriers of the lines in our data, as described in Sect.~\ref{sec:lineid}, we performed a basic analysis of all the unambiguous detections of unblended lines. We fit a soft parabola line profile function \citep{Olofsson1993} defined as
\begin{equation}\label{eq:softpara}
F(\upsilon) = F_0 \left( 1 - \left[ \frac{\upsilon - \upsilon_\mathrm{cent}}{\upsilon_\mathrm{w}} \right] ^2\right)^{\gamma/2},
\end{equation}
where the central velocity of the line is $\upsilon_\mathrm{cent}$, the width of the line is $\upsilon_\mathrm{w}$, and the flux at the centre of the line is $F_0$. The fit is done over the free parameters $F_0$, $\upsilon_\mathrm{cent}$, $\upsilon_\mathrm{w}$, and $\gamma$. The soft parabola fit is performed for all extraction apertures in which a given line is detected. In Table \ref{tab:oh30ids} we list the central velocities and velocity widths obtained from these fits and the extraction aperture for which we determined that the line was most clearly detected. This means that for lines with compact emission we fit soft parabolas to spectra extracted for smaller apertures, while for extended emission we used spectra extracted for larger apertures. Generally, we were not able to fit soft parabolas to tentative lines or lines with hyperfine structure. After obtaining the line widths from the soft parabola fits, we used these as a guide to calculate the integrated line fluxes, over velocity ranges a few \kms\ wider than the line widths from the soft parabola fits, to better account for flux that may be in line wings. For each individual line, the same spectrum (extracted for the same aperture) was used to find both the central velocity and the integrated flux. The integrated fluxes are also listed in Table \ref{tab:oh30ids}.

Some examples of lines with their corresponding soft parabola fit are shown in Fig.~\ref{fig:softpara}, where we plot CO ($J=2-1$), low energy lines of SiS and \so2, and a higher energy ($\varv=2$) line of NaCl. The fits to the SiS and \so2 lines produce larger line widths than for CO: $\upsilon_\mathrm{w} \sim 22$~\kms\ compared with $\upsilon_\mathrm{w} = 18$~\kms\ for CO. This is because the automated fit does not account for the blue wing of the CO emission where flux is strongly detected in the channels 76.5--81.5~\kms, despite there not being a clear CO detection in the channel at 84.1~\kms. The reason for this phenomena is discussed further in Sect.~\ref{sec:codiscussion}. To better account for the full velocity width of the CO line, and hence the expansion velocity of the CSE, we calculate the full width of the CO line from the channel maps (Fig.~\ref{fig:cochans}. By considering the reddest and bluest channels with CO detections centred on the continuum peak with a certainty $>30\sigma$, we determine the expansion velocity to be 23~\kms.

By averaging the central velocities found from the soft parabola fits, we calculate a systemic velocity of $\upsilon_\mathrm{LSR} = 100.3\pm1.2$~\kms. This value is in agreement with the LSR velocity of 100.0~\kms\ found by \cite{De-Beck2010} from CO observations. 
The most significant outlier in our sample is the Na$^{35}$Cl ($\varv=1,\,J=8 - 7$) line at 103.415 GHz in Band 3, which has a central velocity of 103.7~\kms. We judge the line unlikely to be misidentified since several other NaCl lines are detected in our observations, including vibrationally excited lines. It is possible that this line might be blended with another, unidentified line. However, we note that the other vibrationally excited NaCl lines tend to lie redwards of the systemic velocity, while the lines in the ground vibrational state lie closer to the systemic velocity (within 1~\kms\ but on the blue side). We also find that the KCl lines overall tend to lie slightly redwards of the systemic velocity (with the exception of one $^{41}$K$^{35}$Cl line). The CO lines also lie 0.6 to 1.6~\kms\ redwards of the systemic velocity. In contrast, the SiS, \so2, and CS lines tend to lie bluewards of the systemic velocity.
These differences are likely because of differences in excitation conditions and regions and certainly contribute to the moderate uncertainty on our averaged systemic velocity.

\section{Detected molecules}\label{sec:results}


In the following subsections we discuss the detections of the various molecular species seen towards \ohth\ in our observations. The observations, including the number of detected lines and the range of lower level energies they cover, are summarised in Table \ref{tab:overview}.
Of particular note are the large number of NaCl and KCl lines (discussed in Sect.~\ref{sec:nacl-kcl-det}) and the NS lines detected (discussed in Sect.~\ref{sec:ns-det}). After carefully checking our data, we were not left with any unidentified lines.

\begin{table}
	\centering
	\caption{An overview of all the molecules detected towards \ohth.}
	\label{tab:overview}
	\begin{tabular}{lxxxccccc}
		\hline
		Molecule  & $E_\mathrm{low}$ [K] & $N$ && $E_\mathrm{low}$ [K] & $N_\mathrm{lines}$ & $N_\mathrm{iso}$ & Sect. & Ref.\\
		\hline
	CO & 16 & 3 && 0--6 & 3 & 2 & \ref{sec:co-det} & 1\\
	SiS & 145--5431 & 11 && 66--1134 & 8 & 4& \ref{sec:sis-det} & 2\\	
	CS & 49 & 1 && 24 & 1 &1& \ref{sec:cs-det} & 3\\	
\ce{SO2} & 19--152 & 4 && 3--108 & 10 &3& \ref{sec:so-so2-det} & 4\\	
	NS &53--373 & 3* && 3 & 2* &1& \ref{sec:ns-det} & 5\\
	NaCl & 199--2263 & 10 && 17--2143 & 9 &2& \ref{sec:nacl-kcl-det} &6\\
	KCl & 333--1563 & 17 && 39--977 & 9 &3& \ref{sec:nacl-kcl-det} & 7\\
		\hline
	\end{tabular}
\begin{flushleft}
\textbf{Notes.} The $E_\mathrm{low}$ columns give the range of lower level energies among the detected lines of that molecule. The $N_\mathrm{lines}$ column indicates the total number of lines from that molecule, including isotopologues but excluding hyperfine components, that were detected. The $N_\mathrm{iso}$ column indicates the total number of isotopologues that were detected. The Sect column gives the reference to the section in this work where observations of that molecule are discussed in more detail. The Ref column gives the reference for the line frequencies included in this work.
(*) indicates that the molecular lines are made up of multiple (hyperfine) components which are not counted as separate lines.
\textbf{References.} 1: \cite{Winnewisser1997}, 2: \cite{Muller2007}, 3: \cite{Gottlieb2003}, 4: \cite{Belov1998,Klisch1997,Muller2005a} 5: \cite{Lee1995} 6: \cite{Caris2002}, 7: \cite{Caris2004}.
\end{flushleft}
\end{table}

\subsection{CO}\label{sec:co-det}

\begin{figure*}
	\includegraphics[width=\textwidth]{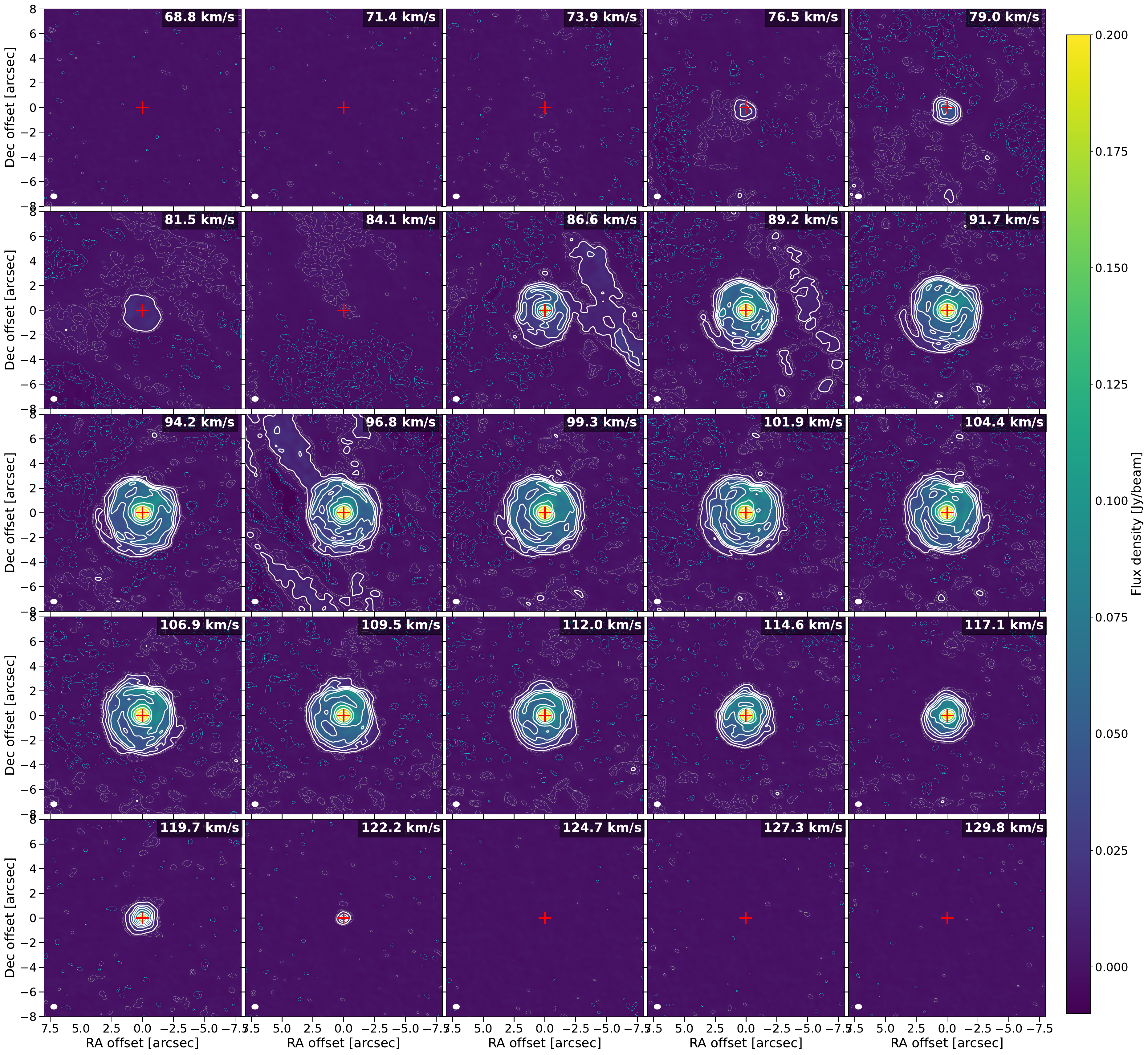}
    \caption{Combined ALMA observations of CO $J=2-1$ towards \ohth. The thicker white contours are drawn at levels of 10, 30, 50, 70, 100, 150, 200, 300, and 500$\sigma$, the thinner grey contours at levels of 3 and 5$\sigma$, and the dashed cyan contours at levels of $-3$ and $-5\sigma$. The red crosses indicate the position of the continuum peak. The velocity of each channel is shown in the top right corner of each panel and the beam is shown as a white ellipse in the bottom left corners. North is up and east is left. Note that there is significant interstellar contamination, especially for channels with velocities lower than 100~\kms.}\label{fig:cochans}
\end{figure*}

\begin{figure*}
	\includegraphics[width=0.32\textwidth]{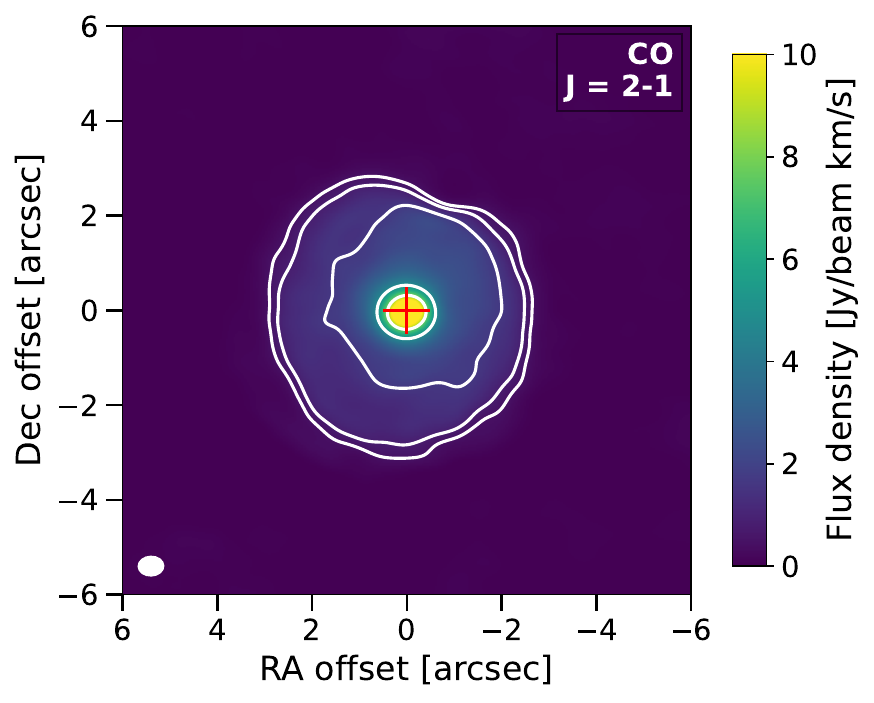}
	\includegraphics[width=0.32\textwidth]{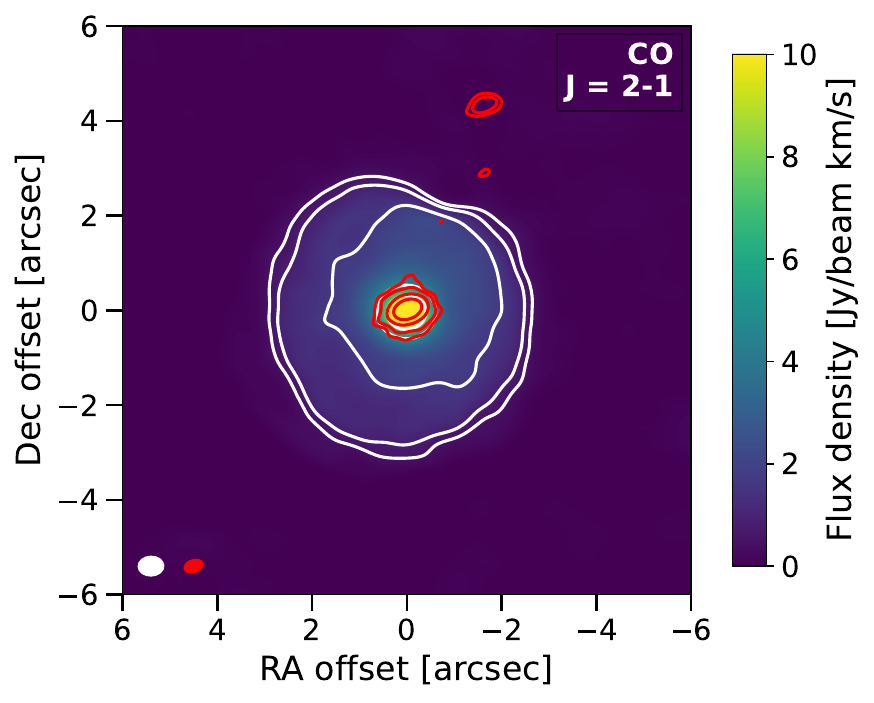}
	\includegraphics[width=0.32\textwidth]{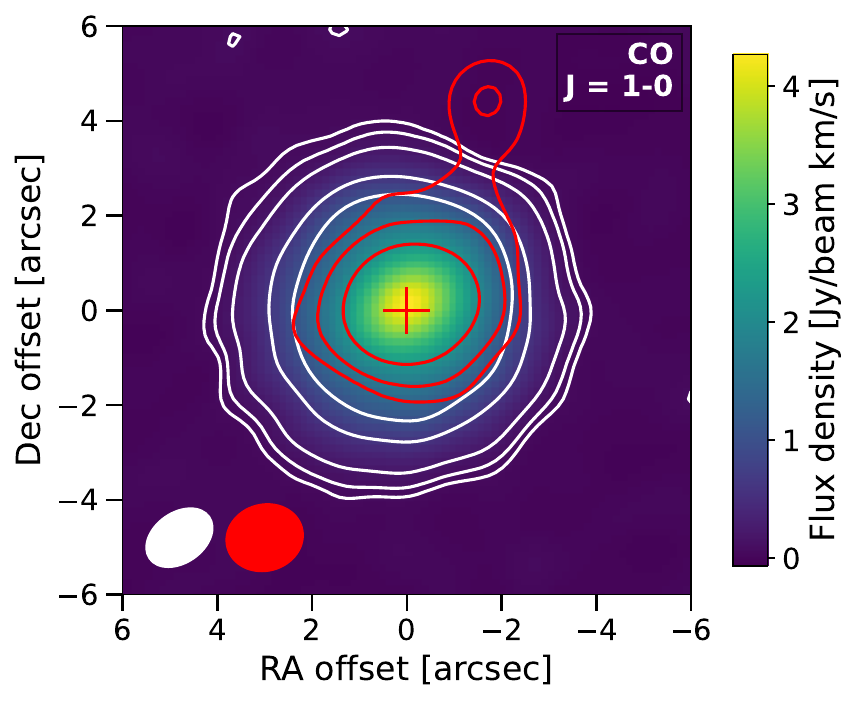}
    \caption{\textit{Left:} The zeroth moment map of CO $J=2-1$ from the combined data towards \ohth. The contours indicate flux levels of 5, 10, 30 and $50\sigma$. The continuum peak is indicated by a red cross and the beam is shown as a white ellipse in the bottom left corner. North is up and east is left. \textit{Centre:} The same CO $J=2-1$ emission overplotted with red contours showing the continuum emission in Band 6 at levels of 5, 10, 30 and 100$\sigma$. The white ellipse in the bottom left corner corresponds to the CO beam and the red ellipse to the continuum beam. \textit{Right:} The intensity map of CO $J=1-0$ integrated over the red channels (from 101.7 to 124.6~\kms, refer to Fig.~\ref{oh30-co-b3-chan}), overplotted with the Band 6 compact array continuum emission. The white contours indicated levels of 3, 5, 10, 30, and 50$\sigma$ in the CO flux and the red contours indicated the 5, 10, and 30$\sigma$ levels of the continuum flux. The white ellipse in the bottom left corner corresponds to the CO beam and the red ellipse to the continuum beam.}\label{fig:oh30cocont}
\end{figure*}


CO is a well-studied density tracer of circumstellar envelopes. 
%
In Fig.~\ref{fig:cochans} we plot the combined channel maps of the $J=2-1$ emission towards \ohth. Because of the large amount of interstellar contamination, we highlight contours at and above $10\sigma$ and also show significant (3 and $5\sigma$) negative contours, which arise from resolved out (interstellar) flux. The CO emission originating from \ohth\ has an almost circular distribution across channels, with the diameter of the emission shrinking towards more extreme red and blue velocities, as is typical of emission from a fairly spherical, radially expanding envelope. The most significant deviation from this is the absence of circumstellar emission in the channel at 84.1~\kms. {This could be the result of CO self-absorption or a background ISM source with a similar flux level visually blended with the CSE and resulting in the CSE being resolved out in this channel. Single-dish observations \citep[e.g.][and Fig.~\ref{fig:co-contamination}]{Heske1990,Justtanont2013} show significant contamination around this channel.} There are also some internal arcs contributing to density structures within the CSE, notably in the channels from 86.6 to 114.0 \kms. The internal structures are more clearly seen in the (overall slightly less sensitive) extended configuration only channel maps, shown in Fig.~\ref{fig:oh30-co-TE-chan}, which also show less contamination owing to a smaller maximum recoverable scale. 
The $J=1-0$ channel maps are shown in Fig.~\ref{oh30-co-b3-chan} and show a similar almost-circular distribution. However, the internal structures are not resolved, owing to a larger beam size ($1.5\arcsec\times1.1\arcsec$, see Table \ref{tab:oh30obs}).

The emission in the central channel of $J=2-1$, which we defined as the channel with LSR velocity $\upsilon_\mathrm{LSR}=99.3$~\kms\ because it is the closest to the systemic velocity, extends out to separations of 2.5 to 3.9\arcsec\ (9800 to 15000 au) from the continuum peak when considering the $3\sigma$ contour. The minimum extent occurs to the north northwest, while the maximum extent is in the southeasterly direction (see Fig.~\ref{fig:oh30cocont}). For the Band 3 data of CO $J=1-0$, the channel closest to the systemic velocity (99.2~\kms) suffers from a large amount of interstellar contamination, so we measured the extent for the channel with $\upsilon_\mathrm{LSR}=101.7$~\kms, and found that the smallest and largest extents were 3.3 and 4.4\arcsec, in the same directions as for the $J=2-1$ emission (see Fig.~\ref{oh30-co-b3-chan}).

In Fig.~\ref{fig:oh30cocont} we plot integrated intensity maps of CO $J=2-1$ and $J=1-0$ emission. For the $J=2-1$ map we also overplot the combined continuum emission. For the $J=1-0$ map we overplot the continuum emission from the Band 6 compact array, which is of a comparable resolution and is sensitive to feature B, unlike the Band 3 continuum. For the $J=1-0$ emission we only integrate over the red part of the CO line (i.e. for $\upsilon \geq 101.7$~\kms) because the ISM contamination is too large and confounds the data at lower velocities.
The comparison in Fig.~\ref{fig:oh30cocont} shows that the ``indentation'' seen in the CO emission to the north-northwest, corresponds well to the location of feature B in the continuum emission. The indentation is most clearly seen in the CO $J=2-1$ emission in the velocity channels from 91.7 to 114.6~\kms\ (Fig.~\ref{fig:cochans}). In the CO $J=1-0$ emission the indentation is less apparent, owing to the larger beam size of those observations. Nevertheless, there is a slight asymmetry in the direction of feature B. 
This phenomenon is discussed in more detail in Sect.~\ref{sec:oh30-sec-peak}.

One line of CO in the first vibrationally excited state was covered in our observations but not detected. In addition to the main isotopologue of CO, we tentatively detect the \ce{C^17O} $J=1-0$ line in the spectrum. The line is not clearly seen in the zeroth moment map and it is possible that there is some interstellar contamination around that frequency.
No lines of \ce{^13CO} or \ce{C^18O} were covered and we note that detections of \ce{C^18O} were not expected, as previous studies have shown evidence of HBB \citep{Justtanont2015}, which destroys \ce{^18O} \citep{Karakas2016}.

\subsection{SiS}\label{sec:sis-det}

\begin{figure*}
	\includegraphics[width=\textwidth]{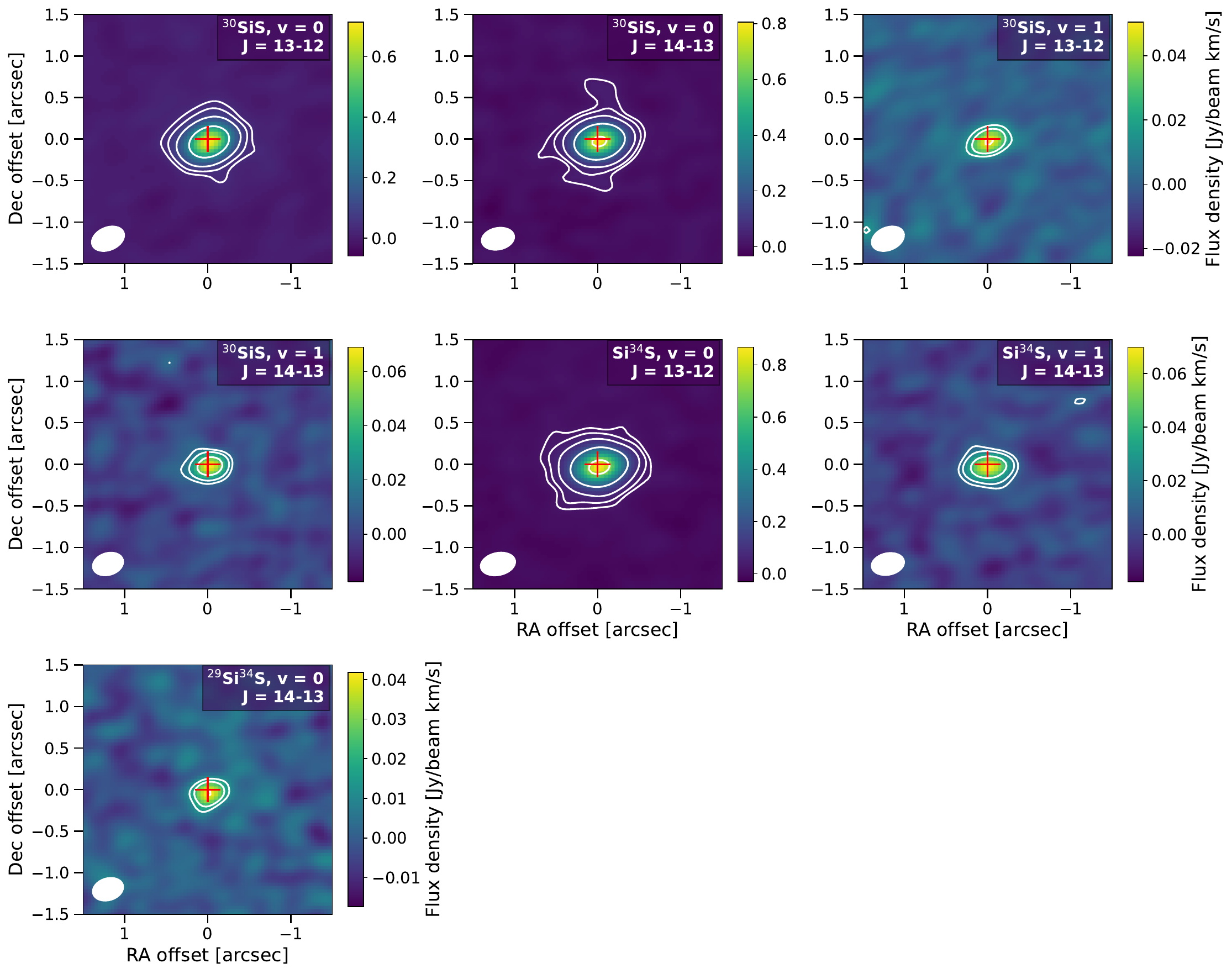}
    \caption{Zeroth moment maps of SiS towards \ohth. The details of the transitions are given in  Table \ref{tab:oh30ids}. The transitions are indicated in the top right of each panel with the details of the transitions given in Table \ref{tab:oh30ids}. The contours are drawn at levels of 3, 5, 10 and 30$\sigma$. The beam is shown as a white ellipse in the bottom left corner and the location of the continuum peak is indicated by the red cross. North is up and east is left.}
    \label{fig:oh30sis}
\end{figure*}

Our spectral coverage of SiS is uneven and misses lines from the most abundant isotopologue, \ce{^28Si^32S} (hereafter we drop the isotope labels for \ce{^28Si} and \ce{^32S}). Therefore, although we detect several lines from isotopologues of SiS, we do not perform an analysis of isotopic ratios in this work.

We detect all the covered lines of \ce{^29SiS}, \ce{^30SiS} and \ce{Si^34S} in the ground and first vibrationally excited states and note that these lines were not covered for \ce{Si^33S}. The vibrational ground state \ce{^29Si^34S} line covered in our Band 6 observations was also detected. No lines in the lowest two vibrationally excited states were covered by our Band 3 observations. The Band 3 data were not sensitive enough to detect any lines of doubly substituted isotopologues.

Zeroth moment maps of our SiS detections are plotted in Fig.~\ref{fig:oh30sis}. The lines in the ground vibrational state of the singly-substituted isotopologues are relatively featureless, with only a few small protrusions seen for the \ce{^30SiS} $\varv=0,\,J=14-13$ line at a level of $3\sigma$. The lines in the vibrational ground state extend out to 0.3 to 0.7\arcsec\ (1200 to 2700~au) from the continuum peak, and the peaks of the SiS zeroth moment maps tend to be offset slightly from the continuum peak in the southwest direction.
The SiS lines in vibrationally excited states and for the doubly-substituted isotopologue are either spatially unresolved or marginally resolved and do not exhibit any structure, though some are still slightly offset to the southwest of the continuum peak. 

\subsection{CS}\label{sec:cs-det}

\begin{figure}
	\centering
	\includegraphics[width=0.35\textwidth]{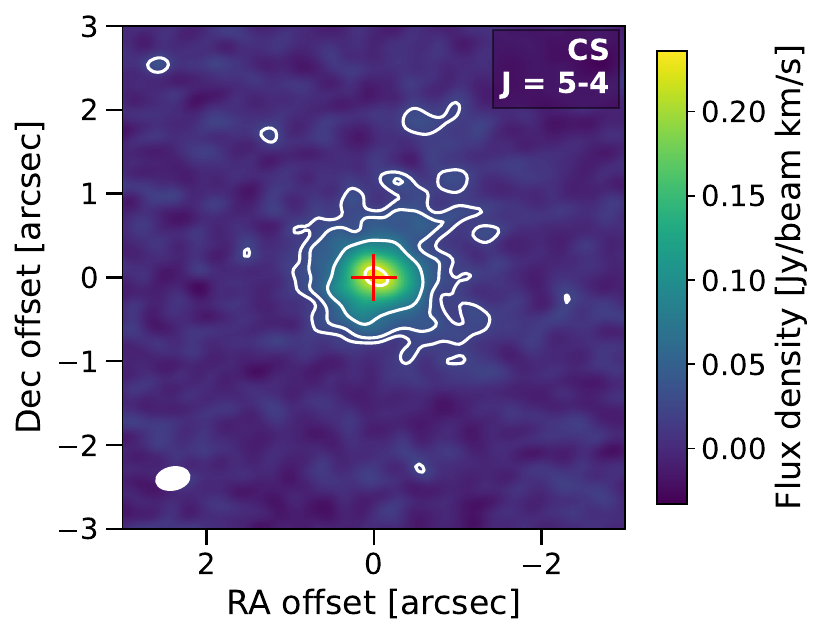}
    \caption{Zeroth moment map of CS towards \ohth. The details of the transition are given in  Table \ref{tab:oh30ids}. The contours are drawn at levels of 3, 5, 10 and 30$\sigma$. The beam is shown as a white ellipse in the bottom left corner and the location of the continuum peak is indicated by the red cross. North is up and east is left.}
    \label{fig:oh30cs}
\end{figure}

One line of \ce{^12C^32S} (hereafter, CS) in the vibrational ground state was detected. The zeroth moment map for CS is shown in Fig.~\ref{fig:oh30cs}. The emission is centrally peaked and the smoothed $3\sigma$ contour extends out to $\sim1\arcsec$ (3900~au). We smoothed the contour by binning and averaging it over $60\deg$ sections, to reduce the ragged edges of the contour. We also neglected isolated islands of emission for this measurement.

No lines in the vibrational ground state of \ce{^13CS} or \ce{C^34S} were covered in our frequency setups. One line of \ce{C^33S} in the ground vibrational state was covered and not detected, as expected given the likely isotopic ratio and the intensity of the corresponding \ce{C^32S} line.

\subsection{\ce{SO2}}\label{sec:so-so2-det}

\begin{figure}
	\centering
	\includegraphics[width=0.49\textwidth]{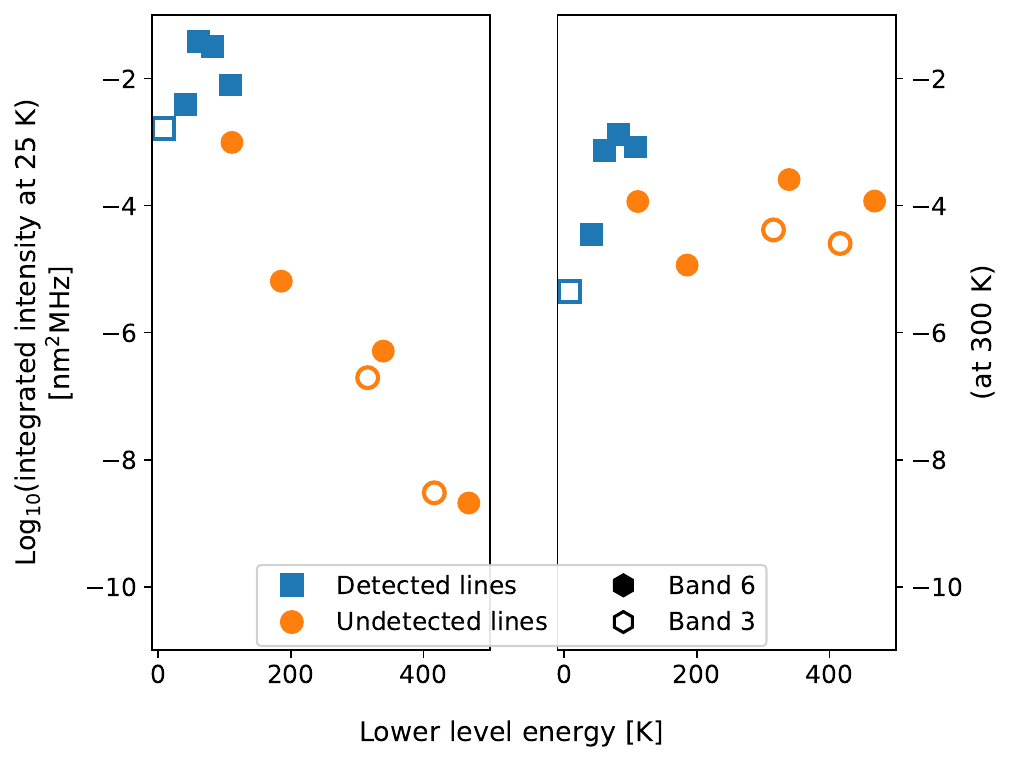}
    \caption{The predicted relative integrated intensities for the \so2 lines covered by our observations with lower level energies $<500$~K. The detected lines are shown as blue squares and the other covered lines as orange circles. Filled markers are lines that were covered in Band 6 and unfilled markers in Band 3. On the left we show the predictions for a gas temperature of 25~K and on the right we show the same for a gas temperature of 300~K.}
    \label{fig:coveredso2}
\end{figure}

\begin{figure*}
	\includegraphics[width=\textwidth]{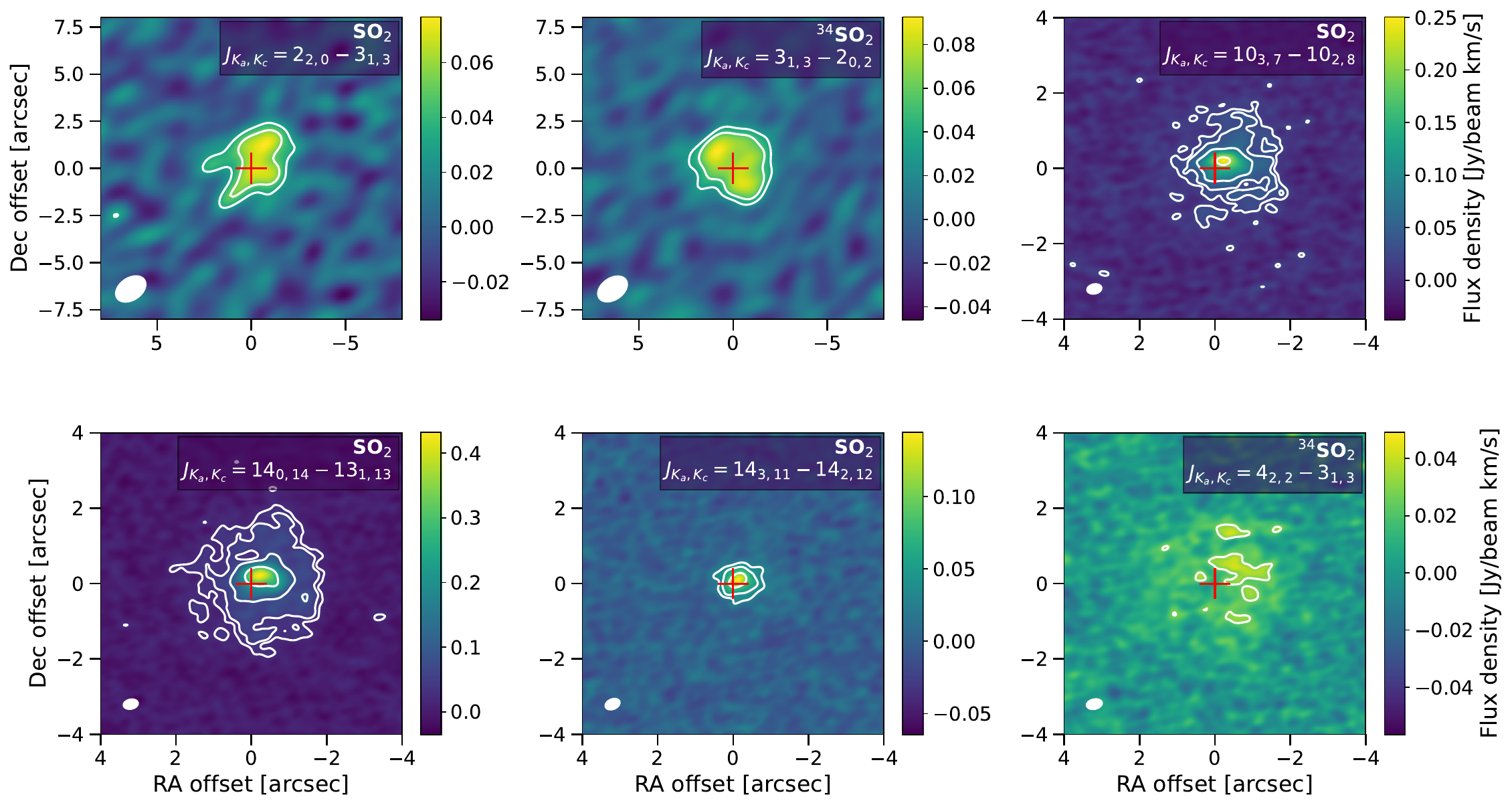}
    \caption{Zeroth moment maps of \so2 towards \ohth. The top left and centre lines were observed with ALMA band 3 and the remaining lines were observed with ALMA band 6. Note the different angular scales and resolutions for the two sets of observations. The transitions are indicated in the top right of each panel with the details of the transitions given in Table \ref{tab:oh30ids}. The contours are drawn at levels of 3, 5, 10 and 30$\sigma$. The beam is shown as a white ellipse in the bottom left corner and the location of the continuum peak is indicated by the red cross. North is up and east is left.}
    \label{fig:oh30so2}
\end{figure*}

\ce{SO2} was detected towards \ohth\ with a shell-like distribution, as expected for a star with a high mass-loss rate and a dense stellar wind {\citep[][see also Sect.~\ref{previousso2}]{Danilovich2016,Wallstrom2024}}. In general, the detected transitions span relatively low energy levels ($E_\mathrm{low}< 110$, see Table \ref{tab:overview}). This is despite higher-energy transitions being covered by our observations, in line with \so2 mainly being formed and excited further out in the wind. A notable non-detection was the \so2 $J_{K_a,K_c}=11_{5,7}-12_{4,8}$ line, which has a lower level energy of 111~K, only 3~K above our highest-energy detected line ($J_{K_a,K_c}=14_{3,11}-14_{2,12}$, $E_\mathrm{low}=108$~K). This result is in line with the trend reported by \cite{Wallstrom2024} for higher mass-loss rate AGB stars to exhibit lower-energy \so2 lines, and lower mass-loss rate AGB stars to exhibit higher-energy \so2 lines.

In Fig.~\ref{fig:coveredso2} we plot the predicted line intensities of all the \so2 lines (taken from CDMS) with lower level energies $<500$~K. If we consider the predicted line intensities at 300~K (see right panel in Fig.~\ref{fig:coveredso2}), then all the covered lines with lower level energies between 100 and 500~K are expected to be brighter than at least one or two lines that were detected. We checked the predicted line intensities for several other temperatures and found that the undetected $J_{K_a,K_c}=11_{5,7}-12_{4,8}$ line is only predicted to be fainter than all the detected lines for temperatures $\lesssim 25$~K. The distribution for predicted line intensities at 25~K is shown in the left panel of Fig.~\ref{fig:coveredso2}. This suggests a surprisingly low excitation temperature for \so2 around \ohth. This result from the most abundant isotopologue is in agreement with the exclusively very low energy transitions (tentatively) detected for \ce{^34SO2} and \ce{^33SO2}.

%

In Fig.~\ref{fig:oh30so2} we plot zeroth moment maps of the most clearly detected \so2 lines. In total, five lines of SO$_2$ were detected, one of which we do not plot because it is blended with a line of KCl, and four lines of $^{34}$\so2 were detected, two of which are bright enough to plot in Fig.~\ref{fig:oh30so2}. We tentatively detected one $^{33}$SO$_2$ line, which we do not plot because the emission is too faint in the zeroth moment map (but is tentatively detected in the spectrum). The two plotted Band 3 lines of \so2 and $^{34}$\so2 coincidentally have the lowest level energies with $E_\mathrm{low}<10$~K. The lines are resolved and show some internal structure in the zeroth moment maps, with visible emission peaks offset from the continuum peak. We find that the Band 3 emission extends out to a minimum of 1.5\arcsec\ and a maximum of 2.5\arcsec\ from the continuum peak for the \so2 line (measured after binning the $3\sigma$ contours, as described for CS), and out to a minimum of 1.7\arcsec\ and a maximum of 2.2\arcsec\ for the $^{34}$\so2 line. The slight difference between these extents and between the corresponding flux distribution patterns shown in Fig.~\ref{fig:oh30so2} are most likely a result of different sensitivities and the slightly different gas temperatures being probed by the lines.
The Band 6 observations of \so2 show well-resolved emission and cover slightly higher energy lines (40--108~K for \so2 and 8--82~K for $^{34}$\so2). The $J_{K_a,K_c} = 14_{0,14} - 13_{1,13}$ line is the brightest covered line ($E_\mathrm{low}=82$~K) and exhibits well-resolved extended emission. As can be seen in the channel maps plotted in Fig.~\ref{fig:so2chans}, the emission peak lies to the northwest of the continuum peak in most channels. The smoothed $3\sigma$ contour in the zeroth moment map (Fig.~\ref{fig:oh30so2}) extends out to 1.1--2.1\arcsec, comparable to the Band 3 emission. The $J_{K_a,K_c} = 10_{3,7}  -  10_{2,8}$ ($E_\mathrm{low}=61$~K) has a very similar distribution, including the offset emission peak, and a slightly smaller extent for the $3\sigma$ contour, ranging from 0.7--1.9\arcsec. This difference most likely arises because the full (outer) extent of the weaker line is below the noise. The highest-energy detected line is $J_{K_a,K_c} = 14_{3,11}  -  14_{2,12}$ with $E_\mathrm{low} = 108$~K. The emission peak of this line is also offset slightly to the northwest. As can be seen in Fig.~\ref{fig:oh30so2}, it has the smallest extent of the well-detected \so2 lines, out to only 0.4--0.8\arcsec. This could be a result of observational sensitivity and the line being intrinsically fainter, or could be an indication that the slightly warmer emission is not distributed as widely as for the {lower-energy} lines. Given our low estimate for the \so2 gas temperature, the former explanation is more likely. 

We also note that the central velocities found for the sufficiently bright \so2 lines (see Table~\ref{tab:oh30ids}) tend to fall on the blue side of the average systemic LSR velocity by 1--2~\kms\ for \so2, and by 0.4~\kms\ for $^{34}$\so2. The two brightest Band 6 \so2 lines are the most blueshifted.
We do not calculate isotopic ratios here because we do not clearly detect any two pairs of lines with the same quantum numbers across isotopologues.

%
%



\subsection{NS}\label{sec:ns-det}

\begin{figure*}
	\includegraphics[height=5cm]{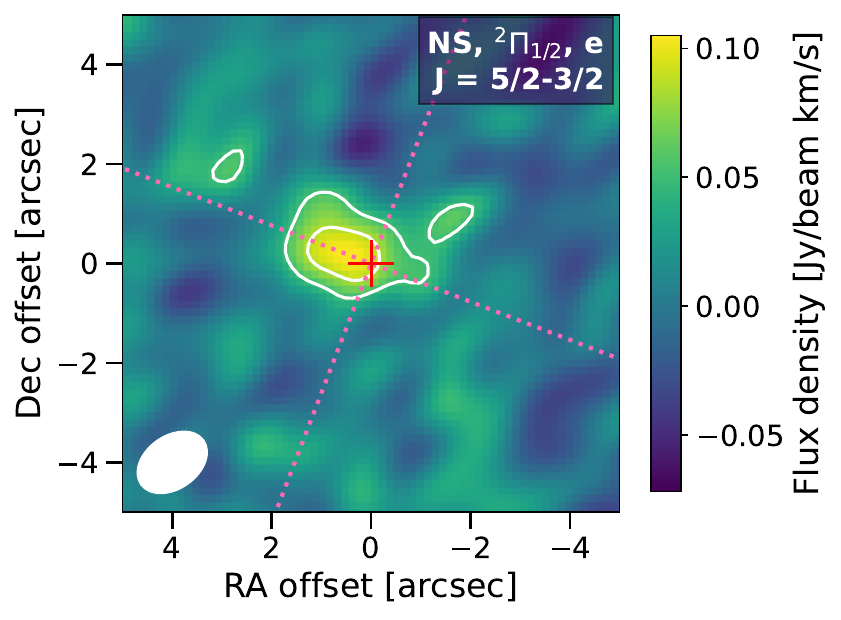}
	\includegraphics[height=5cm]{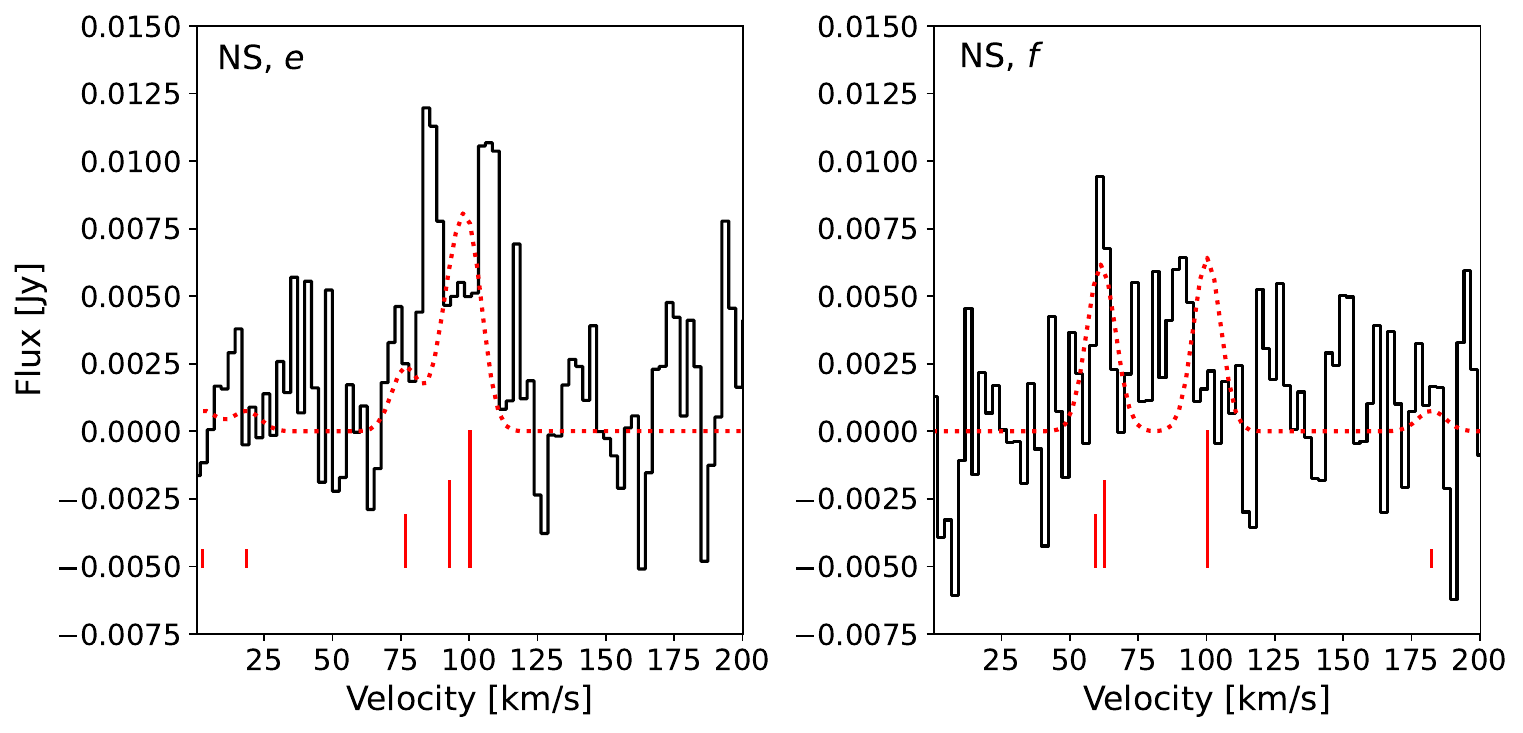}
    \caption{NS emission towards \ohth. The details of the transitions are given in Table \ref{tab:oh30ids}. \textbf{Left:} Zeroth moment map of $e$ component of NS towards \ohth. The contours are drawn at levels of 3 and 5$\sigma$. The pink dotted lines are drawn such that one connects the AGB continuum peak (red cross) with the peak of feature B (not shown) and the second line is perpendicular to this, passing through the AGB continuum peak. The beam is shown as a white ellipse in the bottom left corner. North is up and east is left. \textbf{Right:} Spectra extracted from centred circular apertures with radii of 2\arcsec\ for NS lines observed towards \ohth. The positions of the hyperfine components are shown as vertical red lines with heights corresponding to their relative intensities. The dotted red lines show the predicted line profiles for Gaussian hyperfine components of width 5~\kms. The velocities are calculated with respect to the brightest hyperfine component.}
    \label{fig:oh30ns}
\end{figure*}

NS was only covered by the less sensitive Band 3 observations. Two lines in the $^2\Pi_{1/2}$ ladder were covered, though only the $e$ line was clearly detected. Spectra for both lines are shown in Fig.~\ref{fig:oh30ns} with the hyperfine components marked. In Table~\ref{tab:oh30ids} we list the three brightest hyperfine components for each line. For the $f$ line, the brightest hyperfine component is separated from the next two brightest components by 38 and 41~\kms, further than the half-width of single-component lines towards this star. Whereas for the $e$ line, the brightest hyperfine component is only 7.5 and 24~\kms\ from the next two brightest components. The overlap of the two brightest components means that their fluxes combine to make the $e$ line more detectable. We plot a theoretical Gaussian profile with the spectra (red dotted line in Fig.~\ref{fig:oh30ns}, assuming Gaussian components with widths of 5~\kms) to emphasise the theoretical line shapes. The observed $e$ component has a more double-peaked profile than the predicted Gaussian profile, suggesting that emission might originate in a shell around the star, rather than in a spherical region centred on the star.

The zeroth moment map in Fig.~\ref{fig:oh30ns} shows that the NS emission is mostly located to the east and north of \ohth, and is only marginally spatially resolved. The emission is highly asymmetric; the averaged $3\sigma$ contour of the emission extends out to only  0.7\arcsec\ to the south-southwest and out to 1.8\arcsec\ to the east-northeast.
The major axis of the emitting region lies perpendicular to the projected axis joining the AGB continuum peak and feature B. This is emphasised by overplotting two dotted lines on the zeroth moment map in Fig.~\ref{fig:oh30ns}.
Although the signal to noise of the NS emission is rather low in individual channels (summing to a $5\sigma$ detection in the zeroth moment map), we constructed a position-velocity diagram along the major axis ($69\deg$ east of north, as illustrated in Fig.~\ref{fig:oh30ns}), which shows emission offset from the continuum peak by around $\sim1.5$ to 2\arcsec\ and $\sim5$~\kms\ (see Fig.~\ref{fig:ns-pv}).
The potential astrochemical implications of the NS distribution will be discussed in Sect.~\ref{sec:nschem}.

\subsection{NaCl and KCl}\label{sec:nacl-kcl-det}

\begin{figure*}
	\includegraphics[width=\textwidth]{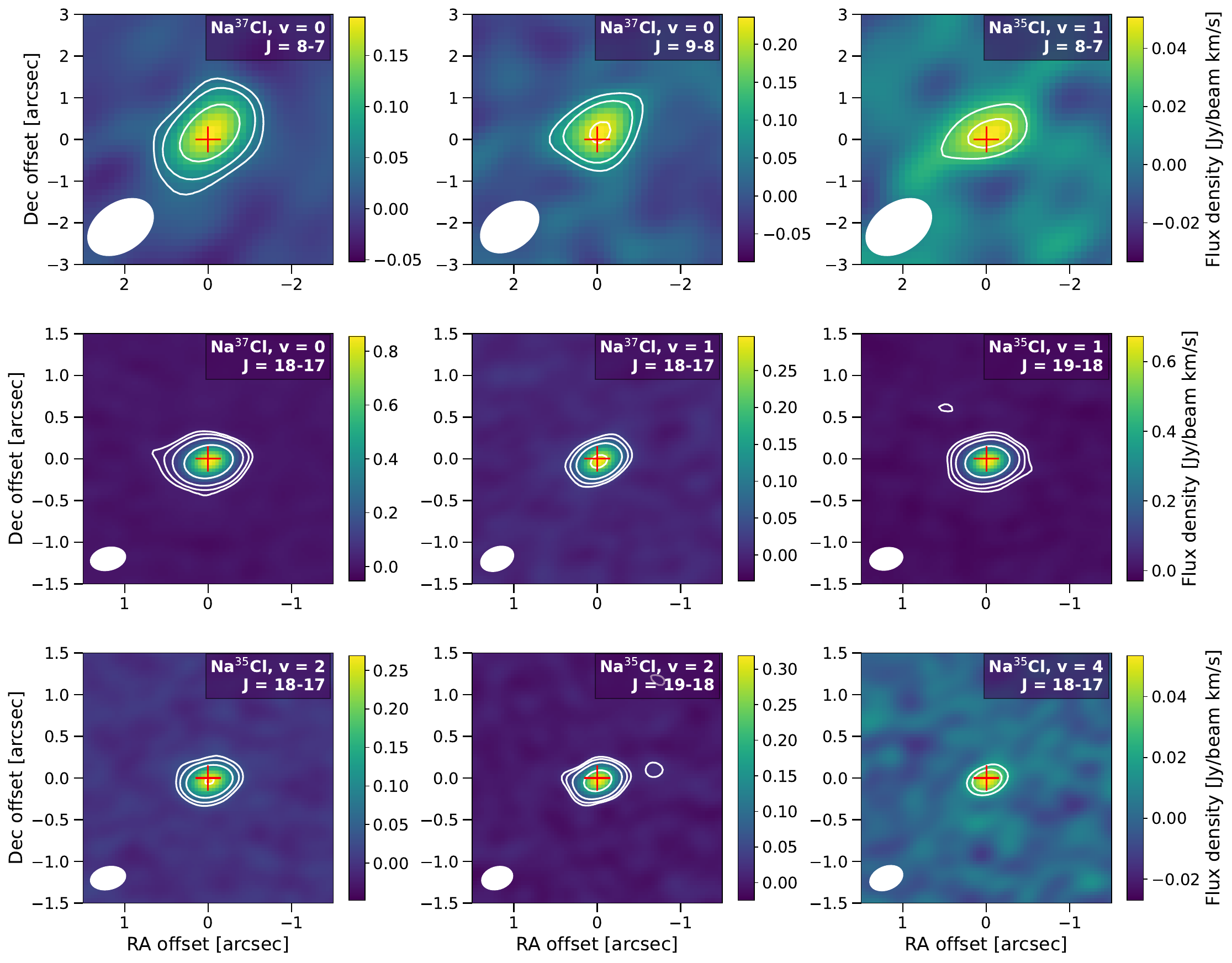}
    \caption{Zeroth moment maps of NaCl towards \ohth. The lines in the top row were observed with ALMA Band 3 and the middle and bottom rows were observed with ALMA Band 6. Note the different angular scales and resolutions. The details of the transitions are given in  Table \ref{tab:oh30ids}. The contours are drawn at levels of 3, 5, 10 and 30$\sigma$. The beam is shown as a white ellipse in the bottom left corner and the location of the continuum peak is indicated by the red cross. North is up and east is left.}
    \label{fig:oh30nacl}
\end{figure*}

\begin{figure*}
	\includegraphics[width=\textwidth]{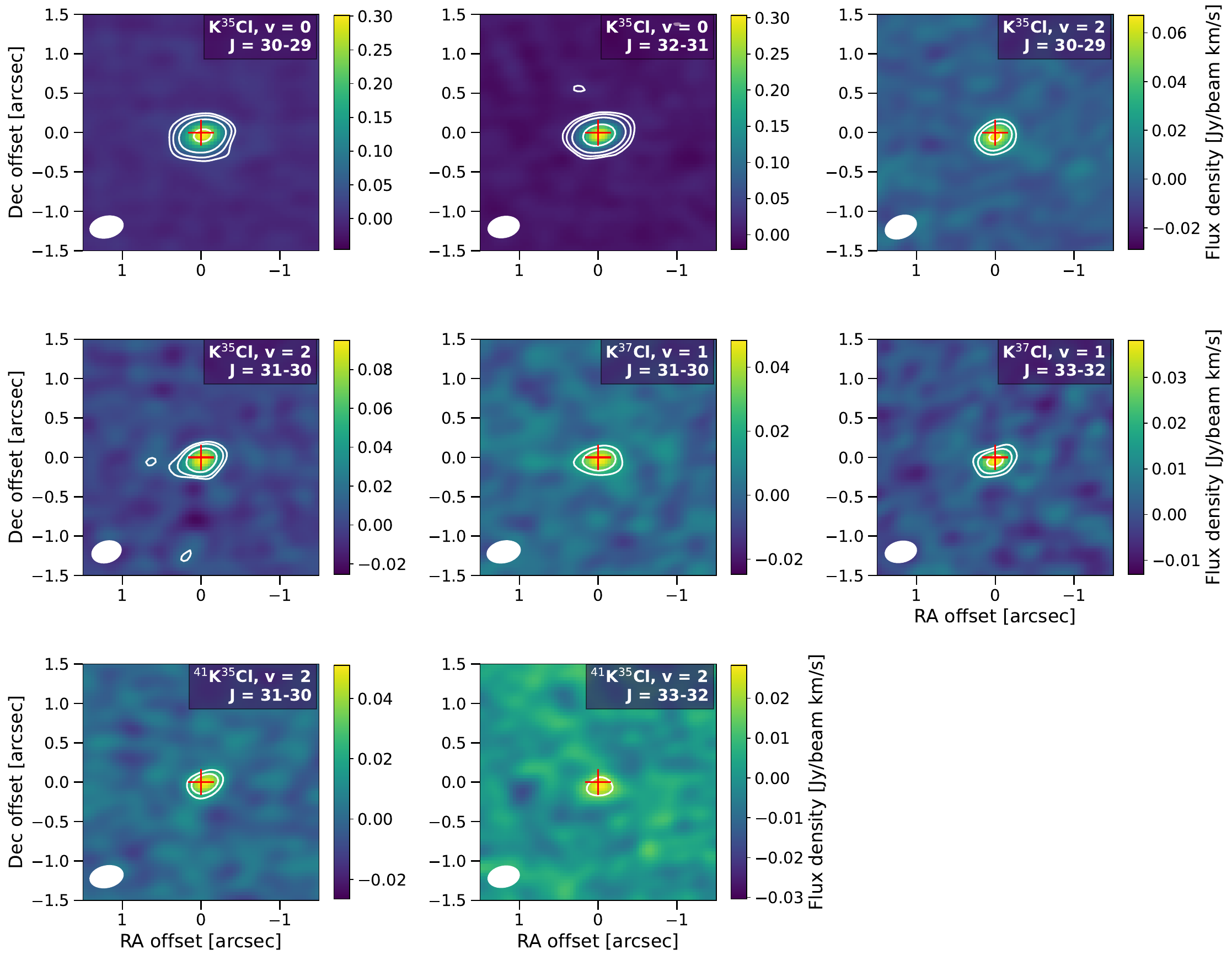}
    \caption{Zeroth moment maps of KCl towards \ohth. The details of the transitions are given in  Table \ref{tab:oh30ids}. The contours are drawn at levels of 3, 5, 10 and 30$\sigma$. The beam is shown as a white ellipse in the bottom left corner and the location of the continuum peak is indicated by the red cross. North is up and east is left.}
    \label{fig:oh30kcl}
\end{figure*}

We detect several lines of NaCl and KCl, including lines from the \ce{^35Cl} and \ce{^37Cl} isotopologues of both species. We detect lines up to the third and fourth vibrationally excited states of \ce{Na^35Cl} and up to the third vibrationally excited state of \ce{K^35Cl}. These data represent the first detections of highly excited vibrational states of both NaCl and KCl towards AGB stars.
Zeroth moment maps of NaCl are given in Fig.~\ref{fig:oh30nacl}, and of KCl in Fig.~\ref{fig:oh30kcl}. The emitting regions of both molecules are generally featureless and uniform, even for the lower-energy lines which are spatially resolved.


No lines of \ce{Na^35Cl} in the ground vibrational state were covered by our observations. However, we {clearly} detect three lines of \ce{Na^37Cl} in the ground vibrational state. {All lines of both isotopologues covered in Band 6 were detected up to $\varv=4$. Owing to the lower sensitivity of the Band 3 data, only one \ce{Na^35Cl} line in the first vibrationally excited state was detected.}

The $\varv=0,\,J=8-7$ line of Na$^{37}$Cl is appears to be marginally spatially resolved, with the $3\sigma$ contour extended out to 1--1.6\arcsec\ from the continuum peak. This is likely an over-estimate of the emission extent owing to the larger beam for the Band 3 observations compared with the Band 6 observations. We note that the $J=9-8$ line appears slightly smaller, despite the line being a little brighter, but this is almost certainly a sensitivity effect since the line lies in a slightly noisier part of the spectrum. Most of the NaCl lines covered in Band 6 appear to be more spatially resolved than the Band 3 lines (with the exception of the $\varv=4$ line). However, we note that even the two brightest lines, the Na$^{37}$Cl $\varv=0,\,J=18-17$ and $\varv=1,\,J=19-18$ lines, which appear several times larger than the beam, still have a slightly oblong shape along the same axis as the beam, {suggesting that they are not truly resolved}. The smoothed $3\sigma$ contour for the Na$^{37}$Cl $\varv=0,\,J=18-17$ line (the only ground vibrational state line observed in Band 6) extends out to 0.3--0.6\arcsec.



Our Band 3 data were only sensitive enough to detect the single covered \ce{^39K^35Cl} line in the ground vibrational state, which is partly blended with CO, and none of the lines in higher vibrational states. In the Band 6 data, we detect all covered lines in the ground, first and second vibrational states of \ce{^39K^35Cl} and \ce{^39K^37Cl}. We detect two lines of \ce{^41K^35Cl} in the second vibrationally excited state, one tentatively, and we note that no lines in the ground or first vibrationally excited state were covered in the Band 6 observations.

As can be seen in Fig.~\ref{fig:oh30kcl}, the KCl lines are not spatially resolved. Similar to the observed NaCl emission, the two KCl lines in the ground vibrational state appear larger than the beam but are still slightly elongated along the same axis as the beam. The 3$\sigma$ contours for both lines extend out to 0.2--0.5\arcsec\ from the continuum peak.


The significance of detecting so many (high-energy) lines of NaCl and KCl are discussed in Sect.~\ref{sec:clchem}, as are the potential chemical formation pathways. 
Based on the data presented here, we constructed population diagrams for \ce{Na^35Cl} and \ce{K^35Cl}. The resultant plot is given in Fig.~\ref{fig:saltpop}. We derive very similar temperatures for both species: approximately 600~K. This is consistent with the emission originating from the inner wind, unlike the {low-energy} \so2 emission (see Sect.~\ref{sec:so-so2-det}). We also derive column densities of $1.4\e{12}$~cm$^{-2}$ for Na$^{35}$Cl and $1.6\e{11}$~cm$^{-2}$ for K$^{35}$Cl, resulting in Na$^{35}$Cl/K$^{35}$Cl = 8.8. This is higher than expected based on chemical equilibrium and may be an indication that the abundance ratio of Na/K is higher than solar \citep[{in agreement with models} for intermediate mass stars, especially if \ohth\ had a higher initial metallicity than solar,][]{Cinquegrana2022}. This will be explored in more detail in a follow up study.

\section{Discussion}\label{sec:discussion}

\subsection{The circumstellar chemistry of \ohth}\label{sec:astrochem}

Several of the molecules detected towards \ohth\ are commonly seen around oxygen-rich AGB stars generally. These include CO, SiS, and CS. In the case of oxygen-rich AGB stars, SiS, CS are  commonly seen for those with high mass-loss rates \citep[or, equivalently, high wind densities;][]{Danilovich2018}. \so2 is very commonly detected around oxygen-rich AGB stars (but has not been detected for carbon-rich or S-type stars), {with a shell-like distribution for high mass-loss rate stars \citep{Danilovich2020,Wallstrom2024}}. The observations of all of these molecules towards \ohth\ agree with previously established trends, {as explained in more detail in Appendix \ref{app:previousmol}.}

NS, NaCl and KCl have been observed less frequently and hence studied in less detail thus far. It is on these three molecules that we will focus our discussion in the following subsections. {Some additional background on past observations of these molecules is also given in Appendix \ref{app:previousmol}.}

\subsubsection{NS}\label{sec:nschem}

{Previous observations of the NS molecule towards AGB stars include} the oxygen-rich AGB star IK~Tau \citep{Velilla-Prieto2017}, where resolved ALMA observations show its emitting region to be close to the continuum peak \citep[within 170 au,][]{Decin2018}, and the S-type AGB star W~Aql \citep{Danilovich2024}. For W~Aql, \cite{Danilovich2024} found that NS may have been formed when the main-sequence (F9V) companion to the AGB star passed through the dense inner circumstellar envelope, temporarily driving UV photochemistry in that region. This conclusion is based on the chemical models of \cite{Van-de-Sande2022}, who found that the formation of NS was enhanced in the presence of UV photons from a companion star. 
Although NS towards IK~Tau has not been analysed in detail \citep[only some basic measurements are given in][]{Decin2018}, the recent analysis of NaCl in the same star found evidence that a companion may be driving or enhancing the formation of NaCl \citep{Coenegrachts2023}. It is also possible that the formation of NS around IK~Tau could be driven by that companion, albeit primarily through UV chemistry rather than the shock-heating proposed by \cite{Coenegrachts2023} for NaCl.

The radial extent of NS in the case of \ohth\ is much larger than the extents seen for IK~Tau or W~Aql. The integrated NS emission towards W~Aql is irregular but is detected no more than 0.4\arcsec\ $\approx160$~au from the star \citep{Danilovich2024}. For IK~Tau, the angular width is reported to be 490--650~mas \citep{Decin2018}, which corresponds to radial extents of $\sim130$ to 170~au. In contrast, for \ohth\ the most distant NS emission within the 3$\sigma$ contour is seen at 1.8\arcsec. This corresponds to 7000~au, while the southeastern extent of 0.7\arcsec\ corresponds to 2700~au. These distances come with the caveat that the Band 3 emission is only marginally resolved, so the radial extents may be overestimated and only represent upper limits (for example, see the difference between the Band 3 and 6 emission extents for NaCl in Sec.~\ref{sec:nacl-kcl-det}).
Taken at face value, this emission appears to be 20--40 times further from the AGB star than the most distant NS emission seen around W~Aql and IK~Tau.
{The theoretical study of \cite{Van-de-Sande2022} found that higher mass-loss rates result in NS peaks further from AGB star and that a relatively warm companion star is required to form NS; i.e. a solar-type or white dwarf companion will drive NS formation but a red dwarf will not.}

The asymmetric NS emission in the zeroth moment map (Fig.~\ref{fig:oh30ns}) is reminiscent of the SiN emission around W~Aql \citep{Danilovich2024}, which was determined to lie on the side of the AGB star that aligned with the periastron of the binary orbit. If a similar scenario can account for the NS around \ohth, then the alignment of the NS emission can help constrain the projection of the orbital plane on the sky. {Another possibility is a companion on} a circular but wider orbit such that NS is preferentially formed on the side of the CSE where the companion is presently located. 

\subsubsection{Chlorides}\label{sec:clchem}


%

There have been enough adequately sensitive observations of a variety of AGB stars that we can firmly say that NaCl and KCl are not ubiquitously present in circumstellar envelopes. This is most clearly illustrated by the detection of one or both molecules in 3 AGB stars out of a sample of 14 in the ATOMIUM survey \citep{Wallstrom2024}. Some physical or chemical characteristic must set these stars apart from their unsalted counterparts. Similarly, something should also set stars like \ohth, with a large number of NaCl and KCl lines, apart from stars like IK~Tau, with  detections only in the ground vibrational state \citep{Velilla-Prieto2017,Decin2018}. 

\cite{Coenegrachts2023} proposed that an unseen companion could be contributing to the formation of NaCl around IK~Tau, since the motion of the companion through the CSE would be supersonic, hence generating a shock. As discussed by \cite{Decin2019}, {and evidenced by our detection of NS,} \ohth\ is also thought to have a companion. Since the \ohth\ shows evidence of hot bottom burning \citep{Justtanont2015}, we can conclude it had an intermediate initial mass, $\gtrsim 4~\msol$. In this mass range, the binary fraction is thought to be close to one or higher \citep[that is, stars with masses from 4--$8~\msol$ are expected to have one or more companions with mass ratios $q>0.1$,][]{Moe2017}. In light of this, and the results of \cite{Coenegrachts2023}, it is {highly} likely that a companion is present around \ohth\ and that such a companion could be driving the formation of NaCl and KCl in a similar way. 
Unlike the case of IK~Tau,
we see many vibrationally excited NaCl and KCl lines around \ohth. This is consistent with the different temperatures found for NaCl around the two stars: 14 to 70 K for IK Tau \citep{Velilla-Prieto2017,Coenegrachts2023} and 600 K for \ohth\ (see Sect.~\ref{sec:nacl-kcl-det} and Fig.~\ref{fig:saltpop}). This may be a result of more extreme shocks owing to the higher wind density of \ohth\ and possibly a higher orbital velocity of the companion.

{The formation pathways of NaCl and KCl are discussed in detail in Appendix \ref{app:chlorideform}. We conclude that}
a companion star hotter than around $\sim12000$~K would cause too much photoionisation of Na and K to reconcile our observations, unless densities are very high in that region of the wind. Hence, we can hypothesise that the companion to \ohth\ cannot be hotter than a late B-type main sequence star.




%
%

\subsection{Our evolving understanding of \ohth}\label{sec:physics}

\ohth\ has been classified as an extreme OH/IR star, which is to say, a dusty AGB star with a high mass-loss rate and bright OH masers \citep{Justtanont2006}. It is an example of a large-amplitude
variable star with one of the longest periods known \citep[$P = 2170$~days $\sim6$ years\footnote{Obtained from its OH maser light curve:
\url{https://hsweb.hs.uni-hamburg.de/projects/nrt_monitoring/stockert.html}},][]{Engels2024}. Such stars are thought to have evolved to the end of the AGB phase \citep[e.g.][]{Olofsson2022}. From \textsl{Herschel} observations of \h2O isotopologues, \ohth\ was determined to have an intermediate initial mass \citep[$\gtrsim 4~\msol$,][]{Justtanont2015}. This is in contrast with several other OH/IR stars which have been found to have lower initial masses $\sim 1.5~\msol$ \citep{Olofsson2022}.
{An overview of past observational difficulties, including a discussion on the reliability of observations in light of ISM contamination towards \ohth, are given in Appendix \ref{sec:pointings}.} 

%

\subsubsection{Past results from the literature}\label{sec:pastresults}

\ohth\ was included in the CO study of \cite{Heske1990}, which was the first to observe a sample of OH/IR stars in CO emission. {\cite{Heske1990} found that their observed intensity ratios of the CO $J=1-0$ and $J=2-1$ lines} could be explained by variable mass-loss rates at different eras, which could be the reason why mass-loss rates derived from CO observations resulted in an order of magnitude lower mass-loss than those from dust. 
This idea, first dubbed the ``superwind'' scenario by \cite{Renzini1981a}, was taken up and developed in subsequent work \citep[e.g.][]{Justtanont1992,Delfosse1997,Justtanont2013}. The idea, as it developed, being that in the last portion of the AGB phase, mass loss increases substantially (i.e. the stellar wind becomes a superwind) during a relatively short period immediately before the completion of the AGB phase.

\cite{Decin2019} presented the ALMA CO observations of \ohth\ (from the dataset also presented here, though we have performed a new reduction of the ALMA data using self-calibration, see Sect.~\ref{sec:datared}) and of another extreme OH/IR star, \ohtw. A key argument of \citeapos{Decin2019} work was that it is statistically unlikely that a dozen extreme OH/IR stars should all be observed in the midst of the relatively short superwind phase. Rather, they interpret the structures within the CO emission of \ohth\ (and the spiral structure seen for \ohtw) as interactions with a companion star. The ultimate result of these interactions, they argue, is an equatorial density enhancement; that is, the companion pulls some of the wind into a toroidal structure of denser gas and dust around the star. This equatorial density enhancement could then mimic the superwind by making the mass-loss rate appear high because of the line of sight alignment of this region. 

Low $^{12}$CO/$^{13}$CO ratios $\sim3.5$ were found by both \cite{Delfosse1997} and \cite{Justtanont2013} for \ohth, using different observational datasets and a similar approach involving radiative transfer modelling. 
This low $^{12}$CO/$^{13}$CO ratio provides further evidence of hot bottom burning in \ohth, and hence that it must be an intermediate mass star, with an initial mass $4~\msol \lesssim M_i \lesssim 8~\msol$. 

The recent study of \cite{Marini2023} modelled the SED of \ohth\ and compared various properties such as luminosity and pulsation period of the star to theoretical models. Based on their results (and considering the distance derived by \cite{Engels2015}), their determination of the initial mass of \ohth\ is likely between 4 and 5~$\msol$. However, if the luminosity is higher (i.e. $5\e{4}~\lsol$ {based on the luminosity adopted by \cite{Justtanont2013}, scaled to our adopted distance of 3.9 kpc}) then \ohth\ could have an initial mass of $6~\msol$.

\subsubsection{New understanding from ALMA CO}\label{sec:codiscussion}

Our ALMA observations show that the angular extent of the CO emission for both the $J=2-1$ and $J=1-0$ are essentially the same. Our measured maximum angular extents (see Sect.~\ref{sec:co-det}) are $3.5\arcsec$ for $J=2-1$ from the combined data and $4.2\arcsec$ for $J=1-0$. Although it may appear from this that the $J=1-0$ extent is larger than the $J=2-1$ extent, this is primarily a function of the different beam sizes (see Table \ref{tab:oh30obs}). If we instead compare the data from only the compact array configuration for $J=2-1$ (with a comparable beam to the Band 3 data), then we find a maximum angular extent of $4.2\arcsec$, as for the $J=1-0$. This unexpected result suggests that both the emission from CO $J=1-0$ and $J=2-1$ trace the outer extent of the CO envelope.
Note also that in the channel maps close to the LSR velocity (and ignoring the regions with the most contamination) the lowest few contours are very close together (3, 5, and $10\sigma$ for $J=1-0$, Fig.~\ref{oh30-co-b3-chan}, and 5, 10, 30, and $50\sigma$ for $J=2-1$, Fig.~\ref{fig:cochans}), indicating a sharp drop-off in CO emission and likely also CO abundance. This is also seen in the reddest and bluest channels (more clearly for CO $J=2-1$ because of higher contamination for CO $J=1-0$) where the CO emission is detected above $30\sigma$, then not detected in the subsequent channel at all. This behaviour is in contrast with other spatially resolved observations of AGB stars which typically have less abrupt edges, with gradually decreasing emission across the velocity and angular axes (see for example \cite{Cernicharo2015a}, \cite{Kim2017} and the channel maps in the supplementary material of \cite{Decin2020} for a selection of varied AGB stars). 

If we take half the maximum velocity extent in the CO channel maps as the wind speed (23~\kms) and consider the well-constrained distance of 3.9 kpc, we find that it took around 2800 years for the gas to expand to $3.5\arcsec$ ($2\e{17}$~cm = 14000~au) from the star. This is around a factor of 3 longer than the superwind phase obtained for this star by \cite{Delfosse1997} and a factor of 8 larger than the superwind radius derived by \cite{Justtanont2013}. 
We also do not detect any significant emission beyond this radius, aside from the background contamination, which is not evenly distributed around the star. 
In conjunction with the sharp drop-off in CO emission discussed above, this suggests that any earlier mass-loss must have been at a much lower rate, if it was present at all. 
If we consider the photodissociation radius of CO based on the mass-loss rate, expansion velocity and interstellar radiation field, and take the measured envelope size to be the half-abundance radius for CO destruction, then the mass-loss corresponding to the observed envelope size is $\dot{M} \approx 2\e{-5}\spy$ \citep{Mamon1988,Saberi2019}. Essentially, any earlier mass-loss at this rate or lower would not be detectable, because that CO would have photodissociated before expanding to larger radii. So if there was a lower mass-loss rate ($\leq 2\e{-5}\spy$) before the current high mass-loss rate period commenced 2800 years ago, it would not be detectable.

If we take the mass-loss rate of $\dot{M} = 1.8\e{-4}\spy$ computed by \cite{Justtanont2013}, and assume it has not changed over the past 2800 years, we calculate that $0.5~\msol$ has been ejected by the star in this time. If we instead use a lower rate such as $\dot{M}=2\e{-5}\spy$, close to the value found by \cite{Heske1990} and in agreement with the findings of \cite{Decin2019}, we correspondingly find less mass ejected in that time frame: only 
$0.06~\msol$. Clearly the mass-loss rate will significantly affect the longevity of this object.
A further analysis of the CO properties of \ohth\ would require a detailed radiative transfer analysis and is beyond the scope of the present work, but will be the subject of a follow-up paper.

\subsubsection{Evidence from other molecular species}

As discussed in Sect.~\ref{sec:astrochem}, the circumstellar chemistry of \ohth, especially the detection of NS, strongly suggests the presence of a companion star emitting UV in the inner wind. However, while the companion star should emit enough UV to facilitate the formation of NS through the photodissociation of \ce{N2} (it should not be a red dwarf), the companion should not be too hot ($<12000$~K) to avoid photoionising Na and K and inhibiting the formation of NaCl and KCl.
Hence, the companion ought not to be a massive star and is likely to be a late B- to G-type star. This would put a companion in the mass range of $\sim1$ to $4~\msol$, {which also fulfils the criteria of being less massive than the AGB star, since the AGB star has evolved off the main sequence first}. 

{Based on the temperature we calculated for the \so2 lines (see Fig.~\ref{fig:coveredso2} and Sect.~\ref{sec:so-so2-det}), we can make an estimate of the average radial temperature profile of the circumstellar envelope. For this we take the \so2 temperature to be 25~K and the average outer radius of \so2 to be $1\e{17}$~cm (corresponding to $\sim 1.7\arcsec$, close to the average maximal extent of the \so2 observations, see Fig.~\ref{fig:oh30so2}, and in good agreement with the radius of the outer \so2 shell visible in Fig.~\ref{fig:so2chans}). We take the inner temperature to be 2000~K close to the surface of the star (at $1.3\e{14}$~cm). Assuming a power law radial temperature profile, $T(r) = T_\star \left( {r}/{R_\star}\right)^{-\epsilon}$, we then calculate the exponent, $\epsilon = 0.66$. This is very close to the values found for lower mass-loss rate oxygen-rich AGB stars such as R Dor \citep[0.65,][]{Van-de-Sande2018a} and W Hya \citep[0.65,][]{Khouri2014}. If we try to include the average NaCl and KCl temperature (600 K, Fig.~\ref{fig:saltpop}) at the apparent size of the emitting region, we find an implausibly high value of $\epsilon=2.4$. This is most likely because the NaCl and KCl lines are not spatially resolved in our observations, as noted in Sect.~\ref{sec:nacl-kcl-det}. If we instead use the estimated radial temperature profile with $\epsilon = 0.66$ to calculate the average radius of the salt extents, we find $R_\mathrm{salt}=8\e{14}$~cm, which corresponds to $\sim0.014\arcsec$ (and $\sim50$~au), much smaller than our restoring beams (Table \ref{tab:oh30obs}).}

{The companion is most likely driving the formation of the salt molecules. Hence, the above indicates} that the separation between the AGB star and the likely main sequence companion is {significantly} smaller than our minimum resolution of $\sim0.3\arcsec = 1200$~au $\approx 2\e{16}$~cm. 
{From our estimate of the average salt extent, we predict that observations with an angular resolution of 5~mas or better are required to resolve the structures in the inner wind. Further observations}
targeting molecular tracers of UV-emitting companions such as NS and SiN \citep{Van-de-Sande2022}, will help us to understand the {nature and orbit of such a companion}. 

Finally, we note that as well as affecting the circumstellar chemistry and gravitationally shaping the wind, a companion may also contribute to dust formation. Even wide binary companions (with semimajor axes on the order of hundreds of au) are essentially guaranteed to have orbital velocities in excess of the local sound speed, resulting in additional shocks in the wind. \cite{Freytag2023} found, in the context of stellar pulsations, that dust forms in the wake of shocks. Some evidence for this was seen by \cite{Coenegrachts2023} who found that an asymmetric dust distribution in the inner CSE of IK~Tau correlated with a clumpy spiral of NaCl emission that may have formed partly as a result of shocks and other interactions with an unseen binary companion. Indeed, a more extreme version of companion-enhanced dust formation has long been known to take place for Wolf-Rayet stars, where dust forms at the nexus of colliding winds from massive stars \citep{Usov1991}. 
The lack of prior direct detection of such a companion is not evidence of its absence, since the optically thick circumstellar dust would very strongly attenuate the signal at optical wavelengths \citep[e.g. see the strong silicate absorption in Fig.~1 of][]{Justtanont2013}.

%
%
%
%

\subsection{On the nature of feature ``B''}\label{sec:oh30-sec-peak}

Feature B is detected in the continuum and lies $4.6\arcsec$ north and slightly west of the primary continuum peak. The lower limit on the separation is $1.8 \e{4}$~au. We extracted spectra around this peak, using a circular aperture with a radius of $0.3\arcsec$, and did not detect any line emission above the noise aside from CO. As shown in our channel maps (Figs.~\ref{fig:cochans} and \ref{fig:oh30-co-TE-chan}), this emission is most likely dominated by contamination from the background molecular cloud associated with W43. Inspecting the channel maps, there are no localised features in the CO emission associated with the position of feature B. As can be clearly seen in the channel maps and in the integrated intensity (zeroth moment) map in Fig.~\ref{fig:oh30cocont}, feature B lies beyond any of the molecular emission associated with the AGB star. The size of feature B, when considering the $5\sigma$ contours, is approximately $0.8\times0.5\arcsec$.

In Sect.~\ref{sec:specind} we calculated the spectral indices of both the primary continuum peak and feature B. The spectral index of the primary peak, $2.8\pm0.3$ is in general agreement with past results for \ohth. For example \cite{Nicolaes2018} found spectral indices ranging from 2.13 at shorter (far-IR) wavelengths to 3.42 at long wavelengths, based on \textsl{Herschel}/PACS and \textsl{Herschel}/SPIRE observations. The lower limit of 3.6 for feature B is very high compared with commonly found spectral indices \citep[see for example][who find values around 1--2 for cold cores]{Juvela2015}. \cite{Nicolaes2018} found some comparably steep slopes for other evolved stars in their sample, notably at longer wavelengths ($>100\micron$) and for stars classed as OH/IR stars, similar to \ohth. When it comes to dust sources not associated with evolved stars, \cite{Kuan1996} found very steep spectral indices (up to 4.6) for Sagittarius B2(N). They conclude that this is most likely a result of a cold dust temperature and ice-coated grains and that Sgr B2(N) is a young object in an early stage of star formation. Whether our feature B is a similar background source to Sgr B2(N) or more closely associated with \ohth, cold ice-coated dust grains are a feasible explanation for the spectral index. Given that \cite{Justtanont2006} found evidence of \h2O ice in the dust around \ohth, the high spectral index of feature B could indicate a similar dust composition (i.e. ice-coated grains) but at a lower temperature since the feature is located further from the AGB star with much of the stellar light obscured by the circumstellar dust.


\subsubsection{A direct association with \ohth?}

If we draw a line connecting the primary continuum peak and feature B, we find it lies $20.5\deg$ west of north (clockwise on the observed maps in Fig.~\ref{fig:continua}). 
The CO emission appears mostly round in shape (Fig.~\ref{fig:cochans}), however, there is a slight indentation in the direction of feature B. The centre of the indentation sits about 2.4\arcsec\ from the primary continuum peak whereas on the opposite side the $5\sigma$ contour of the CO emission extends to around 3.1\arcsec\ from the primary continuum peak. This alignment could be coincidental, but it is also possible that feature B is contributing to the shaping of the CO emission or was formed through the same interaction as the CO indentation. This would imply that feature B is an object at the same distance as \ohth.
Given the large separation between the CO indentation and feature B (around 2.2\arcsec), it is unlikely that the source of feature B is (at present) contributing to significant gravitational shaping of the AGB circumstellar environment. \cite{Decin2019} suggested that the indentation in the CO was a knot caused by a companion and its density wake. 
Another possibility is that feature B is causing the indentation through non-gravitational means, perhaps through high energy UV photons, although this is difficult to reconcile with the cold dust implied by the spectral index. 

A recent study of the S-type AGB star, W~Aql, found that the F-type main sequence companion, located around 200~au from the AGB star, contributed to the enhanced photodissociation of several molecular species, including SiO, SiS, CS and HCN. The channels exhibiting such photodissociation bear some resemblance to the CO towards \ohth, although the CO indentation is less extreme. At 6000~K, the companion to W~Aql did not produce enough energetic UV photons to significantly photodissociate CO. However, a hotter stellar companion could potentially affect the CO distribution. For example, when analysing the impact of increasing the interstellar radiation field on the size of the CO envelopes of AGB stars, \cite{Saberi2019} found that, for stars with similar mass-loss rates and velocities to \ohth, doubling the strength of the interstellar radiation field reduced the size of the CO envelope by 14--20\%. High mass-loss rates result in smaller effects because of CO self-shielding. For example, \cite{Van-de-Sande2022} found that including a hot 80,000 K (white dwarf) inner companion in their model did not significantly impact the CO distribution for higher mass-loss rates. If feature B is a hot star, it could act as a localised increase in the UV radiation field.
Counter to this and unlike a companion located within the dust envelope of \ohth, it is unclear why such a UV source would be too attenuated to be detected from Earth but not too attenuated to affect the \ohth\ CO envelope. Or why such a hot source would not destroy ice-covered dust grains.
We also note that if feature B does correspond to a companion, this would not negate the potential presence of a closer companion that does shape the wind, even though feature B is too distant to be such a companion itself.

A more likely possibility is that feature B is a clump of dust (and perhaps undetected gas) that was ejected from \ohth\ itself, perhaps as a pre-cursor to the present-day mass loss (beginning 2800 years ago, according to the CO extent) or owing to a complex binary interaction. The fact that the extent of the NS emission lies perpendicular to the projected separation between the AGB star and feature B lends some weight to this scenario. If feature B were ejected with the same velocity as the observed CO envelope, that would suggest an ejection around 3700 years ago. However, if the ejection took place at the same time that mass loss was initiated, that would imply a velocity of 30~\kms, which is not implausibly high. If feature B is rather a high velocity bullet, such as those seen for the carbon star V Hya \citep[which has a relatively close companion on an eccentric orbit,][]{Sahai2016}, then feature B could have been ejected more recently at a higher velocity.
In both of these cases we are assuming that feature B was moving in the plane of the sky, making our time and distance estimates lower limits.
An ejection scenario could, however, account for the apparent bridge between the primary continuum peak and feature B, if some ejecta were not accelerated as strongly.

\subsubsection{A background source?}

It is possible that feature B is a coincident background (or foreground) source. Given the line of sight proximity of \ohth\ to the star-forming region W43, that is the most likely origin of background sources. The distance to W43 has been determined to be $5.5\pm0.3$~kpc \citep{Zhang2014}, significantly further than the 3.9~kpc to \ohth\ \citep{Engels2015}. At the distance of W43, the extent of feature B would correspond to around 2800--4400~au. Placing it at the distance of \ohth\ gives an extent of around 2000--3000~au. Both of these sizes are upper limits as they are not significantly larger than the beam size.

\section{Conclusions}\label{sec:conclusions}

We present newly reduced ALMA observations of the intermediate-mass AGB star \ohth, including continuum and line emission. In the continuum maps, we identified a secondary peak (feature B), detected with a certainty of $>10\sigma$, and we speculate that it is {most likely ejecta associated with the AGB star, but} may be a background source or a companion object. With a high lower limit for its spectral index, feature B is likely composed of cold dust with ice covering the dust grains. We do not conclusively find any molecular emission associated with the feature, aside from some weak CO emission which may originate from background contamination rather than from that source.

We identified all the molecular emission present in the ALMA maps and spectra, finding rotational transitions arising from CO, SiS, CS, \so2, NS, NaCl, and KCl. {In addition to} transitions in the ground vibrational state, we find several rotational transitions in vibrationally excited states from SiS (up to $\varv=1$), NaCl (up to $\varv=4$) and KCl (up to $\varv=2$). {Based on a simplified molecular line excitation analysis,} we estimate the radial temperature profile as a power law with a similar index to other oxygen-rich AGB stars.

{We interpret the presence of NS as indicative of a UV-emitting} main sequence companion in the inner regions of the \ohth\ wind, with a separation smaller than our resolution limit of $\sim1200$~au. The distribution of the NS emission suggests an eccentric binary orbit with an orbital plane that may be perpendicular to the direction of feature B. While the presence of NS suggests a UV emitting companion, the presence of NaCl and KCl require that the companion not be too hot. From this we determine that the companion cannot be hotter than a late B-type main sequence star, but may be as cool as a G-type star, and that a red dwarf star would not be able to trigger adequate NS formation.

Past observations and studies of \ohth\ have indicated optically thick dust and CO emission --- which our observations support --- and have suggested that the AGB star may have entered a superwind phase at the tail end of the AGB. Our observations reveal a very steep drop off at the edges of the CO emission, both in the plane of the sky and in velocity space.
This suggests that the bounds of the CO envelope are not truncated by the interstellar radiation field, as is usually the case. Instead it seems that the observed CO corresponds to a period of high mass loss beginning around 2800 years ago. This is a longer period than predicted by the superwind scenario, and a detailed explanation is deferred to a later study. We also conclude that any earlier period of lower mass loss (with a mass-loss rate $\dot{M}\leq 2\e{-5}\spy$) would not be detectable in light of the recent high mass loss. We suggest companion-enhanced dust production, leading to a higher mass-loss rate, as a possible explanation for the current high mass-loss rate. 

Additional higher angular resolution observations are required to fully understand \ohth\ and its high mass loss and dense wind. In particular, we recommend observations with angular resolution of {5} mas or better to resolve possible companion interactions in the inner wind.

\section*{Acknowledgements}

The authors thank Elvire De Beck for sharing the original data from her 2010 paper \citep{De-Beck2010}.
TD is supported in part by the Australian Research Council through a Discovery Early Career Researcher Award (DE230100183).
This research is supported in part by the Australian Research Council Centre of Excellence for All Sky Astrophysics in 3 Dimensions (ASTRO 3D), through project number CE170100013.
MVdS acknowledges the Oort Fellowship at Leiden Observatory.
TJM's research at QUB is supported by grant number ST/T000198/1 from the STFC.
HSPM thanks the Deutsche Forschungsgemeinschaft (DFG) for support through the collaborative research center SFB~1601 (project ID 500700252) subprojects A4 and Inf.
KJ acknowledges the support from the Swedish National Space Agency (SNSA).
JMCP was supported by STFC grant number
ST/T000287/1.
SHJW acknowledges support from the Research Foundation Flanders (FWO) through grant 1285221N.
LD acknowledges support from the KU Leuven C1 MAESTRO grant C16/17/007, the FWO research grant G099720N, and the KU Leuven C1 BRAVE grant C16/23/009.
This paper makes use of the following ALMA data: ADS/JAO.ALMA\#2016.1.00005.S. ALMA is a partnership of ESO (representing its member states), NSF (USA) and NINS (Japan), together with NRC (Canada), MOST and ASIAA (Taiwan), and KASI (Republic of Korea), in cooperation with the Republic of Chile. The Joint ALMA Observatory is operated by ESO, AUI/NRAO and NAOJ.
We acknowledge excellent support from the UK ALMA Regional Centre (UK ARC), which is hosted by the Jodrell Bank Centre for Astrophysics (JBCA) at the University of Manchester. The UK ARC Node is supported by STFC Grant ST/P000827/1.
Based on observations with the Atacama Pathfinder EXperiment (APEX) telescope with project ID: O-0102.F-9301B. APEX is a collaboration between the Max Planck Institute for Radio Astronomy, the European Southern Observatory, and the Onsala Space Observatory. Swedish observations on APEX are supported through Swedish Research Council grant No 2017-00648.
This research made use of Astropy \citep{Astropy2013,Astropy2018}, SciPy \citep{Virtanen2020}, Matplotlib \citep{Hunter2007} and NumPy \citep{Harris2020}.

\section*{Data Availability}

The observational data used here are openly available through the data archives for ALMA (\url{https://almascience.nrao.edu/aq/}) and ESO for the APEX data (\url{http://archive.eso.org}). Custom ALMA data products that were produced for this study are available from TD or AMSR upon reasonable request.



\bibliographystyle{mnras}
\bibliography{master} 




\appendix

\section{Additional tables and plots}

\begin{figure*}
	\includegraphics[width=0.32\textwidth]{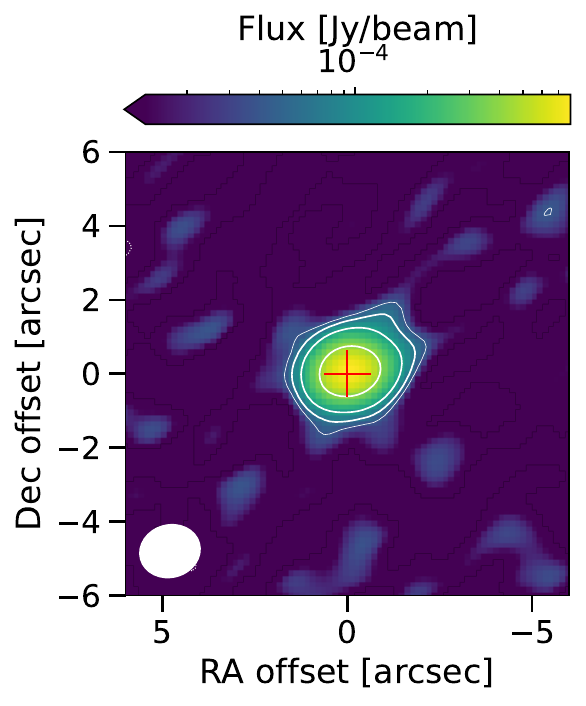}
	\includegraphics[width=0.32\textwidth]{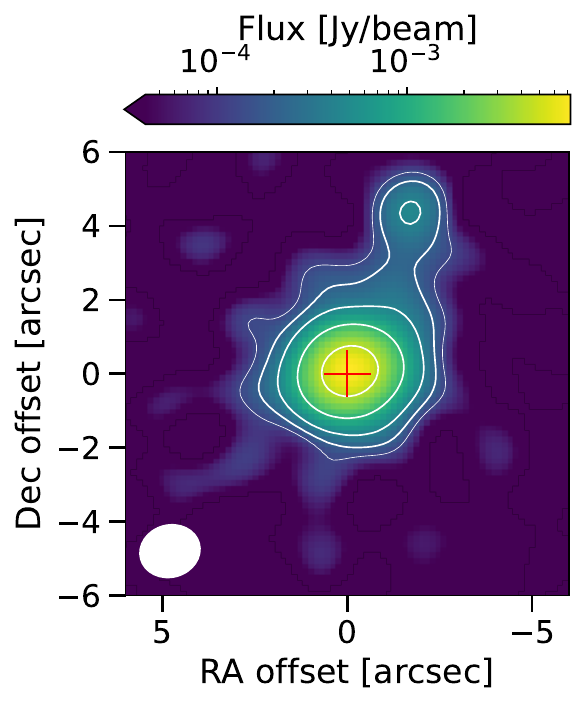}
	\includegraphics[width=0.32\textwidth]{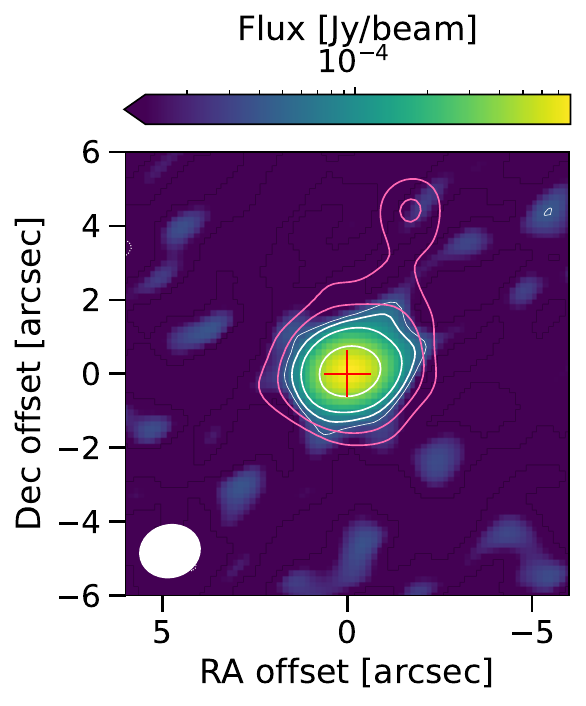}
    \caption{The continuum plots for \ohth\ observed with ALMA Band 3 (\textit{left}) and the compact configuration of ALMA Band 6 (\textit{centre}), plotted on a logarithmic flux scale. The Band 3 image has been convolve to have the same beam as the Band 6 image. The contours indicate levels of 3$\sigma$ (thinnest white line), and 5, 10, 30 and $100\sigma$ (thicker white lines). On the \textit{right} we plot the Band 3 continuum with the 5 and 10$\sigma$ contours of the compact Band 6 continuum overplotted in pink. In all panels, the red crosses indicate the continuum peak. The beams are shown as white ellipses in the bottom left corners. North is up and east is left.}\label{fig:oh30-b3-cont}
\end{figure*}

{In this section we present additional tables and plots which provide additional details of our observations. Table \ref{tab:oh30obs} provides details of the ALMA line observations and Table \ref{tab:continuum} provides details of the continuum images. Additional continuum images, with a focus on the Band 3 data, are shown in Fig.~\ref{fig:oh30-b3-cont}. The molecular line IDs are given in Table \ref{tab:oh30ids} and the full spectra are shown in Fig.~\ref{fig:oh30specb6} for the Band 6 spectra (with one spectral window given in Fig.~\ref{fig:lineidexample}) and in Fig.~\ref{fig:oh30specb3} for the Band 6 spectra.}

{The additional figures in this section are supplemental to the results presented in the body of the paper. In Fig.~\ref{fig:softpara} we show some examples of soft parabola fits, as were discussed in Sect.~\ref{sec:lines}. Figures \ref{fig:oh30-co-TE-chan} and \ref{oh30-co-b3-chan} present channel maps of the CO emission as respectively observed with the extended array configuration in Band 6 and with Band 3. Fig.~\ref{fig:so2chans} shows the channel maps of the brightest observed \so2 line, which is discussed in Sect.~\ref{sec:so-so2-det}. In Fig.~\ref{fig:ns-pv} we give a PV diagram of the NS line detected in Band 3, discussed in Sect.~\ref{sec:ns-det}, and in Fig.~\ref{fig:saltpop} we present the population diagrams of Na$^{35}$Cl and K$^{35}$Cl, as discussed in Sect.~\ref{sec:nacl-kcl-det}.}

\begin{table*}
	\centering
	\caption{Parameters of Band 3 and Band 6 ALMA observations for OH~30.1$-$0.7 from project 2016.1.00005.S.}
	\label{tab:oh30obs}
	\begin{tabular}{lccccccc} 
		\hline
		Band, configuration 	&	Frequencies		&		Restoring beam				&	Beam PA	&	\multicolumn{2}{c}{Channel separation}& MRS$^{a}$ &	Date observed	\\
		& [GHz] & [$\arcsec$] & \multicolumn{1}{c}{[$\deg$]} & [\kms] & \multicolumn{1}{c}{[MHz]} & [$\arcsec$]\\
		\hline
Band 3	&	100.028	--	101.899	&	$	1.756	\times	1.179	$	&	$	-56.0	$	&	2.88	&	0.98	&	15.0	&	2016-12-05	\\
Band 3	&	101.830	--	103.701	&	$	1.731	\times	1.155	$	&	$	-55.7	$	&	2.83	&	0.98	&	15.0	&	2016-12-05	\\
Band 3	&	111.947	--	113.818	&	$	1.576	\times	1.078	$	&	$	-54.6	$	&	2.62	&	0.98	&	15.0	&	2016-12-05	\\
Band 3	&	113.827	--	115.698	&	$	1.535	\times	1.059	$	&	$	-54.9	$	&	2.57	&	0.98	&	15.0	&	2016-12-05	\\
Band 6, extended	&	226.086	--	227.957	&	$	0.390	\times	0.265	$	&	$	-67.2	$	&	2.57	&	1.95	&	4.5	&	2017-05-05	\\
Band 6, compact	&	226.086	--	227.957	&	$	1.714	\times	1.500	$	&	$	-76.6	$	&	2.57	&	1.95	&	12.8	&	2017-03-28	\\
Band 6, combined	&	226.090	--	227.957	&	$	0.412	\times	0.277	$	&	$	-67.1	$	&	2.57	&	1.95	&	-	&	-	\\
Band 6, extended	&	229.085	--	230.956	&	$	0.395	\times	0.256	$	&	$	-78.4	$	&	2.54	&	1.95	&	4.5	&	2017-05-05	\\
Band 6, compact	&	229.085	--	230.956	&	$	1.711	\times	1.480	$	&	$	-75.8	$	&	2.54	&	1.95	&	12.8	&	2017-03-28	\\
Band 6, combined	&	229.089	--	230.956	&	$	0.427	\times	0.273	$	&	$	-78.0	$	&	2.54	&	1.95	&	-	&	-	\\
Band 6, extended	&	242.073	--	243.944	&	$	0.355	\times	0.255	$	&	$	-73.5	$	&	2.42	&	1.95	&	4.5	&	2017-05-05	\\
Band 6, compact	&	242.073	--	243.943	&	$	1.626	\times	1.397	$	&	$	-75.9	$	&	2.42	&	1.95	&	12.8	&	2017-03-28	\\
Band 6, combined	&	242.073	--	243.940	&	$	0.379	\times	0.269	$	&	$	-73.1	$	&	2.42	&	1.95	&	-	&	-	\\
Band 6, extended	&	244.086	--	245.957	&	$	0.373	\times	0.251	$	&	$	-78.0	$	&	2.40	&	1.95	&	4.5	&	2017-05-05	\\
Band 6, compact	&	244.086	--	245.957	&	$	1.617	\times	1.384	$	&	$	-76.2	$	&	2.40	&	1.95	&	12.8	&	2017-03-28	\\
Band 6, combined	&	244.086	--	245.953	&	$	0.401	\times	0.265	$	&	$	-78.6	$	&	2.40	&	1.95	&	-	&	-	\\
		\hline
	\end{tabular}
\begin{flushleft}
\textbf{Notes.} ($^{a}$) MRS is the maximum recoverable scale.
\end{flushleft}
\end{table*}

\begin{table*}
	\centering
	\caption{Parameters of continuum images of OH~30.1$-$0.7 constructed from the observations of project 2016.1.00005.S.}
	\label{tab:continuum}
	\begin{tabular}{lcccccc} 
		\hline
		Band, configuration 	&	Restoring beam				&	Beam PA	& rms & Peak flux & Dynamic range	\\
		& [$\arcsec$] & \multicolumn{1}{c}{[$\deg$]} & [mJy/beam] & [mJy/beam] \\
		\hline
Band 3					& $1.661 \times 1.105$ & $-55.9$ & $1.27\e{-2}$ & \phantom{0}$0.746$ & 59\\
Band 3, resolution matching Band 6, compact	& $1.630 \times 1.400$ & $-75.0$ & $1.34\e{-2}$ & \phantom{0}$0.788$ & 59 \\
Band 6, extended				& $0.372 \times 0.239$ & $-71.7$ & $1.40\e{-2}$ & $5.22$ & 372 \\
Band 6, compact					& $1.628 \times 1.399$ & $-75.0$ & $4.21\e{-2}$ & $7.26$ & 172 \\
Band 6, combined, full resolution		& $0.397 \times 0.252$ & $-71.6$ & $1.41\e{-2}$ & $5.26$ & 374\\
Band 6, combined, taper $=0.5$	& $0.662 \times 0.536$ & $-88.3$ & $1.51\e{-2}$ & $6.47$ & 427\\
Band 6, combined, taper $=0.8$	& $0.986 \times 0.829$ & $-87.7$ & $1.78\e{-2}$ & $7.19$ & 403\\
		\hline
	\end{tabular}
\end{table*}

\begin{table*}
	\centering
	\caption{Lines identified towards \ohth. Spectra are shown in Figures \ref{fig:oh30specb6} and \ref{fig:oh30specb3}.}
	\label{tab:oh30ids}
	\begin{tabular}{lcrlccccl}
		\hline
Species	&	Freq 	&	E$_\mathrm{low}$ &  Transition	& $\upsilon_\mathrm{cent}$ & $\upsilon_\mathrm{w}$ &  Flux & Ap.$^{a}$ &	Notes	\\
	& [GHz]	& [K]\,\, & &[\kms]	& [\kms]	&	[Jy~\kms] & [\arcsec]	\\
		\hline
SO$_2$	&	100.878	&	7.8	&$	\nu=0	,	J_{K_a,K_c} = 2_{2,0}  -  3_{1,3}	$ &	99.3	&	18.9	&	0.32	&	3.0	&		\\
Na$^{37}$Cl	&	101.962	&	17.1	&$	\varv=0	,	J=8 - 7	$ &	100.2	&	19.0	&	0.24	&	2.0	&		\\
$^{34}$SO$_2$	&	102.032	&	2.7	&$	\nu=0	,	J_{K_a,K_c} = 3_{1,3}  -  2_{0,2}	$ &	99.9	&	19.0	&	0.40	&	3.0	&		\\
$^{33}$SO$_2$	&	102.998	&	2.8	&$	\nu=0	,	J_{K_a,K_c} = 3_{1,3}  -  2_{0,2}	$ &	...	&	...	&	...	&	...	&	Tentative and hyperfine	\\
Na$^{35}$Cl	&	103.415	&	537.4	&$	\varv=1	,	J=8 - 7	$ &	103.7	&	12.0	&	0.06	&	2.0	&		\\
C$^{17}$O	&	112.359	&	0.0	&$	\varv=0	,	J=1 - 0	$ &	...	&	...	&	...	&	...	&	Tentative	\\
Na$^{37}$Cl	&	114.701	&	22.0	&$	\varv=0	,	J=9 - 8	$ &	99.4	&	23.2	&	0.19	&	1.0	&		\\
NS	&	115.154	&	3.3	&$	^2\Pi_{1/2} , \varv=0	,	e,\,  J=5/2-3/2,\, F = 7/2 - 5/2	$ &	...	&	...	&	...	&	...	&	Brightest hyperfine component	\\
NS	&	115.157	&	3.3	&$	^2\Pi_{1/2} , \varv=0	,	e,\,  J=5/2-3/2,\, F = 5/2 - 3/2	$ &	...	&	...	&	...	&	...	&	Hyperfine	\\
NS	&	115.163	&	3.3	&$	^2\Pi_{1/2} , \varv=0	,	e,\,  J=5/2-3/2,\, F = 3/2 - 1/2	$ &	...	&	...	&	...	&	...	&	Hyperfine	\\
CO	&	115.271	&	0.0	&$	\varv=0	,	J=1 - 0	$ &	100.9	&	17.0	&	48.20	&	5.0	&		\\
K$^{35}$Cl	&	115.292	&	38.8	&$	\varv=0	,	J=15 - 14	$ &	...	&	...	&	...	&	...	&	Blend with CO; tentative	\\
NS	&	115.556	&	3.4	&$	^2\Pi_{1/2} , \varv=0	,	f,\,  J=5/2-3/2,\, F = 5/2 - 3/1	$ &	...	&	...	&	...	&	...	&	Brightest hyperfine, tentative	\\
NS	&	115.571	&	3.4	&$	^2\Pi_{1/2} , \varv=0	,	f,\,  J=5/2-3/2,\, F = 5/2 - 3/2	$ &	...	&	...	&	...	&	...	&	Hyperfine, tentative	\\
NS	&	115.572	&	3.3	&$	^2\Pi_{1/2} , \varv=0	,	f,\,  J=5/2-3/2,\, F = 3/2 - 1/2	$ &	...	&	...	&	...	&	...	&	Hyperfine, tentative	\\
SO$_2$	&	226.300	&	108.2	&$	\nu=0	,	J_{K_a,K_c} = 14_{3,11}  -  14_{2,12}	$ &	99.2	&	19.0	&	0.79	&	2.0	&		\\
$^{30}$SiS	&	226.502	&	1118.2	&$	\varv=1	,	J=13 - 12	$ &	99.7	&	11.7	&	0.05	&	0.3	&		\\
Na$^{35}$Cl	&	227.351	&	2143.1	&$	\varv=4	,	J=18 - 17	$ &	101.2	&	7.6	&	0.05	&	0.5	&		\\
K$^{35}$Cl	&	227.497	&	954.4	&$	\varv=2	,	J=30 - 29	$ &	100.3	&	7.0	&	0.07	&	0.5	&		\\
Na$^{37}$Cl	&	227.560	&	607.5	&$	\varv=1	,	J=18 - 17	$ &	101.8	&	12.6	&	0.33	&	0.5	&	Blend with wing of $^{30}$SiS	\\
$^{30}$SiS	&	227.590	&	65.6	&$	\varv=0	,	J=13 - 12	$ &	99.8	&	22.0	&	1.04	&	1.0	&	Blend with wing of Na$^{37}$Cl	\\
Na$^{37}$Cl	&	229.246	&	93.6	&$	\varv=0	,	J=18 - 17	$ &	100.2	&	18.1	&	1.11	&	1.0	&		\\
Si$^{34}$S	&	229.501	&	66.1	&$	\varv=0	,	J=13 - 12	$ &	98.7	&	22.8	&	1.34	&	1.0	&		\\
$^{41}$K$^{35}$Cl	&	229.682	&	952.3	&$	\varv=2	,	J=31 - 30	$ &	98.6	&	14.6	&	0.06	&	0.3	&		\\
K$^{37}$Cl	&	229.819	&	559.6	&$	\varv=1	,	J=31 - 30	$ &	100.6	&	9.5	&	0.07	&	0.5	&		\\
$^{34}$SO$_2$	&	229.858	&	7.6	&$	\nu=0	,	J_{K_a,K_c} = 4_{2,2}  -  3_{1,3}	$ &	99.9	&	22.6	&	1.54	&	3.0	&		\\
K$^{35}$Cl	&	230.321	&	160.5	&$	\varv=0	,	J=30 - 29	$ &	101.6	&	13.4	&	0.36	&	0.5	&		\\
$^{29}$SiS	&	230.510	&	1128.6	&$	\varv=1	,	J=13 - 12	$ &	...	&	...	&	...	&	...	&	Blend with wing of CO	\\
CO	&	230.538	&	5.5	&$	\varv=0	,	J=2 - 1	$ &	101.9	&	17.9	&	214.50	&	5.0	&		\\
Na$^{35}$Cl	&	230.779	&	1129.3	&$	\varv=2	,	J=18 - 17	$ &	101.2	&	11.8	&	0.31	&	0.5	&		\\
$^{29}$Si$^{34}$S	&	242.478	&	75.7	&$	\varv=0	,	J=14 - 13	$ &	...	&	...	&	...	&	...	&		\\
K$^{35}$Cl	&	242.612	&	976.6	&$	\varv=2	,	J=32 - 31	$ &	100.5	&	14.8	&	0.12	&	0.5	&		\\
SO$_2$	&	243.088	&	41.4	&$	\nu=0	,	J_{K_a,K_c} = 5_{4,2}  -  6_{3,3}	$ &	...	&	...	&	...	&	...	&	Tentative. Blend with K$^{37}$Cl	\\
K$^{37}$Cl	&	243.110	&	971.5	&$	\varv=2	,	J=33 - 32	$ &	...	&	...	&	...	&	...	&	Blend with SO$_2$	\\
Na$^{35}$Cl	&	243.574	&	1140.4	&$	\varv=2	,	J=19 - 18	$ &	101.5	&	11.4	&	0.34	&	0.5	&		\\
$^{30}$SiS	&	243.917	&	1129.1	&$	\varv=1	,	J=14 - 13	$ &	100.2	&	6.2	&	0.07	&	0.3	&		\\
SO$_2$	&	244.254	&	82.2	&$	\nu=0	,	J_{K_a,K_c} = 14_{0,14}   -  13_{1,13}	$ &	98.7	&	21.7	&	5.38	&	3.0	&		\\
$^{41}$K$^{35}$Cl	&	244.448	&	974.7	&$	\varv=2	,	J=33 - 32	$ &	...	&	...	&	...	&	...	&	Tentative	\\
$^{34}$SO$_2$	&	244.482	&	81.9	&$	\nu=0	,	J_{K_a,K_c} = 14_{0,14}  -  13_{1,13}	$ &	...	&	...	&	...	&	...	&	Tentative	\\
K$^{37}$Cl	&	244.594	&	582.1	&$	\varv=1	,	J=33 - 32	$ &	100.6	&	8.0	&	0.03	&	0.3	&		\\
CS	&	244.936	&	23.5	&$	\varv=0	,	J=5 - 4	$ &	99.39	&	18.2	&	1.95	&	3.0	&		\\
$^{30}$SiS	&	245.088	&	76.5	&$	\varv=0	,	J=14 - 13	$ &	99.3	&	19.6	&	1.10	&	1.0	&		\\
$^{34}$SO$_2$	&	245.302	&	28.9	&$	\nu=0	,	J_{K_a,K_c} = 6_{3,3}  -  6_{2,4}	$ &	...	&	...	&	...	&	...	&	Tentative	\\
Na$^{35}$Cl	&	245.401	&	626.2	&$	\varv=1	,	J=19 - 18	$ &	101.3	&	14.6	&	0.77	&	0.5	&		\\
SO$_2$	&	245.563	&	61.0	&$	\nu=0	,	J_{K_a,K_c} = 10_{3,7}  -  10_{2,8}	$ &	98.3	&	19.7	&	2.79	&	3.0	&		\\
K$^{35}$Cl	&	245.624	&	183.0	&$	\varv=0	,	J=32 - 31	$ &	101.9	&	12.2	&	0.34	&	0.5	&		\\
Si$^{34}$S	&	245.960	&	1134.2	&$	\varv=1	,	J=14 - 13	$ &	99.7	&	7.7	&	0.07	&	0.5	&		\\
\hline
	\end{tabular}
\begin{flushleft}
\textbf{Notes.} For NS, we list the three brightest hyperfine components (see discussion in Sect.~\ref{sec:ns-det}).
For \ce{C^17O}, we give the frequency and quantum numbers corresponding to the brightest hyperfine component, which happens to also be the central hyperfine component. ($^{a}$) Ap. gives the radius of the circular aperture used to extract the spectrum from which the central velocity ($\upsilon_\mathrm{cent}$), velocity width ($\upsilon_\mathrm{w}$) and integrated flux were calculated. These line parameters were not calculated for known blended lines or lines that were too marginal in the spectra to fit a soft parabola function (see text for details).
\end{flushleft}
\end{table*}

\begin{figure*}
\includegraphics[width=\textwidth]{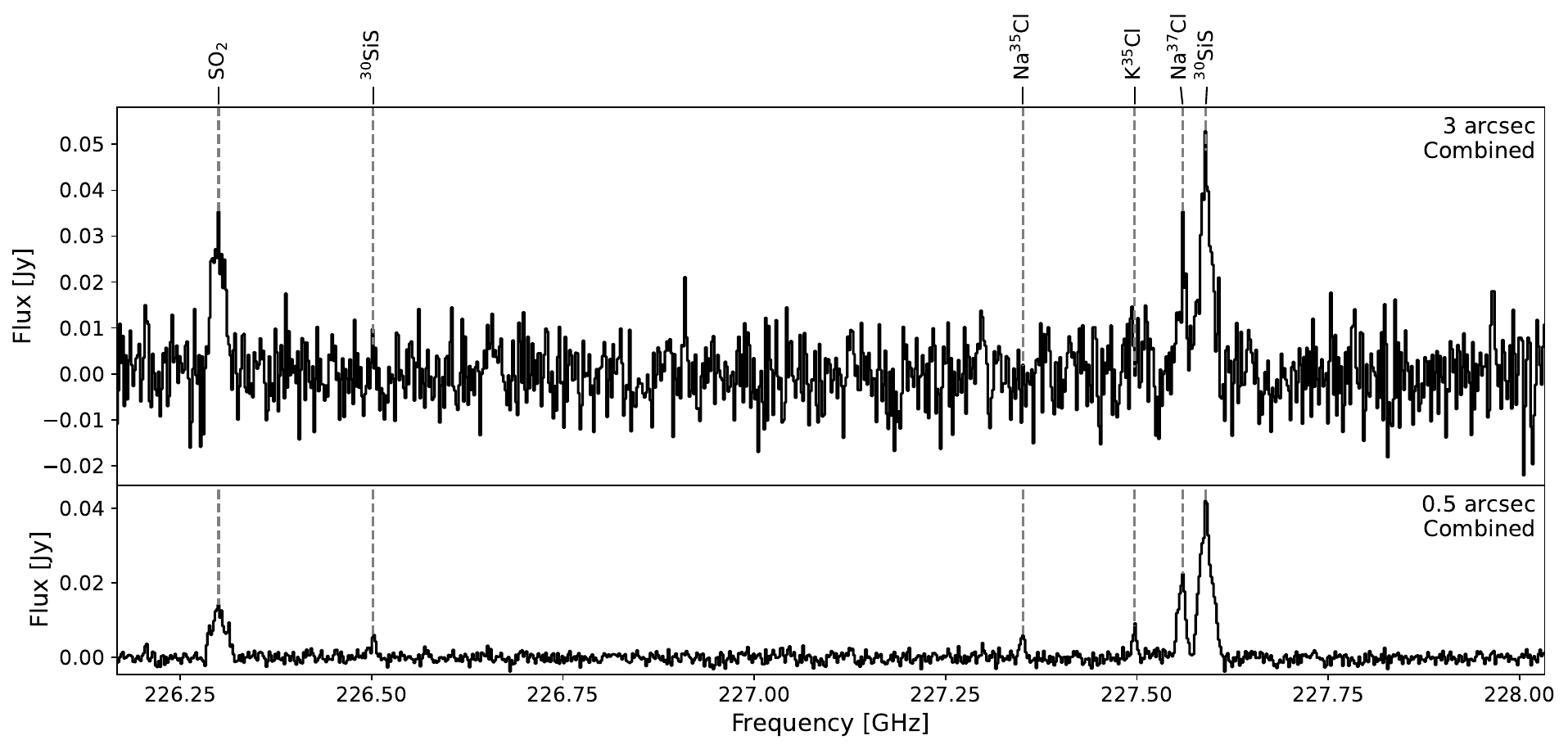}
\includegraphics[width=\textwidth]{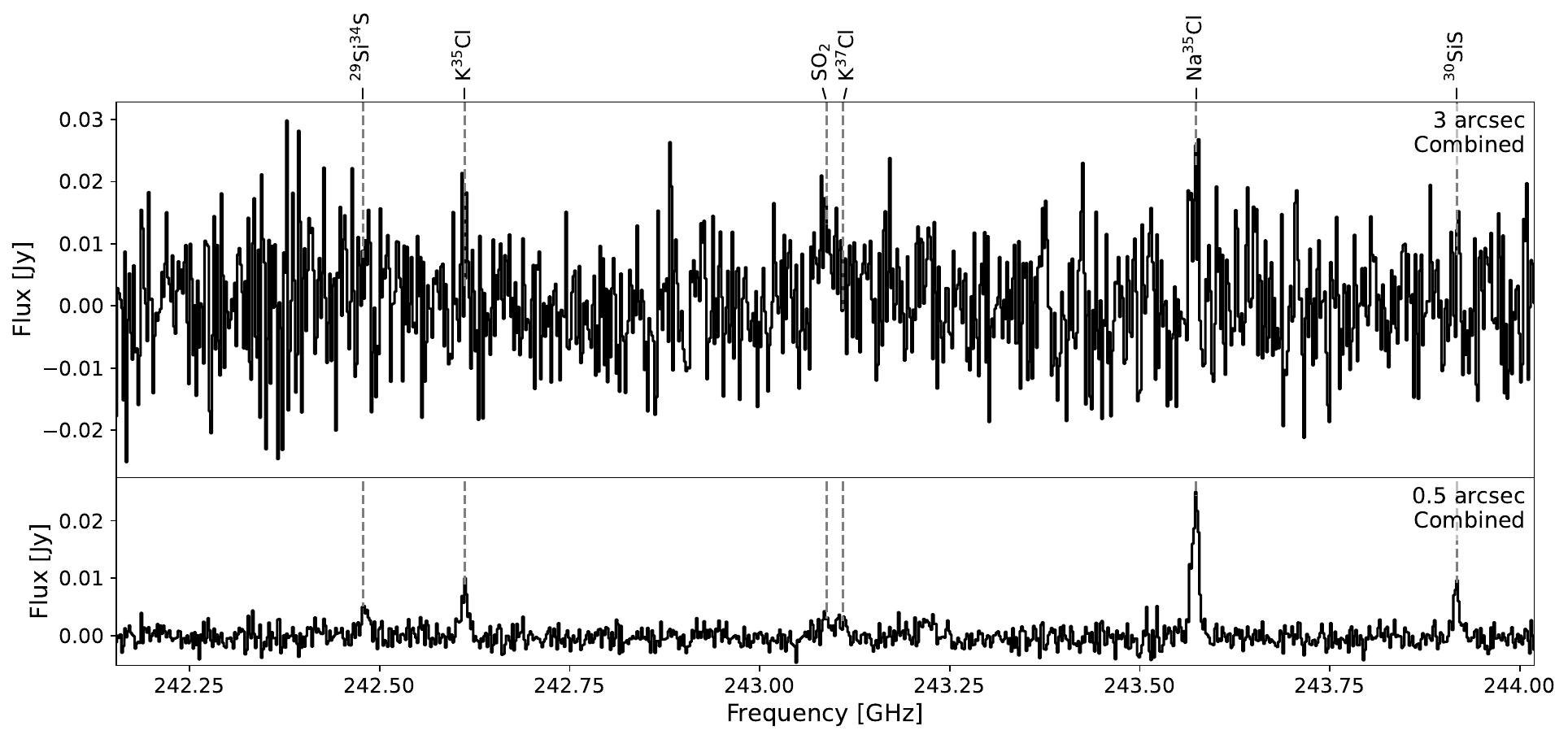}
\caption{ALMA Band 6 spectra for OH~30.1$-$0.7. For each spectral window we show spectra for a large extraction aperture and a small extraction aperture, with the extraction aperture radius and ALMA configuration given in the top right of each subplot. The molecular carriers of various spectral lines are indicated at the top of each plot. See Table \ref{tab:oh30ids} for details.}
    \label{fig:oh30specb6}
\end{figure*}
\addtocounter{figure}{-1}
\begin{figure*}
\includegraphics[width=\textwidth]{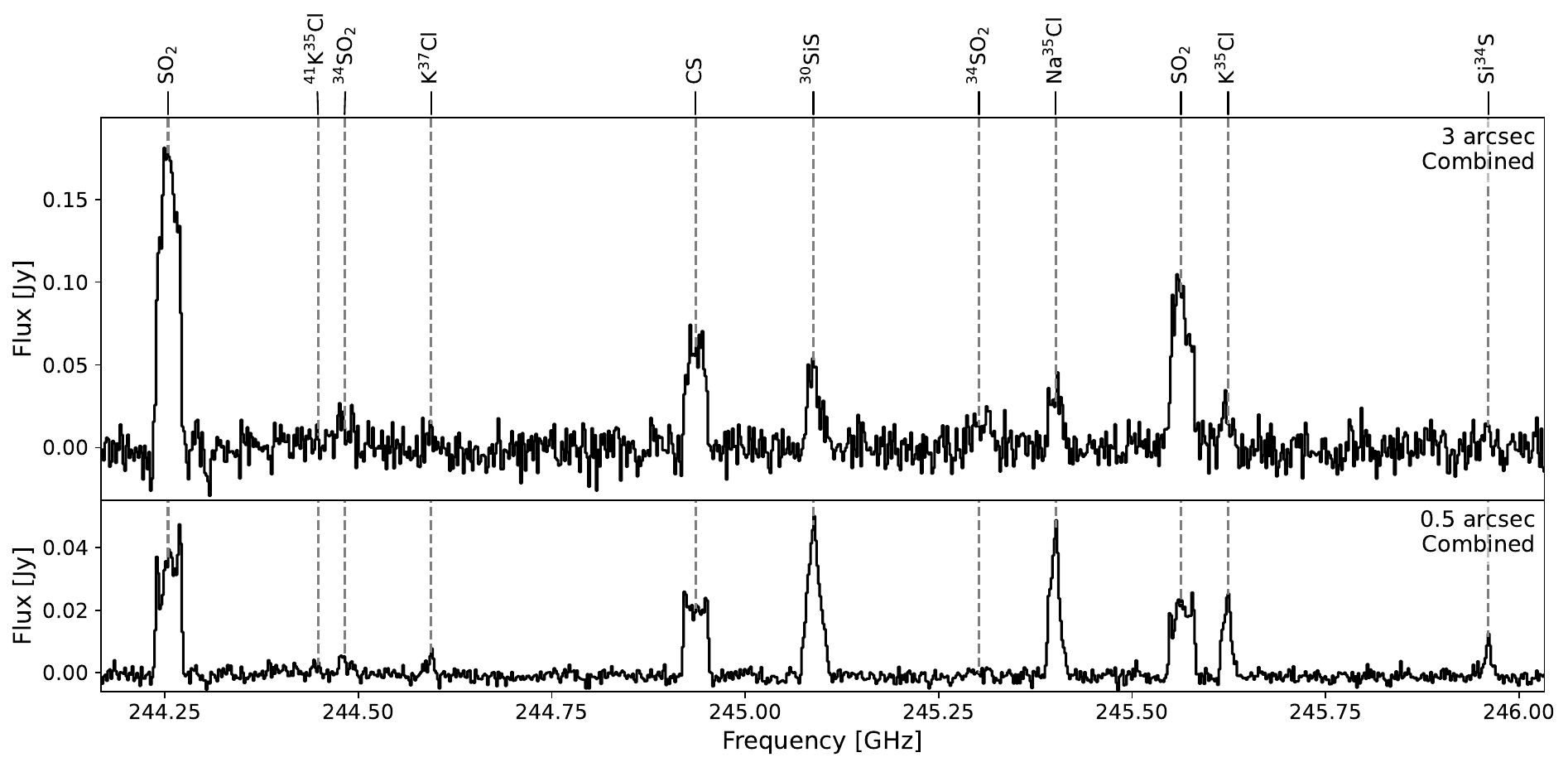}
\caption{continued.}
\end{figure*}

\begin{figure*}
\includegraphics[width=\textwidth]{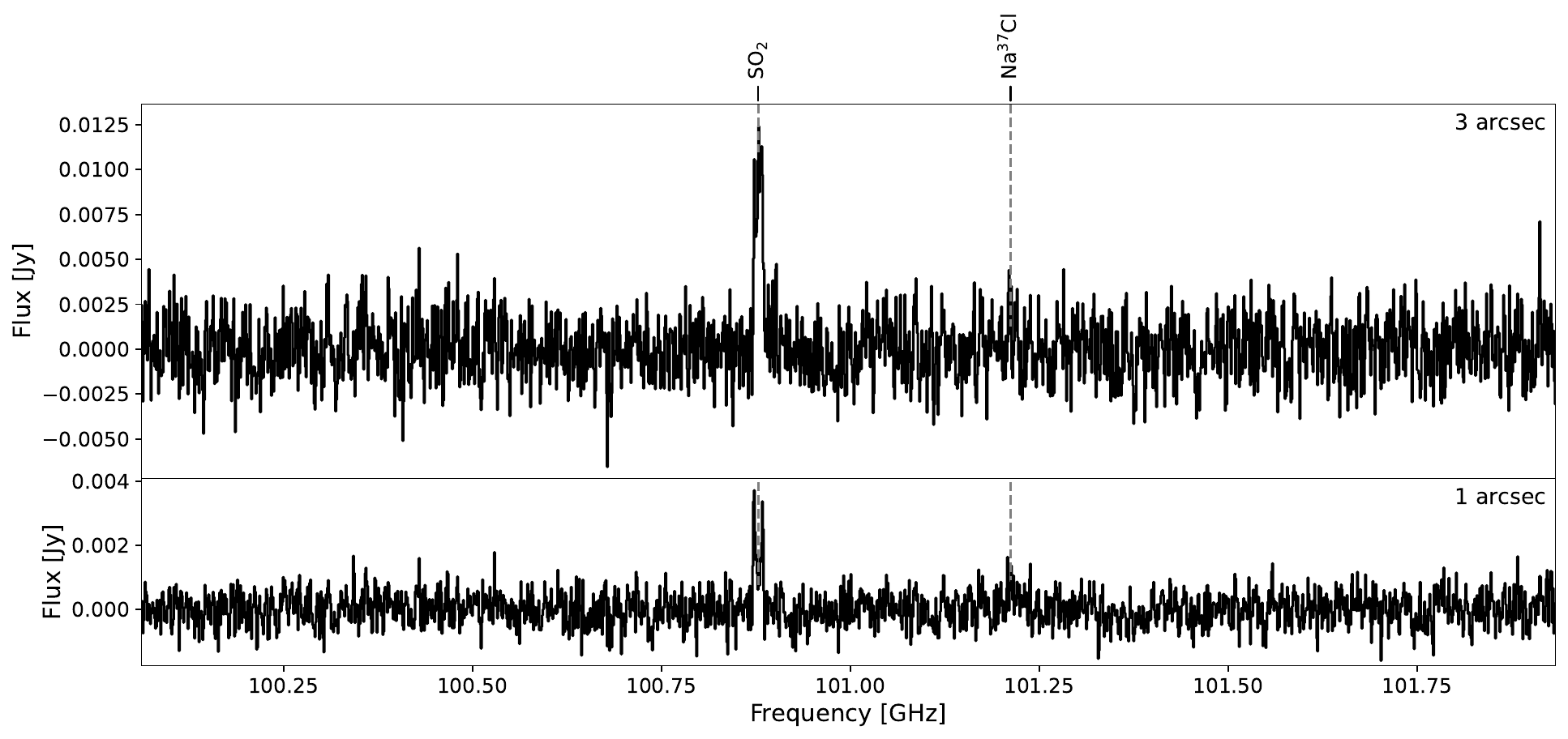}
\includegraphics[width=\textwidth]{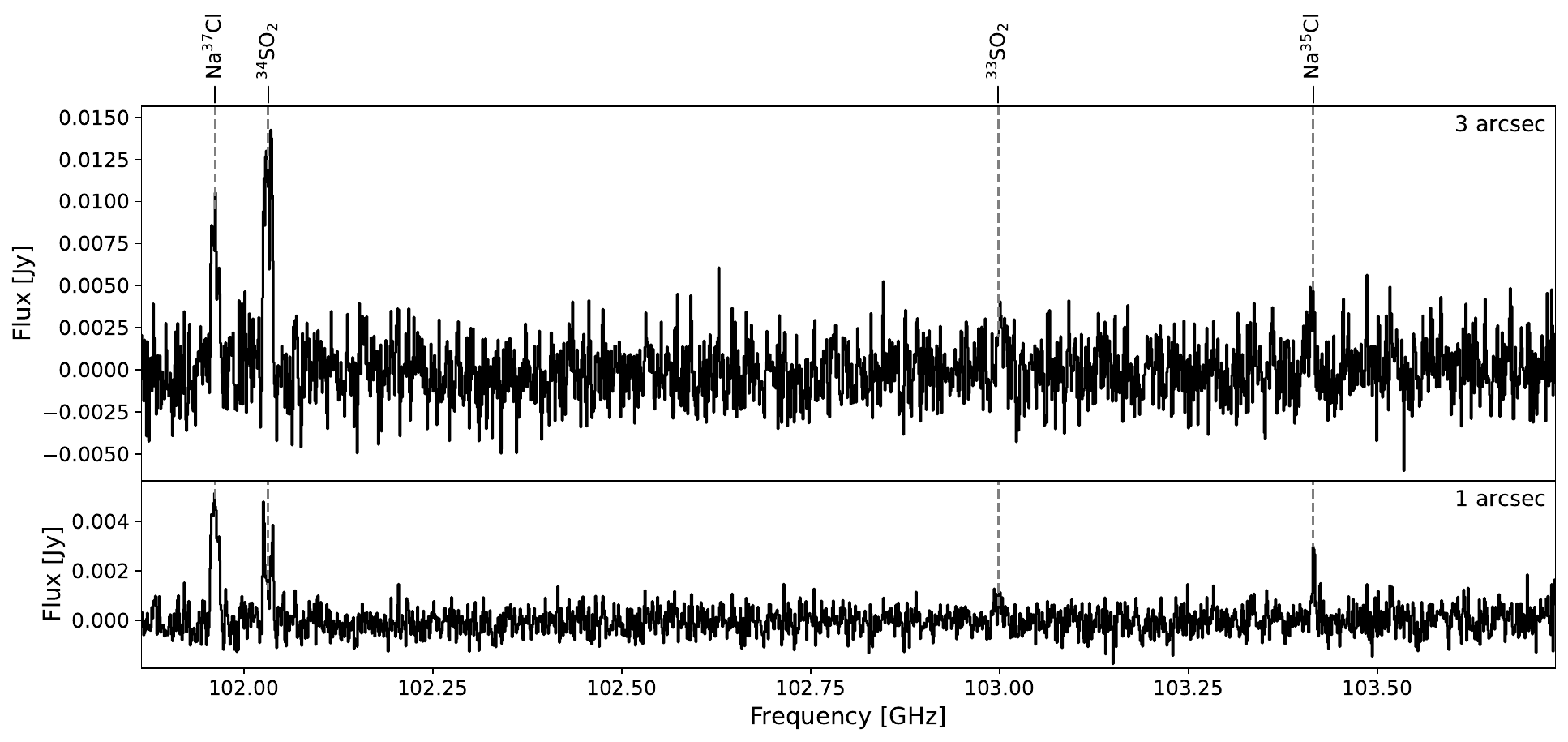}
\caption{ALMA Band 3 spectra for OH~30.1$-$0.7. For each spectral window we show spectra for a large extraction aperture and a small extraction aperture, with the extraction aperture radius in the top right of each subplot. The molecular carriers of various spectral lines are indicated at the top of each plot. See Table \ref{tab:oh30ids} for details.}
    \label{fig:oh30specb3}
\end{figure*}
\addtocounter{figure}{-1}
\begin{figure*}
\includegraphics[width=\textwidth]{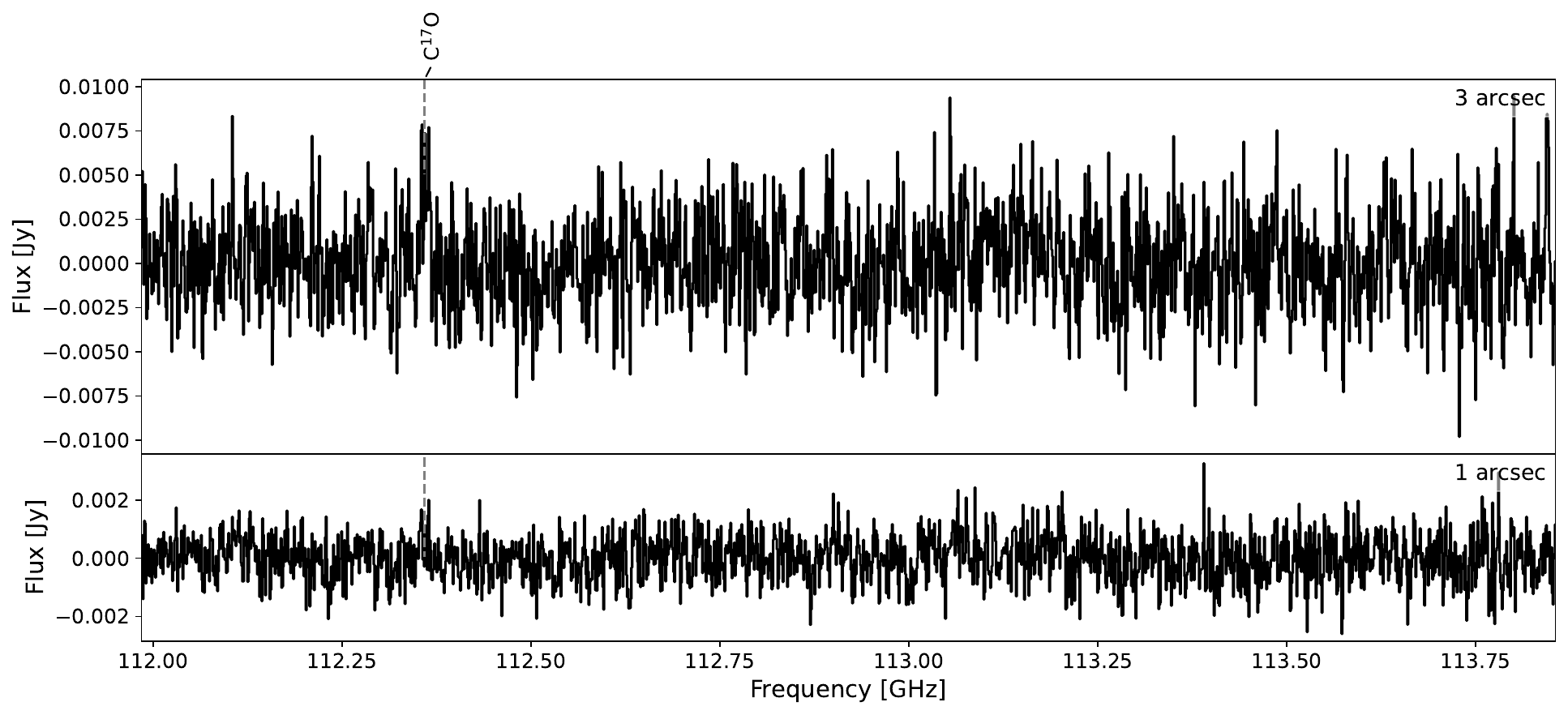}
\includegraphics[width=\textwidth]{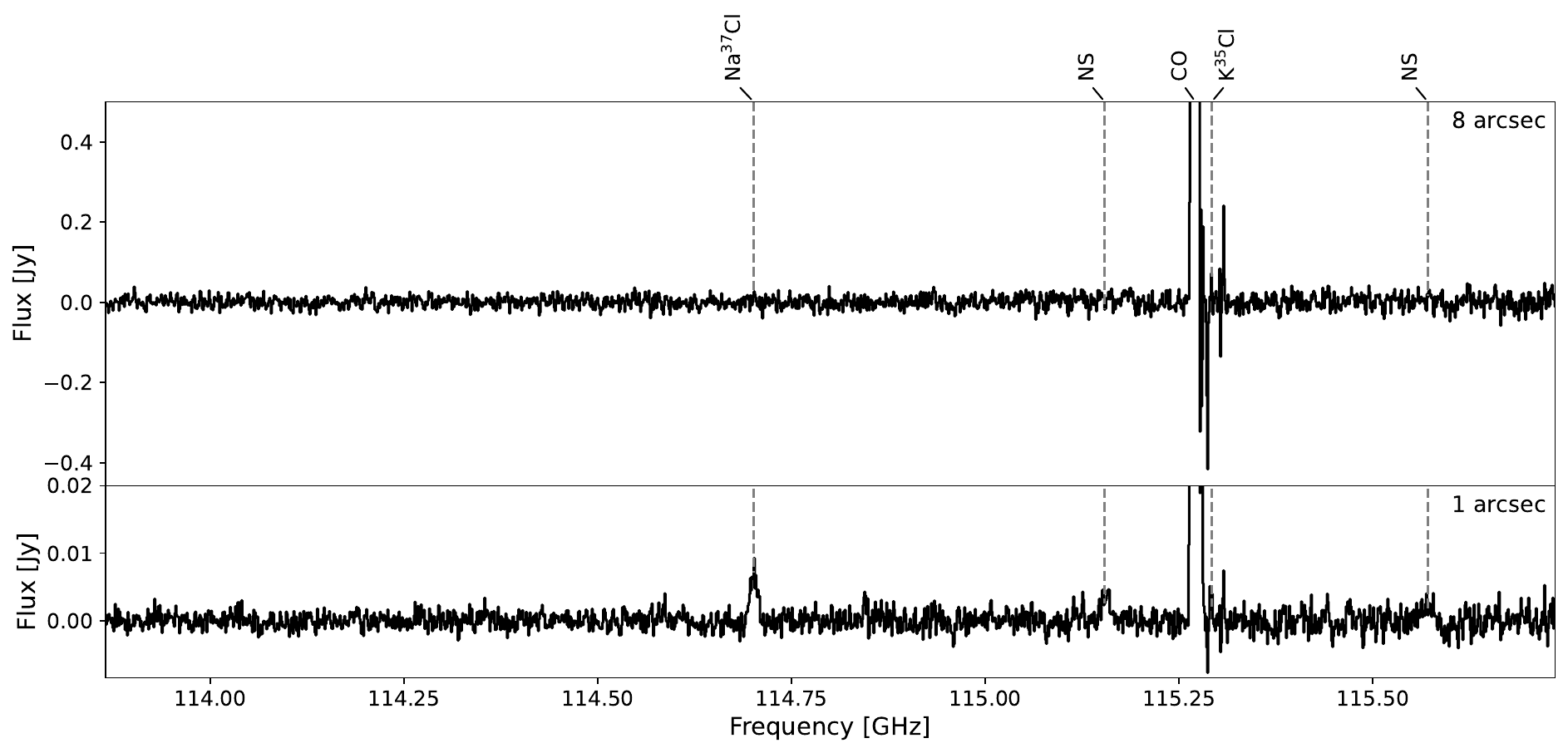}
\caption{continued}
\end{figure*}

\begin{figure*}
	\includegraphics[width=\textwidth]{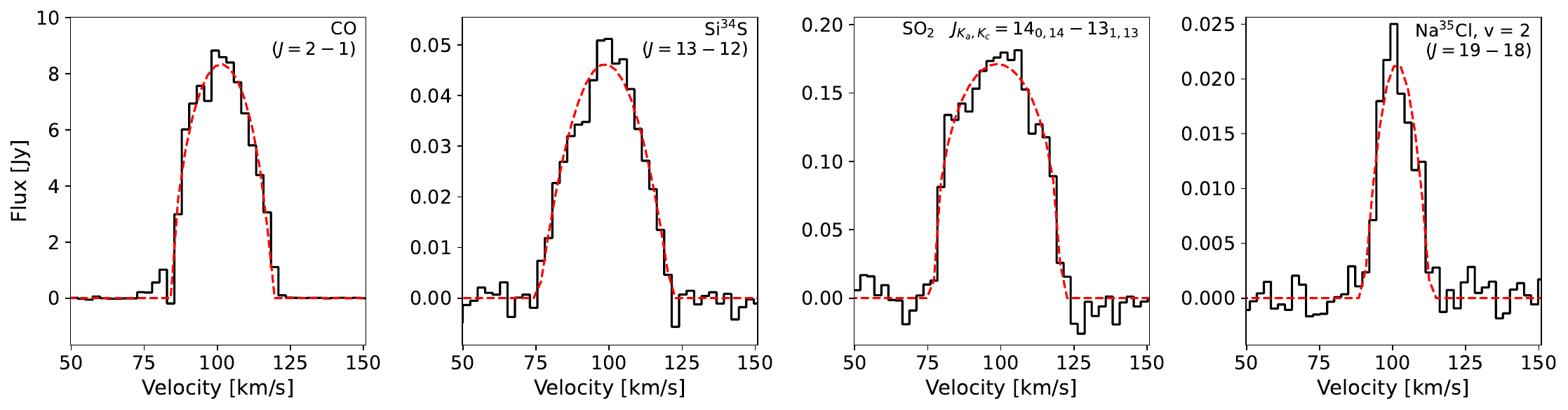}
    \caption{Examples of soft parabola fits to selected lines towards \ohth. Data is plotted as black histograms and the fits are shown as dashed red lines. The molecular carriers and the corresponding transitions are given in the top right corner of each panel. Further details for each line, including the extraction aperture for the spectrum, are given in Table \ref{tab:oh30ids}. Details of the fitting procedure are given in Sect.~\ref{sec:lines}.}\label{fig:softpara}
\end{figure*}

\begin{figure*}
	\includegraphics[width=\textwidth]{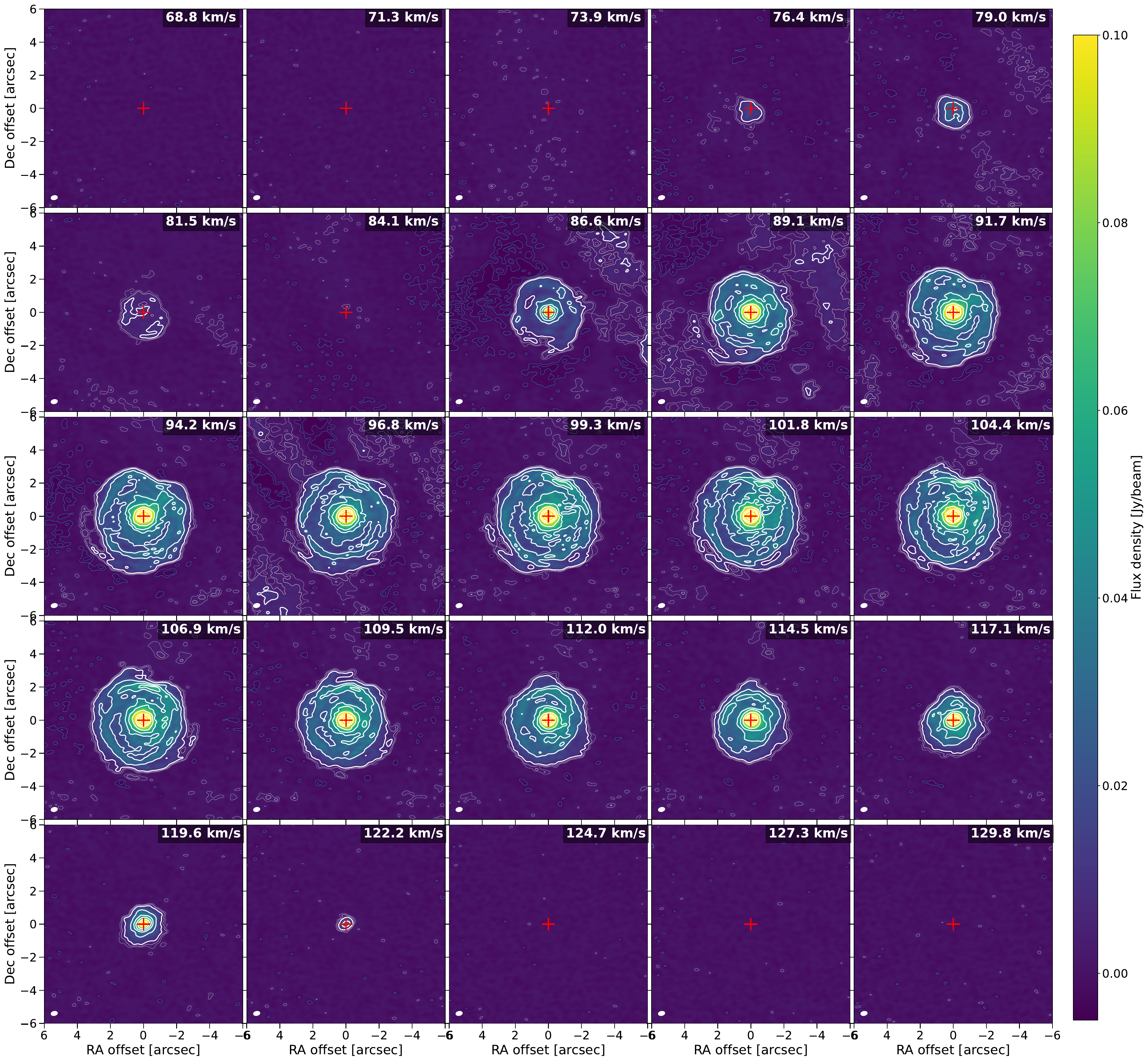}
    \caption{Channel maps for CO $J=2-1$ towards \ohth, as observed with the more extended ALMA configuration. The thicker white contours are drawn at levels of 10, 30, 50, 70, 100, 150, 200, 300, and 500$\sigma$, the thinner grey contours at levels of 3 and 5$\sigma$, and the dashed cyan contours at levels of $-3$ and $-5\sigma$. The red cross indicates the position of the continuum peak. The velocity of each channel is shown in the top right corner of each panel and the beam is shown as a white ellipse in the bottom left corners. North is up and east is left. Note that although some interstellar contamination has been resolved out by ALMA, a significant amount remains, especially for channels with velocities lower than 100~\kms.}\label{fig:oh30-co-TE-chan}
\end{figure*}

\begin{figure*}
	\includegraphics[width=\textwidth]{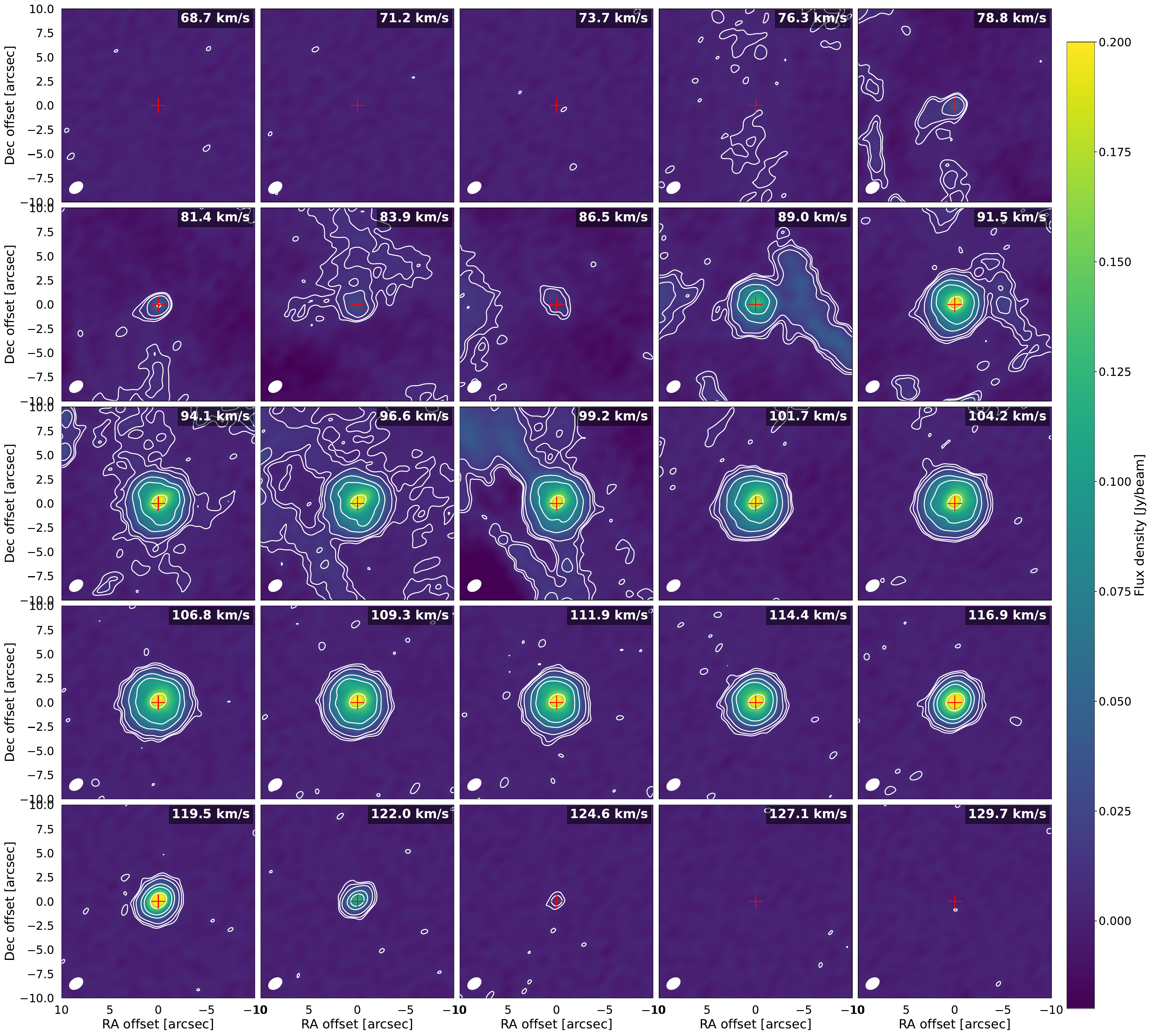}
    \caption{Channel maps for CO $J=1-0$ towards \ohth. The contours are drawn at levels of 3, 5, 10, 30, 50, and 100$\sigma$ and the red cross indicates the position of the continuum peak. The velocity of each channel is shown in the top right corner of each panel and the beam is shown as a white ellipse in the bottom left corners. North is up and east is left. Note that there is significant interstellar contamination for channels with velocities lower than 100~\kms.}\label{oh30-co-b3-chan}
\end{figure*}

\begin{figure*}
	\includegraphics[width=\textwidth]{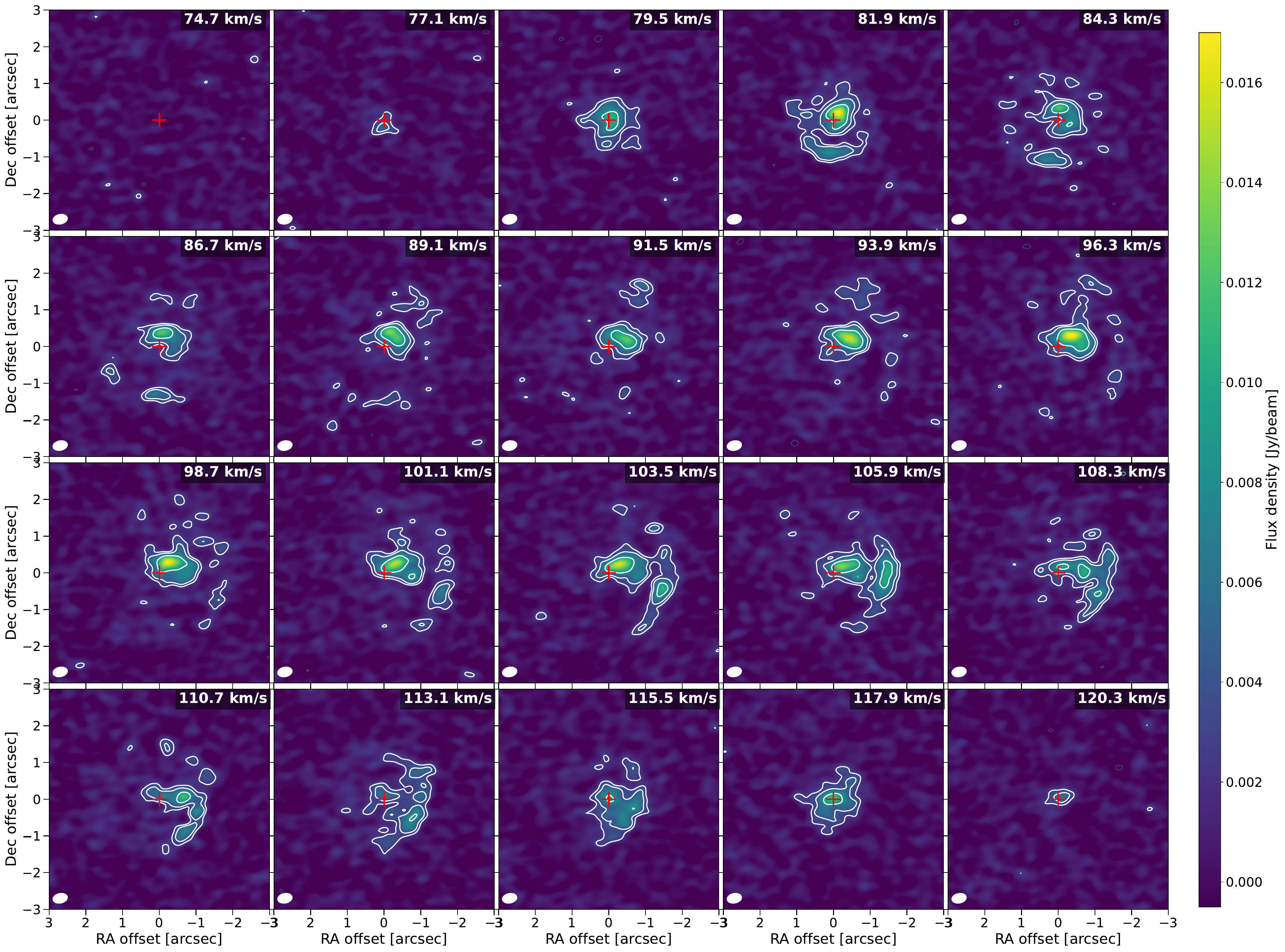}
    \caption{Channel maps for \so2 $J_{K_a,K_c} = 14_{0,14}  \to 13_{1,13}$ at 244.254 GHz towards \ohth. The contours are drawn at levels of 3, 5, and 10$\sigma$ and the red cross indicates the position of the continuum peak. The velocity of each channel is shown in the top right corner of each panel and the beam is shown as a white ellipse in the bottom left corners. North is up and east is left.}\label{fig:so2chans}
\end{figure*}

\begin{figure}
	\includegraphics[width=0.5\textwidth]{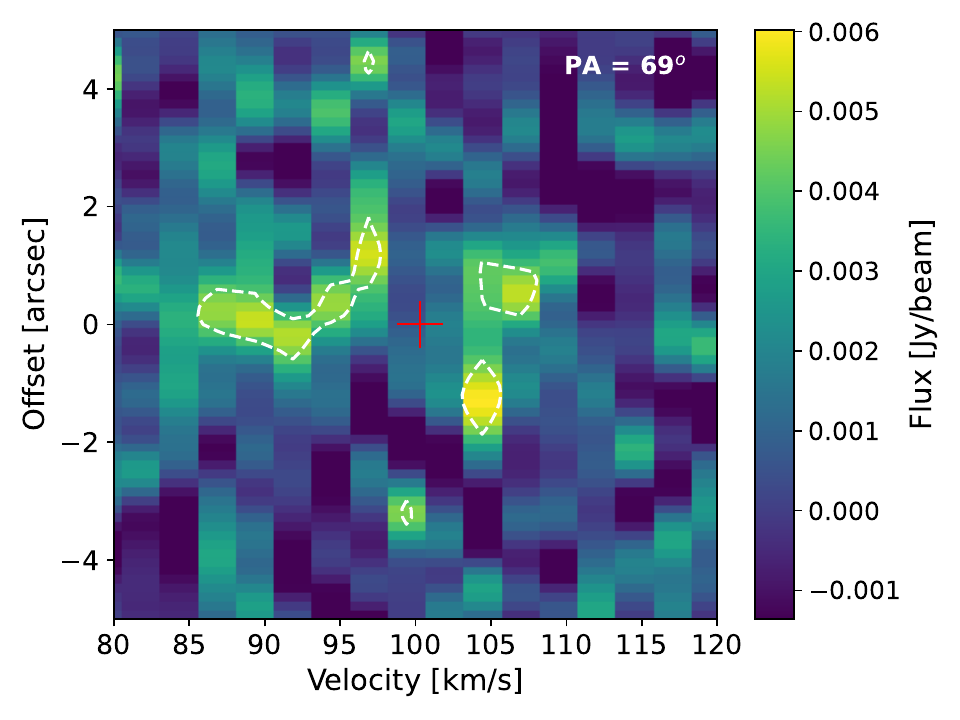}
    \caption{Position-velocity diagram for NS $e$ towards \ohth. The diagram was constructed for line passing through the continuum peak at an angle of $69\deg$ east from north (as shown in Fig.~\ref{fig:oh30ns}). The contours are drawn at levels of $3\sigma$ and the red cross indicates the position and systemic velocity of the AGB star. }\label{fig:ns-pv}
\end{figure}

\begin{figure}
	\includegraphics[width=0.5\textwidth]{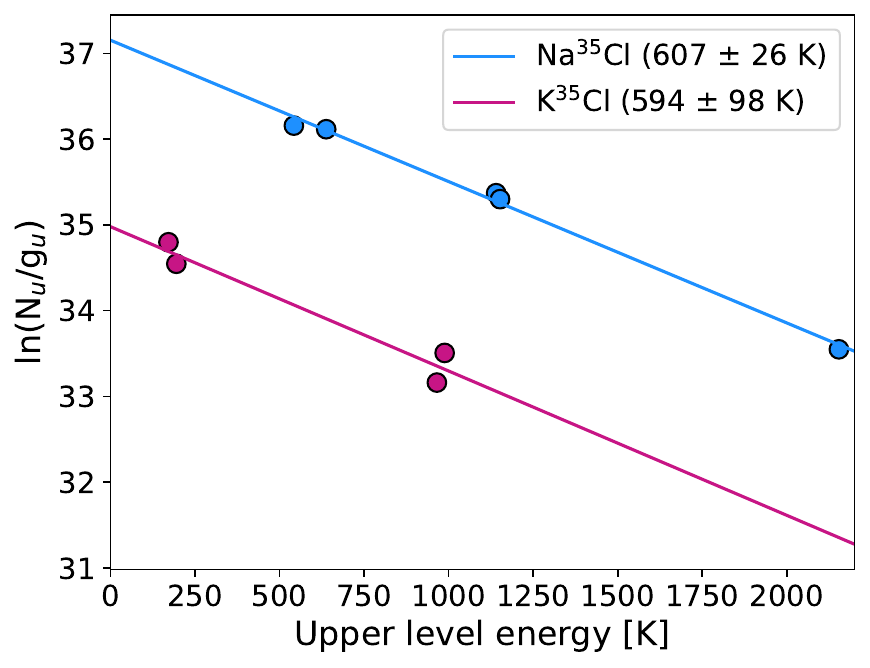}
    \caption{Population diagrams for Na$^{35}$Cl (blue) and K$^{35}$Cl (pink). The lines give the best linear fit to the data. Average gas temperatures are given in the legend and are close to 600 K for both molecules. The source averaged column densities are $1.4\e{12}$~cm$^{-2}$ and $1.6\e{11}$~cm$^{-2}$ for Na$^{35}$Cl and K$^{35}$Cl, respectively.} \label{fig:saltpop}
\end{figure}


\section{Interstellar contamination}\label{app:contam}


\begin{figure*}
	\includegraphics[width=0.32\textwidth]{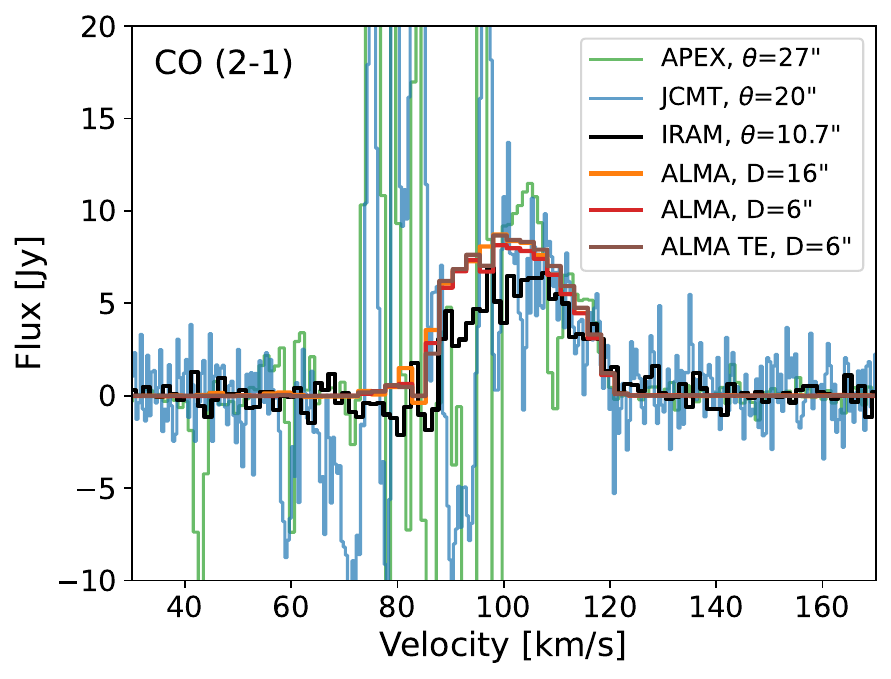}
	\includegraphics[width=0.32\textwidth]{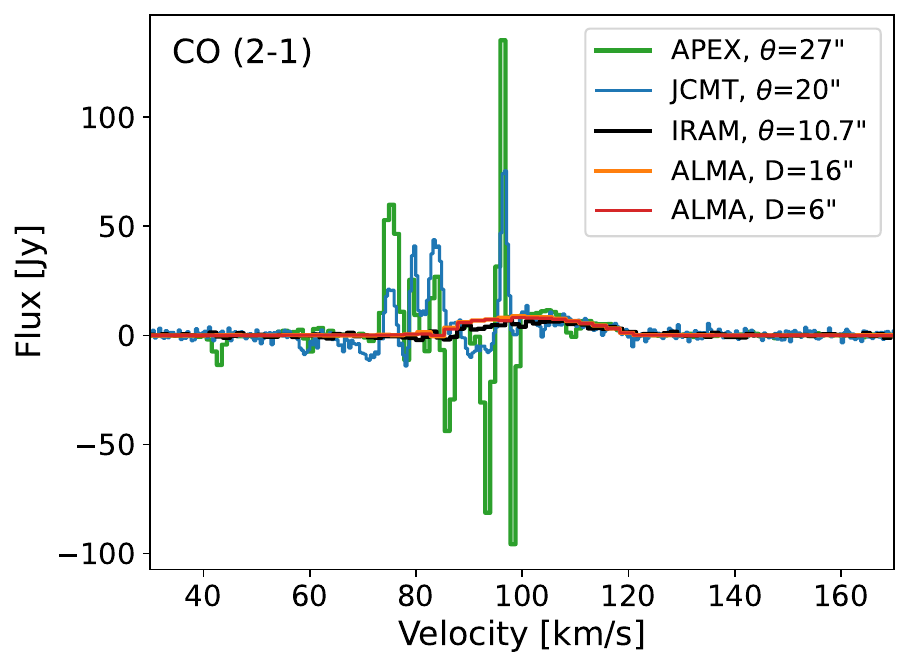}
	\includegraphics[width=0.32\textwidth]{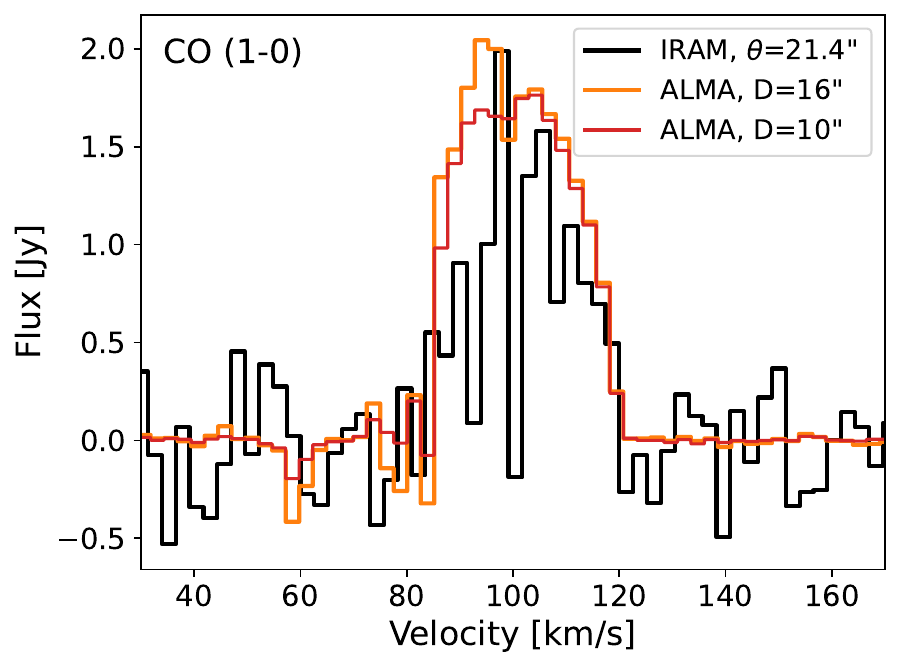}
    \caption{ALMA observations of CO towards \ohth\ compared with APEX, JCMT and IRAM 30m observations of the same lines. The legends give the HPBW, $\theta$, for the single-dish data, and the diameter of the circular extraction apertures, D, for the ALMA data. TE refers to data observed with only the ALMA extended configuration. The single-dish observations show significant line contamination from sources other than the AGB stars. The left and middle panels show CO ($J = 2-1$) with the left panel having a flux scale that allows the CO emission to be clearly seen, while the middle panel has a flux scale that shows the full range of the ISM contamination at velocities $\lesssim 100$~\kms. The right panel shows CO ($J = 1-0$).}\label{fig:co-contamination}
\end{figure*}

{Here we discuss in more detail the interstellar contamination experienced by \ohth\ (see also Sect.~\ref{sec:contam}), which we believe arises from the background star-forming region W43. Contamination can be seen in most previous studies of CO towards this object \citep[e.g.][]{De-Beck2010}, most of which rely on single-dish observations.} Although some clever observational techniques can be employed to improve the single-dish observations of contaminated stars \citep[see for example][]{Heske1990}, these are generally less precise than the spatial filtering of an interferometer.

In Fig.~\ref{fig:co-contamination} we plot the combined ALMA spectra of CO $J=2-1$, extracted for two apertures of different size, superposed with the spectra of the same line observed with the Atacama Pathfinder Experiment \citep[APEX,][]{Gusten2006}, the James Clerk Maxwell Telescope (JCMT), and the IRAM 30m telescope. The JCMT spectrum was first published in \cite{De-Beck2010}, the IRAM spectrum was first published in \cite{Justtanont2013}, and the APEX spectrum is previously unpublished and was observed by the APEX/PI230 instrument\footnote{PI230 is a collaboration between the European Southern Observatory (ESO) and the Max-Planck-Institut f\"ur Radioastronomie (MPIfR).} in August 2018 (project ID: O-0102.F-9301B).
There is considerable contamination in the JCMT observation, especially below velocities of around 107~\kms. The effect is even larger for the APEX observation, which has a larger HPBW ($\theta = 27\arcsec$ for APEX and $\theta = 20\arcsec$ for JCMT). IRAM, which has the smallest beam ($\theta=10.7\arcsec$), suffers from the least amount of contamination, but the contamination still impacts the flux below $\sim107$~\kms. In Fig.~\ref{fig:co-contamination}, we adjust the flux axis of the left panel to best show the CO emission originating from \ohth, while the middle panel shows the full range of the emission/absorption detected by APEX and JCMT. The contamination entirely dominates the APEX and JCMT flux below 100~\kms.
Nevertheless, the ALMA observations are in general agreement with the single-dish observations above 107~\kms, suggesting that the AGB flux is likely not resolved out at those velocities in the Band 6 observations. The fact that the ALMA spectra extracted for two apertures of different sizes (radii of 8\arcsec\ and 3\arcsec) largely agree with each other indicates that the CO flux is mostly confined within 3\arcsec\ of the continuum peak. We also checked the line profile of the ALMA observations taken with only the extended array configuration, and found that the flux (for an extraction aperture with diameter $6\arcsec$) does not deviate from that of the combined data; i.e. even for the extended configuration, there is no resolved-out flux (see also the extended configuration channel maps in Fig.~\ref{fig:oh30-co-TE-chan}). Based on these comparisons, and considering that CO is the most extended observed molecule, we assert that none of the Band 6 observations suffer from resolved out flux.

We also compare the ALMA Band 3 CO $J=1-0$ line with the IRAM observation of the same line first published in \cite{Justtanont2013}, plotted in the right panel of Fig.~\ref{fig:co-contamination}. Since the single-dish beam is larger at the lower frequency, the IRAM line suffers from significant contamination. 
The Band 3 observation has a larger maximum resolvable scale than the Band 6 observation, and hence is also more susceptible to contamination. From inspecting the channel maps, shown in Fig.~\ref{oh30-co-b3-chan}, we can see significant contamination, in the form of diagonal bars, in the channels with velocities below 100~\kms. The contamination can also be seen as spikes in the full spectrum in Fig.~\ref{fig:oh30specb3} and contributes to a difficulty in disentangling a possible \ce{K^35Cl} $\varv=0$ line from both the AGB CO emission and the background contamination (see Table \ref{tab:oh30ids}). However, given the shape of the CO emission and the larger maximum resolvable scale for the Band 3 data, we also assume that there is no resolved out flux in the Band 3 observations.

\section{Comparison with previous studies of molecular detections and distributions}\label{app:previousmol}

\subsection{SiS}

SiS is a commonly observed molecule that has been detected in the envelopes of all chemical types of AGB stars. It is generally found to be more abundant in carbon-rich CSEs than oxygen-rich CSEs \citep{Schoier2007}. However, \cite{Danilovich2018} found that SiS was more likely to be detected for AGB stars with higher mass-loss rates and hence denser circumstellar envelopes, across chemical types. This trend was confirmed by \cite{Massalkhi2019} for a larger sample of carbon stars, although \cite{Danilovich2019} also found that very sensitive ALMA observations could detect low abundances of SiS (down to $\sim10^{-8}$ relative to \h2) for even low mass-loss rate oxygen-rich stars. Since \ohth\ is thought to have a high mass-loss rate, detections of SiS are expected where the lines are covered by our observations. 

\subsection{CS}

Since it is a carbon-bearing molecule, CS has often been associated with carbon-rich CSEs. However, it has also been detected in several oxygen-rich CSEs, as well as towards S-type AGB stars \citep[e.g.][]{Lindqvist1988,Danilovich2018}. Similar to SiS, CS has been found to be more prevalent in higher density oxygen-rich and S-type CSEs \citep{Danilovich2018}, possibly because of a higher prevalence of shocks (e.g. from larger pulsations in Mira variables as compared with semiregular variables) in these CSEs which can liberate C from CO to enable the formation of CS. This may also result in CS being more abundant for Mira variable stars than semiregular variable stars, among the oxygen-rich AGBs \citep{Danilovich2019}. However, for carbon-rich AGB stars, the abundance of CS has generally been found to be one to two orders of magnitude higher than for oxygen-rich stars \citep{Massalkhi2019,Danilovich2018}. Since \ohth\ has a high mass-loss rate and hence high wind density, detections of CS are expected despite it being an oxygen-rich star.

\subsection{\ce{SO2}}\label{previousso2}

\so2 is thought to be formed in circumstellar envelopes from SO. 
Previous studies, mostly focussing on AGB stars known to have low initial masses, have shown that the distributions of SO and \so2 can be divided into two groups of morphological structures in AGB circumstellar envelopes. \cite{Danilovich2016} found, from a small sample, that SO distributions could either be centred on the star or could exhibit a shell-like distribution, with the relative abundance of SO peaking further out in the CSE, rather than being centred on the star. It was found that the stars with lower mass-loss rates (or lower wind densities) exhibited centralised SO distributions, while stars with higher mass-loss rates (or higher wind densities) exhibited shell-like SO distributions. The larger ATOMIUM survey, using spatially resolved ALMA observations, confirmed this trend for SO and also for \so2 \citep{Wallstrom2024}. Earlier data had been insufficient to determine the spatial distribution of \so2. The chemical models of \cite{Van-de-Sande2022} and \cite{Van-de-Sande2023} predict more shell-like SO and \so2 distributions for AGB stars with UV-producing companions and/or with more clumpy CSEs.

\subsection{NS}

Aside from the AGB stars discussed in Sect.~\ref{sec:nschem}, NS has been detected towards an early planetary nebula \citep{Sanchez-Contreras2000}, giant molecular clouds \citep[where it seen to be enhanced in massive star-forming regions such as Orion KL,][]{McGonagle1997}, hot cores \citep{Xu2013}, comets \citep[e.g.][]{Irvine2000}, and a starburst galaxy \citep{Martin2003,Martin2021}. All of these objects contain regions with high UV fluxes (or experience UV flux during part of their orbits, in the case of comets), suggesting that this could be the dominant formation mechanism of astronomical NS. \cite{Viti2001} performed chemical models of shocks in hot cores and found that NS can be enhanced through shock heating, which is broadly similar to the UV-driven chemistry driving the NS formation around W~Aql \citep{Van-de-Sande2022,Danilovich2024}. 
Older chemical models, which do not consider shocks or enhanced UV chemistry were not able to reproduce the observed abundances of NS in hot cores by several orders of magnitude \citep[e.g.][]{Millar1990}.

\subsection{Chlorides}

The first astronomical detections of NaCl and KCl were reported by \cite{Cernicharo1987a} towards the nearby carbon-rich AGB star CW~Leo (along with AlCl). Subsequently, abundances relative to \h2 of $1.8\e{-9}$ for NaCl and $\sim5\e{-10}$ for KCl were derived from more comprehensive observations by \cite{Agundez2012}, who also found the \ce{^35Cl/^37Cl} isotopic ratio to be $2.9\pm0.3$, close to the solar isotopic value of $3.3\pm0.3$. Spatially resolved observations by \cite{Quintana-Lacaci2016a} found elongated NaCl and KCl emission and concluded that the NaCl was likely arranged in a spiral or torus (with the formation of such a structure likely driven by a companion). That study also detected NaCl in the first vibrationally excited state for the first time.

NaCl has also been detected towards a handful of oxygen-rich stars. The low-mass AGB star IK~Tau \citep[initial mass $\sim 1.3~\msol$,][]{Danilovich2017} has been observed to exhibit several different rotational lines of NaCl in the ground vibrational state \citep{Milam2007,Velilla-Prieto2017}. Spatially resolved observations with ALMA found it to have clumpy irregular NaCl ($\varv=0$) emission \citep{Decin2018}. A subsequent 3D analysis by \cite{Coenegrachts2023} found that the emission was roughly arranged in a clumpy spiral and the formation of NaCl may have been driven by a stellar or planetary companion. We note that the nearby low-mass AGB star R~Dor --- observed with the same spectral setup and sensitivity as IK~Tau but not presently thought to have a solar type binary companion \citep{Vlemmings2018} --- did not have any detections of NaCl \citep[or NS,][]{Decin2018}. Additional detections of NaCl and KCl were reported by \cite{Wallstrom2024} in the ATOMIUM survey for the AGB stars IRC$-$10529 (NaCl in $\varv=0,1$ and KCl in $\varv=0$), IRC+10011 (NaCl in $\varv=0,1,2$ and KCl in $\varv=0$) and GY~Aql (NaCl in $\varv=0$). All three of these oxygen-rich AGB stars are thought to have companions shaping their winds, though the precise natures of these companions are not yet known \citep{Decin2020}. The RSGs  VY~CMa, NML~Cyg and VX Sgr also have NaCl 
detected towards them \citep{Milam2007,Kaminski2013c,Decin2016,Singh2022,Quintana-Lacaci2023,Wallstrom2024}. In the spatially-resolved observations of VY~CMa, NaCl is seen to trace high velocity jets \citep{Quintana-Lacaci2023}.

There have also been detections of NaCl towards early post-AGB stars, such as CRL 2688 \citep[the Egg Nebula,][]{Highberger2003} and OH~231.8~+4.2 \citep[the Rotten Egg Nebula][]{Sanchez-Contreras2018,Sanchez-Contreras2022}, both of which have bipolar outflows, with NaCl detected in shell-like structures.
\cite{Highberger2003} used the IRAM 30m telescope to detect several NaCl lines in the ground vibrational state towards CRL~2688. They suggest that the formation of NaCl is likely driven by shocks and clumping in the high-velocity outflows.
\cite{Sanchez-Contreras2022} used ALMA to observe OH~231.8~+4.2, an AGB star which seems to be undergoing a premature evolution into the pre-Planetary Nebula phase, possibly driven by the A0 companion on a relatively close orbit (20~au).
They detect vibrationally excited NaCl up to $\varv=2$ (and tentatively $\varv=3$) and KCl (up to $\varv=1$) in a rotating and expanding cylindrical structure around the central star, coinciding with a circumbinary dust disc. They also find a relatively high vibrational temperature ($T_\mathrm{vib}=1125\pm160$~K), consistent with NaCl forming in hot conditions. In this case the dust disc could also shield the Na and K atoms, preventing excessive photoionisation. \cite{Sanchez-Contreras2022} conclude that NaCl and KCl are good tracers of high-density regions and may be shock tracers. While \ohth\ may have a similar companion to OH~231.8~+4.2, it clearly does not have a bipolar outflow (and is not easily detected in the optical, unlike the Rotten Egg Nebula), suggesting a different orbital configuration.


Another astrophysical environment in which NaCl and KCl have been frequently detected are high-mass young stellar objects \citep{Ginsburg2019,Ginsburg2023}. Dubbed ``brinaries'' because of the strong correlation between NaCl and \h2O detections, \cite{Ginsburg2023} found that the salt detections were neither rare nor ubiquitous. They also find that the NaCl emission is restricted to the relatively inner regions of discs or outflows, which are likely to be the warmer (but not ionised) regions of their sources. This is broadly similar to the centrally confined NaCl and KCl detected here for \ohth, and likely correspond to similar hot and shocked formation conditions.

\subsubsection{The formation of NaCl and KCl}\label{app:chlorideform}

The main formation pathway for NaCl and KCl is thought to be through neutral-neutral reactions between the alkali metal atom and HCl. The reaction for NaCl has a relatively high activation energy of 42~kJ~mol$^{-1}$ \citep{Husain1986,Plane1989}, significantly more than the 12~kJ~mol$^{-1}$ barrier for K + HCl \citep{Helmer1993}. 
Thus formation of NaCl and KCl is favoured at higher temperatures. Furthermore, both reactions are endothermic (by 19.2 and 7.8 kJ mol$^{-1}$, respectively), and the reverse reactions of NaCl and KCl with H are fast with small temperature dependences. This means that NaCl and KCl can only persist if the H density is relatively low (see below).
Another important consideration is that Na and K have low ionisation potentials \citep[5.139 eV for Na, 4.341 eV for K,][which correspond to wavelengths of 241 nm and 286 nm, respectively]{IonisationEnergy} and their singly ionised states, \ce{Na+} and \ce{K+}, are un-reactive closed shell ions. This means that UV photons from a hot star could easily ionise Na and K and subsequent ion-neutral reactions would not readily occur, hence effectively removing Na and K and preventing the formation of their chlorides. 
This dilemma is avoided if the environment is dense enough to provide shielding against ionisation. Given the thick, dusty envelope of \ohth\ \citep[see Sect.~\ref{sec:physics} and][]{Marini2023} and its high mass-loss rate, there should be an extremely high amount of extinction in the inner wind, which would strongly attenuate any UV photons except for in the immediate vicinity of the UV-producing star. For this reason, heating from UV photons alone is unlikely to drive the formation of NaCl or KCl since the UV photons will be highly attenuated and, if they were not, they would instead ionise the metals and prevent the formation of NaCl and KCl. Note that this is the opposite scenario to the formation of NS. NS forms after its precursor \ce{N2} has been photodissociated \citep{Van-de-Sande2022,Danilovich2024}, whereas NaCl and KCl require that their precursors Na and K not be photoionised. This difference could explain why the distributions of NaCl and NS in the Band 3 zeroth moment maps (Fig.~\ref{fig:oh30nacl} and Fig.~\ref{fig:oh30ns}) are not co-located. An interesting additional check on the chemistry would be mapping the SiN emission (which was not covered in our observations) because its formation follows after the photoionisation of Si to \ce{Si+}.

Another possible obstacle for the formation of NaCl and KCl is the H/\h2 ratio in the wind. If this is too high (e.g. if H/\h2~$\gtrsim 10^{-4}$) then the steady state reactions will favour the formation of HCl, Na, and K, rather than NaCl and KCl. Given the observations presented here and the implied high abundances of both NaCl and KCl, we can confidently say that the H/\h2 ratio is low.
High temperatures (e.g. 1000--2000 K) and high densities will also contribute to more rapid formation of NaCl and KCl (as all chemical reactions proceed more rapidly at higher number densities). This fits with previous results finding that NaCl is preferentially formed as a consequence of shocks \citep{Cherchneff2012}.

\subsection{Species not detected or not covered}

Some of the species that we might expect to see towards a star such as \ohth\ were either not covered in our observations or were not detected. To eliminate confusion between these two possibilities, we briefly summarise here which key molecular species were not detected and which were not covered.

SiO and HCN are both seen almost ubiquitously towards AGB stars of all types. Our spectral set up was such that no lines in the ground or first four vibrationally excited states of \ce{^28Si^16O}, \ce{^29Si^16O} or \ce{^30Si^16O} were covered. No lines of HCN or \ce{H^13CN} in the ground vibrational state were covered. Transitions of CN, the photodissociation product of HCN, were covered but not detected above the noise.

SO is generally observed towards the same stars as \so2, with similar distribution patterns in the CSEs \citep{Danilovich2016,Wallstrom2024}. Although we present several detections of \so2 towards \ohth, no lines of SO or its isotopologues (\ce{^32SO}, \ce{^34SO}, or \ce{^33}SO) were covered in our observations.

The recent work of \cite{Baudry2023} identified several highly excited \h2O lines (with upper level energies in the range 3954--9012~K) in ALMA observations of a sample of oxygen-rich AGB and RSG stars. Unfortunately, none of the emission lines detected in the \cite{Baudry2023} sample were covered in our spectral configurations. We found no other lines of \h2O towards \ohth, despite clear detections of {relatively low-energy} \h2O lines observed by \textsl{Herschel}/HIFI \citep{Justtanont2015}. No OH lines in the ground vibrational state were covered in the Band 6 data.

\h2S is associated with high mass-loss rate/high CSE density oxygen-rich AGB stars \citep{Danilovich2017a}. Only two lines of \h2S were covered in our observations and both were high-energy lines ($E_\mathrm{low} > 2000$~K) that have not previously been detected for any AGB stars. One lower-energy line of \h2$^{34}$S was covered ($E_\mathrm{low} = 240$~K) but not detected.

Despite our detections of NaCl and KCl, Al$^{35}$Cl was not detected. However, the only line in the ground vibrational state of Al$^{35}$Cl fell in the less sensitive Band 3 data and could be blended with \ce{^34SO2}. In the case of Al$^{37}$Cl, one line in the ground vibrational state fell in each of the Band 3 and 6 observations. Neither is detected above the noise.

AlF, which was detected alongside AlCl towards W~Aql \citep{Danilovich2021} and has been seen towards various oxygen-rich AGB stars \citep{Saberi2022,Wallstrom2024}, was not detected towards \ohth. This is despite a line in the ground vibrational state being covered by our Band 6 observations. The chemistry behind AlF is not yet fully understood \citep[see][for a more extended discussion]{Danilovich2021}, but it seems to not be directly linked to the other metal halides (i.e. NaCl and KCl) that are so conclusively detected towards \ohth.

{SiN and SiC were identified by \cite{Van-de-Sande2022} as tracers of a UV-emitting companion star, along with NS. Although NS was detected for \ohth,} SiN was not covered in our observations, and two lines of SiC were covered in Band 6 but not detected above the noise.

\section{Previous observational studies of \ohth}\label{sec:pointings}


The study of rotational CO lines towards \ohth\ has been hampered by contamination along its line of sight, as discussed in Sect.~\ref{sec:contam} \citep[see also][]{De-Beck2010}, mainly originating from the massive star-forming region W43, located around 1.6 kpc behind \ohth. The observations taken with APEX and JCMT shown in Fig.~\ref{fig:co-contamination} are too contaminated to allow any meaningful study of the star and, indeed, a determination of the mass-loss rate was not carried out by \cite{De-Beck2010} for this reason. The IRAM 30m telescope has a smaller HPBW and hence appears to suffer less from ISM contamination, especially for the CO $J=2-1$ line, but it is clear from our ALMA observations, especially from inspection of the channel maps (Figs.~\ref{fig:cochans} and \ref{oh30-co-b3-chan}) that there is unavoidable contamination directly along the line of sight to \ohth. Even though a lot of this is resolved out by ALMA, some elements of the contaminating molecular cloud cannot be avoided owing to the small angular sizes of dense elements, comparable in size to the \ohth\ CO envelope extent.
Giant molecular clouds are generally expected to be relatively cold, so they are brighter in the lower-energy CO transitions. However, massive star-forming clumps have been shown to also exhibit high-$J$ CO lines \citep[e.g.][]{Hoang2023}. Without spatially resolved observations for all studied CO transitions, we cannot completely rule out contamination in any CO towards \ohth.

\subsection{Forensic astronomy}

Through a careful examination of past catalogues and other observations of \ohth, we have determined that there has historically been some uncertainty in the precise position of this star. Our ALMA continuum observations are of high astrometric accuracy (see Table \ref{tab:metry}) and represent the best measured position for \ohth. Some examples of past uncertainties are detailed below.

The study of \cite{Justtanont2013} included the IRAM observations shown in Fig.~\ref{fig:co-contamination} and two higher-energy CO lines observed with APEX, $J=7-6$ and $J=4-3$. Both of these higher-energy lines were not detected above the noise but also do not exhibit significant contamination, which is somewhat perplexing, since the $J=9-8$ and $J=5-4$ lines were clearly detected by \textsl{Herschel}/HIFI \citep[with some possible contamination seen in $J=5-4$, see Fig.~10 in][]{Justtanont2013}. 

\begin{figure*}
\includegraphics[width=0.9\textwidth]{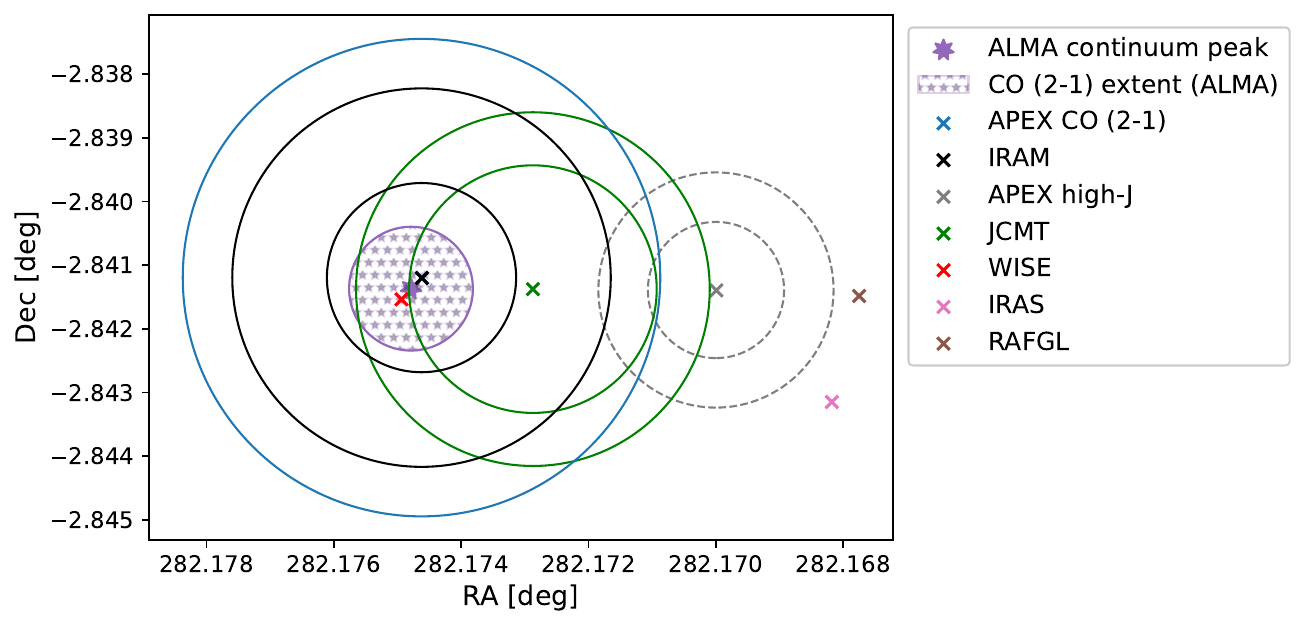}
\caption{Location on the sky of \ohth\ and the pointings of various observations targeting the star. The continuum peak based on our observations is plotted as a purple star and a purple shaded circular region with a radius of $3.5\arcsec$ represents the extent of the CO $J=2-1$ (see Fig.~\ref{fig:cochans}). The coordinates and HPBW of various other observations are also plotted (see text for references). Note that the centres of the APEX CO (2-1) and IRAM beams are at the same coordinates, as are the HIFI observations, the large beams for which are not plotted. Not shown are the coordinates given in the WISE, ISO and GCVS catalogues, which are all within $\sim1\arcsec$ of the ALMA continuum peak.}
    \label{fig:sd-beams}
\end{figure*}

To investigate this conundrum we checked the original observations discussed above and plot the HPBW and right ascension and declination (RA and Dec) of these observations in Fig.~\ref{fig:sd-beams}. We also checked various catalogues and past observations for the coordinates listed for \ohth\ and compared them with the RA and Dec we find for the continuum peak in this work (based on the Band 6 extended continuum, see Table \ref{tab:metry}). The majority of the listed coordinates for this star fall within $\lesssim 1\arcsec$ of our continuum peak, including those from the Wide-field Infrared Survey Explorer \citep[WISE,][]{Cutri2012}, the General Catalogue of Variable Stars \cite[GCVS 5.1,][]{Samus2017}, the Infrared Space Observatory \citep[ISO,][]{Kessler1996} and from the aforementioned IRAM, HIFI \citep{Justtanont2013}, and APEX ($J=2-1$, this work, ID:O-0102.F-9301B) observations. The JCMT CO $J=2-1$ and $J=3-2$ observations \citep{De-Beck2010} are offset from the ALMA continuum peak by about $7\arcsec$, meaning that the telescope beam at $J=2-1$ mostly covers the extent of the emission observed with ALMA, but the smaller $J=3-2$ misses a significant portion of the AGB emission. 

This is illustrated in Fig.~\ref{fig:sd-beams}, where we approximate the ALMA CO emission as a circle of radius $3.5\arcsec$ centred on the continuum peak and plot various telescope beams and their pointings. The relatively small APEX beams covering the CO $J=4-3$ and $J=7-6$ lines are centred on a point $\sim 17\arcsec$ from the ALMA continuum peak and, as shown in Fig.~\ref{fig:sd-beams}, do not overlap with the CO envelope at all. 

We also note that some older catalogues, such as the Infrared Astronomical Satellite \citep[IRAS,][]{IRAS1988} and the Revised Air Force Geophysical Laboratory Infrared Sky Survey \citep[RAFGL,][]{Price1983}, list coordinates even more significantly offset from our continuum peak (by $\sim 25\arcsec$ for both IRAS and RAFGL), most likely owing to resolution limitations. The coordinates listed in the study of \cite{Heske1990}, discussed in Sect.~\ref{sec:pastresults}, agree with our position to within 0.33\arcsec.


%
%
%
%
%

\bsp	
\label{lastpage}
\end{document}